\documentclass{ws-rv9x6}   
\usepackage{bm}
\usepackage{cite}

\newcommand{\la}{\left \langle}
\newcommand{\ra}{\right \rangle}

\newcommand{\nc}{\newcommand}
\nc{\3}{{\ss}}
\nc{\equ}{{\rm equ}}
\nc{\BC}{{_{\rm BC}}}
\nc{\WN}{{_{\rm WN}}}
\nc{\NN}{{_{\rm NN}}}
\nc{\bb}{{\bm{b}}}
\nc{\be}{{\bm{e}}}
\nc{\bs}{{\bm{s}}}
\nc{\bx}{{\bm{x}}}
\nc{\bv}{{\bm{v}}}
\nc{\bp}{{\bm{p}}}
\nc{\bq}{{\bm{q}}}
\nc{\br}{{\bm{r}}}
\nc{\bK}{{\bm{K}}}
\nc{\bKp}{{\bm{K}}_\perp}
\nc{\bvp}{{\bm{v}}_\perp}
\nc{\grad}{{\bm{\nabla}}}
\nc{\pt}{p_\mathrm{T}}
\nc{\kt}{k_\mathrm{T}}
\nc{\mt}{m_\mathrm{T}}
\nc{\Kt}{K_\mathrm{T}}
\nc{\Mt}{M_\mathrm{T}}
\nc{\pL}{p_\mathrm{L}}
\nc{\KET}{\mathrm{KE}_\mathrm{T}}
\nc{\Tcrit}{T_{\rm crit}}
\nc{\scm}{\sqrt{s_{\rm NN}}}
\nc{\Tmunu}{T^{\mu\nu}}
\nc{\gmunu}{g^{\mu\nu}}
\nc{\dmuumu}{\partial_\mu u^\mu}
\nc{\diag}{{\rm diag}}
\nc{\Tdec}{T_{\rm dec}}
\nc{\mT}{m_\mathrm{T}}
\nc{\pT}{p_\mathrm{T}}
\nc{\vT}{v_\mathrm{T}}
\nc{\tauequ}{\tau_{\rm equ}}
\nc{\eequ}{e_{\rm equ}}
\nc{\nequ}{n_{\rm equ}}
\nc{\sequ}{s_{\rm equ}}
\nc{\Tequ}{T_{\rm equ}}
\nc{\edec}{e_{\rm dec}}
\nc{\ecrit}{e_{\rm crit}}
\nc{\eqgp}{e_{\rm QGP}}
\nc{\dec}{{\rm dec}}
\nc{\const}{{\rm const}}

\nc{\Var}{{\rm Var}}
\nc{\fs}{{\rm fs}}
\nc{\kin}{{\rm kin}}
\nc{\form}{{\rm form}}
\nc{\hydro}{{\rm hydro}}
\nc{\ch}{{\rm ch}}
\nc{\sat}{{\rm sat}}
\nc{\half}{\frac{1}{2}}

\nc{\eq}{{\,=\,}}
\nc{\lla}{\la \la}
\nc{\rra}{\ra \ra}
\nc{\eps}{\epsilon}
\nc{\se}{\section}
\nc{\suse}{\subsection}
\nc{\sususe}{\subsubsection}
\nc{\beq}[1]{\begin{equation}\label{#1}}
\nc{\eeq}{\end{equation}}
\nc{\bea}[1]{\begin{eqnarray}\label{#1}}
\nc{\eea}{\end{eqnarray}}
\nc{\bce}{\begin{center}}
\nc{\ece}{\end{center}}

\renewcommand{\phi}{\varphi}
\renewcommand{\theta}{\vartheta}

\newcommand{\gapp}{\,{\raisebox{-.2ex}{$\stackrel{>}{_\sim}$}}\,}
\newcommand{\lapp}{\,{\raisebox{-.2ex}{$\stackrel{<}{_\sim}$}}\,}

\begin{document}

\setcounter{chapter}{0}

\chapter[Hydrodynamics and transport properties of QCD matter]
{Early collective expansion:\\ 
Relativistic hydrodynamics and the transport properties of QCD matter}

\author{Ulrich Heinz}
\address{Department of Physics, The Ohio State University,\\
         191 West Woodruff Avenue, Columbus, OH 43210, USA\\
         E-mail: \emph{\tt heinz@mps.ohio-state.edu}}

%\maketitle

\vspace*{-1cm}
\tableofcontents 
%
%\newpage

%%%%%%%%%%%%%%%%%%%%%%%%%%%%%%%%%%%%%%%%%%%%%%%%%%%%%%%%%%%%%%%%%%%%%%%%%%%%%
% introduction.tex
% last edited by UH on 1/24/09
% 
%%%%%%%%%%%%%%%%%%%%%%%%%%%%%%%%%%%%%%%%%%%%%%%%%%%%%%%%%%%%%%%%%%%%%%%%%%%%%

%%%%%%%%%%%%%%%%%%%%%%%%%%%%%%%%%%%%%%%%%%%%%%%%%%%%%%%%%%%%%%%%%%%%%%%%%%%%%
\section{Introduction}
\label{sec:introduction}
%%%%%%%%%%%%%%%%%%%%%%%%%%%%%%%%%%%%%%%%%%%%%%%%%%%%%%%%%%%%%%%%%%%%%%%%%%%%%

The idea of exploiting the laws of ideal hydrodynamics to describe
the expansion of the strongly interacting matter that is formed in 
high energy hadronic collisions was first formulated by Landau 
in 1953 \cite{Landau53}.
Because of their conceptual beauty and simplicity, models based 
on hydrodynamic principles have been applied to calculate a large 
number of observables for various colliding systems and over
a broad range of beam energies. 
However, it is by no means clear that the highly excited, but still 
small systems produced in those violent collisions satisfy the 
criteria justifying a dynamical treatment in terms of a macroscopic 
theory which follows idealized laws.
Indeed, the history of using hydrodynamics for high-energy phenomenology
is checkered, with qualitative successes overshadowed by quantitative 
failures.
Only recently, with data from the Relativistic Heavy Ion Collider (RHIC) 
at Brookhaven National Laboratory (c.f. the experimental 
reviews \cite{Arsene:2004fa,Back:2004je,Adams:2005dq,Adcox:2004mh}), came 
striking evidence for a strong collective expansion that is, for the first 
time, in good {\em quantitative} agreement with hydrodynamic predictions, 
at least for the largest collision systems (e.g. Au+Au) at the highest 
collision energies ($\sqrt{s}=200$\ GeV per nucleon pair) near midrapidity 
at small to moderate impact parameters.
The long list of qualifiers towards the end of the preceding sentence points 
to continuing limitations of hydrodynamics, at least in its idealized 
perfect fluid limit: dissipative effects become increasingly important 
for smaller collision systems, lower collision energies, larger impact 
parameters and when one moves away from midrapidity.
However, once well-calibrated ideal fluid dynamical benchmarks have 
been established under appropriate experimental conditions, deviations
from perfect fluid behaviour can be used to explore transport 
properties, such as viscosity and heat conduction, of the QCD matter
created in the collisions. 
Such efforts define the present forefront of research in heavy-ion 
collision dynamics.
The validity of ideal hydrodynamics requires local relaxation times
towards thermal equilibrium that are much shorter than any
macroscopic dynamical time scale.
The significance and importance of rapid thermalization 
of the created fireball matter cannot be over-stressed: 
Only if the system is close to local thermal equilibrium, its 
thermodynamic properties, such as its pressure, entropy density and 
temperature, are well defined. 
And only under these conditions can we pursue to study the equation of
state of strongly interacting matter at high temperatures.
This is particularly interesting in the light of the expected phase 
transition of strongly interacting matter which, at a critical energy
density of about 1~GeV/fm$^3$, undergoes a transition from a 
hadron resonance gas to a hot and dense plasma of color deconfined 
quarks and gluons.
Lattice QCD calculations indicate \cite{EKSM82,Karsch:2003jg,Aoki:2006br,%
Cheng:2006qk} that this transition takes place rather rapidly at a critical 
temperature $\Tcrit$ somewhere between 150 and 190 MeV.
In this article I review and discuss data and calculations that 
provide strong evidence that the created fireball matter reaches 
temperatures above $2\,\Tcrit$ and which indicate short 
thermalization times of order 1--2\ fm/$c$.
After a pedagogical introduction into the foundations of
relativistic hydrodynamics and of the relativistic fluid dynamic
equations for ideal and dissipative fluids, I discuss appropriate
initial and final conditions for the hydrodynamic expansion stage.
I describe a few important aspects of the fireball evolution in
central and non-central heavy-ion collisions and the calculation
of final hadron spectra.
Here the anisotropy of the final momentum spectra in non-central
collisions plays a particularly important role because it provides
both evidence for fast thermalization in RHIC collisions and access
to transport properties of the quark-gluon matter created early in
the collision.
Based on a comparison with experimental data, I delineate our
present knowledge (and its limits) of the properties of QCD matter 
created at RHIC, and outline future opportunities for quantitative 
improvement of our understanding of heavy-ion collision dynamics.   
%

%%%%%%%%%%%%%%%%%%%%%%%%%%%%%%%%%%%%%%%%%%%%%%%%%%%%%%%%%%%%%%%%%%%%%%%%%%%%%
% hydrodynamics.tex
% last edited by UH on 1/27/09
%%%%%%%%%%%%%%%%%%%%%%%%%%%%%%%%%%%%%%%%%%%%%%%%%%%%%%%%%%%%%%%%%%%%%%%%%%%%%

%%%%%%%%%%%%%%%%%%%%%%%%%%%%%%%%%%%%%%%%%%%%%%%%%%%%%%%%%%%%%%%%%%%%%%%%%%%%%
\section{The equations of relativistic hydrodynamics}
\label{sec:hydrodynamics}
%%%%%%%%%%%%%%%%%%%%%%%%%%%%%%%%%%%%%%%%%%%%%%%%%%%%%%%%%%%%%%%%%%%%%%%%%%%%%
\suse{Ideal fluid dynamics for perfect fluids}
\label{sec:ideal}
%%%%%%%%%%%%%%%%%%%%%%%%%%%%%%%%%%%%%%%%%%%%%%%%%%%%%%%%%%%%%%%%%%%%%%%%%%%%%

Any fluid dynamical approach starts from the local conservation laws for 
energy-momentum and any conserved charges:
\begin{eqnarray}
\label{eq2}
\partial_\mu T^{\mu\nu} &=& 0,
\\
\label{eq1}
\partial_\mu N^\mu_i &=& 0, \quad i=1,\dots,k.
\end{eqnarray}
For simplicity we restrict ourselves to $k{\,=\,}1$ (say, $N^\mu=$ 
net baryon number current). One must also ensure the second law of 
thermodynamics
\begin{equation}
\label{eq3}
\partial_\mu S^\mu \geq 0,
\end{equation}
where $S^\mu$ is the entropy current. Ideal fluid dynamics follows from
these equations under the assumption of local thermal equilibrium, i.e.
if the microscopic collision time scale is very much shorter than any 
macroscopic evolution time scale such that the underlying phase-space 
distribution $f(x,p)$ relaxes essentially instantaneously to the local
equilibrium form (upper signs for fermions, lower signs for bosons)
\begin{equation}
\label{eq4}
f_{\rm eq}(x,p) = \frac{1}{e^{[p{\cdot}u(x)+\mu(x)]/T(x)}\pm 1}.
\end{equation}
Here $u^\mu(x)$ is the local fluid velocity at point $x$, $\mu(x)$ is 
the local chemical potential associated with the conserved charge 
$N$ (it enters with opposite sign in the distribution $\bar f$ for
antiparticles), and $T(x)$ is the local temperature. Plugging this into 
the kinetic theory definitions 
\begin{eqnarray}
\label{eq5}
N^\mu(x) &=& \frac{1}{(2\pi)^3}\sum_i n_i \int \frac{d^3p}{E} p^\mu f_i(x,p),
\\
\label{eq6}
T^{\mu\nu}(x) &=& \frac{1}{(2\pi)^3}\sum_i 
                  \int \frac{d^3p}{E} p^\mu p^\nu f_i(x,p),
\\
\label{eq7}
S^\mu(x) &=& - \frac{1}{(2\pi)^3}\sum_i 
               \int \frac{d^3p}{E} p^\mu \Bigl[f_i(x,p)\ln f_i(x,p)
%\\
%\nonumber
%&&\quad
         \pm\bigl(1{\mp}f_i(x,p)\bigr)\ln\bigl(1{\mp}f_i(x,p)\bigr)\Bigr],
\end{eqnarray}  
(where the sum is over all particle species (counting particles and 
antiparticles separately) and $n_i$ is the amount of conserved charge $N$
carried by species $i$) leads to the ideal fluid decompositions
\begin{eqnarray}
\label{eq8}
&&N_{\rm eq}^\mu = n\, u^\mu,
\\
\label{eq9}
&&T_{\rm eq}^{\mu\nu} = e\,u^\mu u^\nu - p\, \Delta^{\mu\nu} \qquad 
(\mbox{with}\ \ \Delta^{\mu\nu}{=}g^{\mu\nu}{-}u^\mu u^\nu),\qquad
\\
\label{eq10}
&&S_{\rm eq}^\mu = s\, u^\mu,
\end{eqnarray}  
where the local net charge density $n$, energy density $e$, pressure $p$
and entropy density $s$ are given by the standard integrals over
the thermal equilibrium distribution function in the local fluid
rest frame. They are related by the fundamental thermodynamic relation
\begin{equation}
\label{eq11}
T\,s = p - \mu\, n + e.
\end{equation}
Inserting Eqs.~(\ref{eq8})-(\ref{eq10}) into Eqs.~(\ref{eq1}) and (\ref{eq2}) 
yields the relativistic ideal fluid equations shown in 
Eqs.~(\ref{eq12})-(\ref{eq14}) below. Using Eq.~(\ref{eq11}) together with 
the Gibbs-Duhem relation $dp=s\,dT+n\,d\mu$, it is easy to prove that, 
in the absence of shock discontinuities, these equations also conserve 
entropy, i.e. $\partial_\mu S^\mu=0$.
Note that the validity of the decompositions (\ref{eq8})-(\ref{eq10}) only
requires local momentum isotropy (i.e. that in the local fluid rest frame 
the phase-space distribution reduces to a function of energy $E$ only, 
$f(x,p)=f\bigl(p{\cdot}u(x);T(x),\mu(x)\bigr)$), but not that the 
distribution function has the specific exponential form (\ref{eq4}) 
that maximizes entropy.
This may have relevance in situations where the time scale for local 
momentum isotropization is much shorter than for 
thermalization \cite{ALM03,BBW04,RRS04} (i.e. it is much easier 
to change the direction of the particles'\ momenta than their energies), 
with the macroscopic hydrodynamic time scale in between.%
\footnote{In the absence of such a clear separation of time scales 
entropy production can not be neglected during the macroscopic 
evolution, and ideal fluid dynamics must be replaced by dissipative 
fluid dynamics. 
Furthermore, rapid longitudinal expansion at early times causes strong 
viscous effects that act {\em against} rapid local isotropization
\cite{Song:2007ux} of the momentum distribution. 
Ideal fluid dynamics becomes valid only after these viscous effects 
have died away.}
In this case the local microscopic states would not maximize entropy, 
and the relation (\ref{eq11}) between the quantities $e,\,p,\,n,$ and 
$s$ defined through Eqs.~(\ref{eq5})-(\ref{eq10}) would not hold.
Still, these quantities would follow ideal fluid dynamical evolution since 
entropy production by microscopic kinetic energy-shifting processes would 
only happen on time scales that are large compared to the macroscopic 
evolution time scales.  
The ideal fluid equations read (with $\theta\equiv\partial{\cdot}u$ denoting
the local expansion rate and $c_s^2=\frac{\partial p}{\partial e}$ the 
squared speed of sound)
\begin{eqnarray}
\label{eq12}
&&\dot n = - n\, \theta,
\\
\label{eq13}
&&\dot e = - (e+p)\, \theta,
\\
\label{eq14}
&&{\dot u}^\mu = \frac{\nabla^\mu p}{e+p} 
               = \frac{c_s^2}{1+c_s^2}\frac{\nabla^\mu e}{e}.
\end{eqnarray}
Here we decomposed the partial derivative 
$\partial^\mu=u^\mu D +\nabla^\mu$ into ``longitudinal'' 
and ``transverse'' components $D=u^\nu\partial_\nu$ and 
$\nabla^\mu=\Delta^{\mu\nu}\partial_\nu$, which in the local 
fluid rest frame reduce to the time derivative $\dot f \equiv Df$ and 
spatial gradient $\bm{\nabla}f$. 
The first two equations describe the dilution of the local baryon and 
energy density due to the local expansion rate $\theta$, while the 
third describes the acceleration of the fluid by the spatial pressure 
gradients in the local rest frame, with the enthalpy $e{+}p$ acting as 
inertia. 
The second equality in Eq.~(\ref{eq14}) exhibits the manifest scale 
invariance of the ideal fluid dynamical equations (the absolute 
normalization of the energy density profile drops out) and demonstrates 
that the dynamical ``pushing power'' of the medium is related to the 
``stiffness'' $\frac{\partial p}{\partial e}$ of its {\em Equation of 
State (EOS)} $p=p(e,n)$, reflected in the (temperature-dependent) 
speed of sound $c_s(T)$.
Together with the EOS, the 5 equations (\ref{eq12})--(\ref{eq14}) form a 
closed set from which the fields $n,\,e,\,p(n,e)$ and $u^\mu$ (with 
$u^\mu u_\mu{\,=\,}1$) can be determined.
%

%%%%%%%%%%%%%%%%%%%%%%%%%%%%%%%%%%%%%%%%%%%%%%%%%%%%%%%%%%%%%%%%%%%%%%%%%%%%
\suse{Dissipative fluid dynamics for viscous relativistic fluids}
\label{sec3}
%%%%%%%%%%%%%%%%%%%%%%%%%%%%%%%%%%%%%%%%%%%%%%%%%%%%%%%%%%%%%%%%%%%%%%%%%%%%
As the hydrodynamic evolution changes the local energy and baryon density,
microscopic processes attempt to readjust the local phase-space distribution 
to corresponding new local temperatures and chemical potentials. If this
does not happen fast enough, the phase-space distribution will
start to deviate from its local equilibrium form (\ref{eq4}): 
$f(x,p)=f_{\rm eq}\bigl(p{\cdot}u(x);T(x),\mu(x)\bigr) + \delta f(x,p)$.
The optimal values for the (readjusted) local temperature and chemical
potential in the first term are fixed by imposing the ``Landau matching 
conditions'' \cite{LL59}
\begin{eqnarray}
\label{eq15}
&&u_\mu\, \delta T^{\mu\nu} u_\nu = \int \frac{d^3p}{E}\, (u{\cdot}p)^2\,
\delta f(x,p)=0,\qquad
%\\
%\label{eq15a}
%&& 
 u_\mu \,\delta N^\mu =  \int \frac{d^3p}{E}\, (u{\cdot}p) \,
\delta f(x,p)=0.
\end{eqnarray}
The remaining deviations $\delta f$ from local equilibrium generate
additional terms in the decompositions of $N^\mu,\,T^{\mu\nu},$ and
$S^\mu$:
\begin{eqnarray}
\label{eq16}
&&N^\mu = N^\mu_{\rm eq} +\delta N^\mu = n\,u^\mu + V^\mu,
\\
\label{eq17}
&&T^{\mu\nu} = T_{\rm eq}^{\mu\nu} + \delta T^{\mu\nu}
           = e\,u^\mu u^\nu -(p+\Pi)\Delta^{\mu\nu}
               + \pi^{\mu\nu}
%\qquad
%\nonumber\\
%&&\qquad\qquad\qquad\qquad\quad
   + W^\mu u^\nu +W^\nu u^\mu,
\\
\label{eq18}
&&S^\mu = S^\mu_{\rm eq} +\delta S^\mu = n\,u^\mu + \Phi^\mu.
\end{eqnarray}
The new terms describe a baryon flow $V^\mu{\,=\,}\Delta^{\mu\nu} N_\nu$ 
in the local rest frame, an energy flow $W^\mu{\,=\,}\frac{e{+}p}{n} 
V^\mu{\,+\,}q^\mu$ (where $q^\mu$ is the ``heat flow vector'') in the 
local rest frame, the viscous bulk pressure 
$\Pi{\,=\,}{-}\frac{1}{3}\Delta_{\mu\nu}T^{\mu\nu}{\,-\,}p$ (which contributes
to the trace of the energy momentum tensor), the traceless viscous
shear pressure tensor $\pi^{\mu\nu}= T^{\langle\mu\nu\rangle} 
\equiv \left[\frac{1}{2}\left(\Delta^{\mu\sigma}
\Delta^{\nu\tau}{+}\Delta^{\nu\sigma}\Delta^{\mu\tau}\right)-\frac{1}{3}
\Delta^{\mu\nu}\Delta^{\sigma\tau}\right]T_{\tau\sigma}$
(where the expression $\langle\mu\nu\rangle$ is a shorthand
for ``traceless and transverse to $u_\mu$ and $u_\nu$'', as defined
by the projector in square brackets), and an entropy flow vector
$\Phi^\mu$ in the local rest frame.
The matching conditions (\ref{eq15}) leave the choice 
of the local rest frame velocity $u^\mu$ ambiguous.
This ambiguity can be used to eliminate either $V^\mu$ from 
Eq.~(\ref{eq16}) (``Eckart frame'' $u^\mu = N^\mu/\sqrt{N{\cdot}N}=N^\mu/n$, 
no baryon flow in the local rest frame \cite{Eckart}), in which case the 
energy flow reduces to the heat flow vector $W^\mu{\,=\,}q^\mu$, or 
$W^\mu$ from Eq.~(\ref{eq17}) (``Landau frame'' $u^\mu = T^{\mu\nu} u_\nu
\Big/\sqrt{u_\alpha T^{\alpha\beta} T_{\beta\gamma} u^\gamma}=
T^{\mu\nu} u_\nu/e$, corresponding to no energy flow in the local rest 
frame, $u_\mu\delta T^{\mu\nu}=0$ \cite{LL59}), in which case there is 
a non-zero baryon flow $V^\mu{\,=\,}-\frac{n}{e{+}p}q^\mu$ due to heat 
conduction in the local rest frame. 
%
%(Intermediate frames are also possible, but yield no practical advantage.) 
%
For systems with vanishing net baryon number (as approximately realized 
in RHIC collisions) the Eckart frame is ill-defined and heat conduction 
disappears as an independent transport effect \cite{Danielewicz:1984ww}, 
so we will use the Landau frame.
Inserting the decomposition (\ref{eq18}) into the conservation law 
(\ref{eq2}) and projecting onto time-like and space-like components 
yields the {\em non-ideal fluid equations} for baryon-free systems in 
the Landau frame
\begin{eqnarray}
\label{eq22}
&&\!\!\!\!\! \dot e = - (e{+}p{+}\Pi)\, \theta 
           +\pi_{\mu\nu}\nabla^{\left\langle\mu\right.}
                             u^{\left.\nu\right\rangle},
\\
\label{eq23}
&&\!\!\!\!\! (e{+}p{+}\Pi)\,{\dot u}^\mu = \nabla^\mu(p{+}\Pi) 
                             -\Delta^{\mu\nu}\nabla^\sigma\pi_{\nu\sigma}
             +\pi^{\mu\nu}{\dot u}_\nu.
\qquad 
\end{eqnarray}
The non-equilibrium decompositions (\ref{eq16})-(\ref{eq18}) involve
1+3+5=9 additional dynamical quantities, the ``dissipative flows''
$\Pi,\,q^\mu$, and $\pi^{\mu\nu}$ (the counting reflects their 
transversality to $u^\mu$ and the tracelessness of $\pi^{\mu\nu}$). 
This means that we need 9 additional dynamical equations which 
should be compatible with the underlying transport theory for the 
non-equilibrium deviation $\delta f(x,p)$. 
For the baryon-free case without heat conduction, 
Eqs.~(\ref{eq22})--(\ref{eq23}), the number of needed additional 
equations reduces to 6.

%%%%%%%%%%%%%%%%%%%%%%%%%%%%%%%%%%%%%%%%%%%%%%%%%%%%%%%%%%%%%%%%%%%%%%%%%%%%
\suse{Transport equations for the dissipative flows}
\label{sec4}
%%%%%%%%%%%%%%%%%%%%%%%%%%%%%%%%%%%%%%%%%%%%%%%%%%%%%%%%%%%%%%%%%%%%%%%%%%%%
%
The key property of the kinetic equation governing the evolution of
the phase-space distribution function $f{\,=\,}f_{\rm eq}{+}\delta f$
is that the collision term satisfies the second law of thermodynamics
(\ref{eq3}), i.e. entropy is produced until the system has reached a 
new state of local thermal equilibrium. Here, we don't want to solve the
kinetic theory; instead, we want to write down a phenomenological 
macroscopic theory which is consistent with the constraints arising from
the underlying kinetic theory, in particular the 2$^{\rm nd}$ law. The
macroscopic theory will be constructed from an expansion of the
entropy production rate in terms of the dissipative flows which 
themselves are proportional to the off-equilibrium deviation $\delta f$
of the phase-space distribution \cite{Israel:1976tn,Israel:1979wp}. 
Assuming the latter to be small, $|\delta f|{\,\ll\,}|f_{\rm eq}|$, 
this expansion will be truncated at some low order in the dissipative 
flows $\delta N^\mu,\,\delta T^{\mu\nu}$. The expansion will involve 
phenomenological expansion coefficients which, in principle, should be 
matched to the kinetic theory \cite{Israel:1979wp,Baier:2006um,York:2008rr}.
In practice, they will often be considered as phenomenological parameters 
to be adjusted to experimental data. In the end, the extracted values 
must then be checked for consistency with the entire approach, by making 
sure that the dissipative corrections are indeed sufficiently small to 
justify truncation of the expansion {\em a posteriori}.
The equilibrium identity (\ref{eq11}) can be rewritten as
\begin{equation}
\label{eq24}
  S_{\rm eq}^\mu = p(\alpha,\beta)\beta^\mu - \alpha N_{\rm eq}^\mu
  +\beta_\nu T_{\rm eq}^{\nu\mu},
\end{equation}
where $\alpha{\,\equiv\,}\frac{\mu}{T},\,\beta{\,\equiv\,}\frac{1}{T},$ and 
$\beta_\nu{\,\equiv\,}\frac{u_\nu}{T}$. The most general off-equilibrium
generalization of this is \cite{Israel:1979wp}
\begin{eqnarray}
\label{eq25}
S^\mu &\equiv& S_{\rm eq}^\mu + \Phi^\mu
%\\\nonumber
= p(\alpha,\beta)\beta^\mu - 
\alpha N^\mu +\beta_\nu T^{\nu\mu} + Q^\mu(\delta N^\mu,\delta T^{\mu\nu}),
\end{eqnarray}
where, in addition to the first order contributions implicit in
the second and third terms of the r.h.s., $Q^\mu$ includes terms which
are second and higher order in the dissipative flows $\delta N^\mu$
and $\delta T^{\mu\nu}$.

The form of the expansion (\ref{eq25}) is constrained by the
2$^{\rm nd}$ law $\partial_\mu S^\mu{\,\geq\,}0$. To evaluate this 
constraint it is useful to rewrite the Gibbs-Duhem relation 
$dp{\,=\,}s\,dT+n\,d\mu$ as
\begin{equation}
\label{eq26}
\partial_\mu\left( p(\alpha,\beta)\beta^\mu\right) = 
N_{\rm eq}^\mu\partial_\mu\alpha - T_{\rm eq}^{\mu\nu}\partial_\mu\beta_\nu.
\end{equation}
With additional help from the conservation laws (\ref{eq1})and (\ref{eq2}), 
the entropy production then becomes
\begin{equation}
\label{eq27}
\partial_\mu S^\mu = - \delta N^\mu\partial_\mu\alpha
+ \delta T^{\mu\nu}\partial_\mu\beta_\nu +\partial_\mu Q^\mu.
\end{equation}
Using Eqs.~(\ref{eq16},\ref{eq17}) to express $\delta N^\mu$ and $\delta 
T^{\mu\nu}$ in terms of the scalar, vector and tensor dissipative flows
$\Pi,\,q^\mu,$ and $\pi^{\mu\nu}$, and introducing corresponding scalar,
vector and tensor thermodynamic forces (in terms of gradients 
of the thermodynamic equilibrium variables) which drive these dissipative 
flows \cite{deGroot},
 $X{\,\equiv\,}{-}\theta{\,=\,}{-}\nabla{\cdot}u$, \ 
$X^\nu{\,\equiv\,}\frac{\nabla^\nu T}{T}-{\dot u}^\nu = {-}\frac{nT}{e{+}p}\,
\nabla^\nu\!\left(\frac{\mu}{T}\right)$, and
$X^{\mu\nu}{\,\equiv\,}\nabla^{\left\langle\mu\right.}
 u^{\left.\nu\right\rangle}$ (note that $X^{\mu\nu}{=}X^{\langle\mu\nu\rangle}$
is traceless and transverse to $u^\mu$), the 2$^{\rm nd}$ law constraint 
can be further recast into
\begin{equation}
\label{eq28}
 T \partial_\mu S^\mu = \Pi X - q^\mu X_\mu +\pi^{\mu\nu} X_{\mu\nu}
+ T \partial_\mu Q^\mu \geq 0.
\end{equation}
%
%Note that the first three terms on the r.h.s. are first order while the 
%last term is higher order in the dissipative flows.

%%%%%%%%%%%%%%%%%%%%%%%%%%%%%%%%%%%%%%%%%%%%%%%%%%%%%%%%%%%%%%%%%%%%%%%%%%
\subsubsection{Standard dissipative fluid dynamics (first-order or 
Navier-Stokes theory)}
\label{sec4a}
%%%%%%%%%%%%%%%%%%%%%%%%%%%%%%%%%%%%%%%%%%%%%%%%%%%%%%%%%%%%%%%%%%%%%%%%%%
%
The standard approach \cite{LL59} neglects the higher order contributions 
to the entropy current and sets $Q^\mu{\,=\,}0$. 
The inequality (\ref{eq28}) can then always be satisfied by postulating 
linear relationships between the dissipative flows and the thermodynamic 
forces (``Navier-Stokes relations''),
\begin{eqnarray}
\label{eq29}
  \Pi=-\zeta\theta,\qquad
  q^\nu=-\lambda \frac{nT^2}{e{+}p}\,\nabla^\nu\!\left(\frac{\mu}{T}\right),
  \qquad
  \pi^{\mu\nu} = 2\, \eta\, \nabla^{\left\langle\mu\right.}
  u^{\left.\nu\right\rangle} \equiv 2\, \eta\,\sigma^{\mu\nu},
\end{eqnarray}
with positive {\em transport coefficients} $\zeta{\,\geq\,}0$ (bulk
viscosity),  $\lambda{\,\geq\,}0$ (heat conductivity), and 
$\eta{\,\geq\,}0$ (shear viscosity):
\begin{equation}
\label{eq30}
T \partial{\cdot}S = \frac{\Pi^2}{\zeta} - \frac{q^\alpha q_\alpha}{2\lambda T}
+ \frac{\pi^{\alpha\beta}\pi_{\alpha\beta}}{2\eta}\geq 0.
\end{equation}
(The minus sign in front of the second term is necessary because $q^\mu$,
being orthogonal to $u^\mu$, is spacelike, $q^2<0$.)
Equations~(\ref{eq29}) are the desired 9 equations for the dissipative flows.
Note that the entropy production rate (\ref{eq30}) is of second order
in the dissipative flows.

Unfortunately, using these relations in the hydrodynamic equations 
(\ref{eq22})-(\ref{eq23}) leads to hydrodynamic evolution with acausal 
signal propagation: if in a given fluid cell at a certain time a 
thermodynamic force happens to vanish, the corresponding dissipative
flow also stops instantaneously. 
This contradicts the fact that the flows result from the forces through 
microscopic scattering which involves relaxation on a finite albeit 
short kinetic time scale.
To avoid this type of acausal behaviour one must keep $Q^\mu$.

%%%%%%%%%%%%%%%%%%%%%%%%%%%%%%%%%%%%%%%%%%%%%%%%%%%%%%%%%%%%%%%%%%%%%%%%%%
\subsubsection{Second-order Israel-Stewart theory}
\label{sec4b}
%%%%%%%%%%%%%%%%%%%%%%%%%%%%%%%%%%%%%%%%%%%%%%%%%%%%%%%%%%%%%%%%%%%%%%%%%%
%
A causal theory of dissipative relativistic fluid dynamics is obtained
by keeping $Q^\mu$ up to terms which are second order in the irreversible 
flows \cite{Israel:1979wp,IMuller}. 
For simplicity we here consider only the baryon-free case 
$n{\,=\,}q^\mu{\,=\,}0$; see \cite{Israel:1979wp,Muronga:2001zk} for a 
general treatment.  
One writes
\begin{equation}
\label{eq31}
  Q^\mu = -\left(\beta_0\Pi^2 + \beta_2\pi_{\nu\lambda}\pi^{\nu\lambda}\right)
  \frac{u^\mu}{2T}
\end{equation}
(with phenomenological expansion coefficients $\beta_0,\,\beta_2$) and 
computes (after some algebra using similar techniques as before) the
entropy production rate as
\begin{eqnarray}
\label{eq32}
  T \partial{\cdot}S &=& \Pi \left[ -\theta -\beta_0 \dot\Pi - \Pi T 
  \partial_\mu\left(\frac{\beta_0 u^\mu}{2T}\right)\right]
%\\\nonumber
  + \pi^{\alpha\beta}\left[\sigma_{\alpha\beta}
  - \beta_2{\dot\pi}_{\alpha\beta} -\pi_{\alpha\beta} T 
  \partial_\mu\left(\frac{\beta_2 u^\mu}{2T}\right)
  \right],
\end{eqnarray}
where $\sigma_{\alpha\beta}$ is the flow shear tensor defined in the
last equation (\ref{eq29}).
From the expressions in the square brackets we see that the thermodynamic 
forces $-\theta$ and $\sigma_{\alpha\beta}$ are now modified by terms 
including the time derivatives (in the local rest frame) of the 
irreversible flows $\Pi$, $\pi_{\alpha\beta}$. 
This leads to dynamical (``transport'') equations for the latter. We can 
ensure the 2$^{\rm nd}$ law of thermodynamics by again writing the entropy 
production rate in the form (\ref{eq30}) (without the middle term), which 
amounts to postulating 
\begin{eqnarray}
\label{eq33}
\dot \Pi &=& -\frac{1}{\tau_{_\Pi}}\left[ \Pi +\zeta \theta +
\Pi\zeta T \partial_\mu\left(\frac{\tau_{_\Pi} u^\mu}{2\zeta T}\right)\right]
 \approx -\frac{1}{\tau_{_\Pi}}\bigl[ \Pi +\zeta \theta\bigr],
\\
\label{eq34}
\Delta_{\alpha\mu}\Delta_{\beta\nu}
\dot \pi^{\mu\nu} &=& -\frac{1}{\tau_\pi}\left[ \pi_{\alpha\beta}
- 2 \eta \sigma_{\alpha\beta} + \pi_{\alpha\beta} \eta T 
\partial_\mu\left(\frac{\tau_\pi u^\mu}{2\eta T}\right)\right]
\approx -\frac{1}{\tau_\pi}\left[ \pi_{\alpha\beta}  
- 2 \eta \sigma_{\alpha\beta} \right].
\end{eqnarray}
Here we replaced the coefficients $\beta_{0,2}$ by the relaxation times 
$\tau_{_\Pi}{\,\equiv\,}\zeta\beta_0$ and $\tau_\pi{\,\equiv\,}2\eta\beta_2$.
In principle both $(\zeta,\eta)$ and $(\tau_{_\Pi},\tau_\pi)$ should be 
calculated from the underlying kinetic theory. 
We will use them as phenomenological parameters, noting that for 
consistency the microscopic relaxation rates should be much larger 
than the local hydrodynamic expansion rate, 
$\tau_{_{\pi,\Pi}}\theta{\,\ll\,}1$.

The approximation in the second equalities in Eqs.~(\ref{eq33},\ref{eq34}) 
neglects terms that are of combined second order in dissipative
flows and gradients of the zeroth-order hydrodynamic 
quantities \cite{Israel:1979wp}.
Generically, it is good at early times $\tau{-}\tau_0 \lapp \tau_\pi,
\,\tau_\Pi$. 
During this time, $\dot\Pi$ and $\dot\pi^{\mu\nu}$ are of first order in 
deviations from equilibrium (i.e. of the same order as $\Pi,\,\pi^{\mu\nu}$ 
themselves as well as $\theta$ and $\sigma_{\alpha\beta}$), and 
Eqs.~(\ref{eq33},\ref{eq34}) describe an exponential relaxation (on time
scales $\tau_\pi,\,\tau_\Pi$) of the dissipative flows towards their 
Navier-Stokes values (\ref{eq29}). (The projectors $\Delta_{\mu\nu}$ on
the l.h.s. of Eq.~(\ref{eq34}) ensure the preservation of tracelessness
and transversality to $u$ of the shear pressure tensor during time evolution 
and can be rewritten as additional source terms on the r.h.s. of this 
equation \cite{Baier:2006um}.)
Once the differences between the dissipative flows and their Navier-Stokes 
limits have dropped enough to become comparable in magnitude to the
second-order terms in Eqs.~(\ref{eq33},\ref{eq34}) (i.e. the third terms
in the square brackets), $\dot\Pi$ and $\dot\pi^{\mu\nu}$ must be
counted as being of second order in deviations from local equilibrium,
and their further evolution is essentially affected by the second-order
driving terms on the right hand sides of Eqs.~(\ref{eq33},\ref{eq34}).
For  $\tau{-}\tau_0 \gg \tau_\pi,\,\tau_\Pi$, the approximations 
indicated in the last equalities in Eqs.~(\ref{eq33},\ref{eq34}) thus
break down. 
Heavy-ion collisions with longitudinally boost invariant initial 
conditions present an anomalous situation in that the longitudinal
expansion rate diverges like $1/\tau$ at early times.
As a result, the third terms in the square brackets of 
Eqs.~(\ref{eq33},\ref{eq34}) cannot even be neglected at early
times \cite{Huovinen:2008te} and must be kept throughout the evolution.
If this is not done, serious deviations are observed when comparing
Israel-Stewart viscous hydrodynamics with microscopic kinetic 
simulations \cite{Huovinen:2008te}, and one risks violating the
second law of thermodynamics.
To get a feeling for the role played by the second-order terms for
the time evolution of the dissipative flows, we rewrite 
Eqs.~(\ref{eq33},\ref{eq34}) (including these terms) as modified 
relaxation equations: 
\begin{eqnarray}
\label{eq35}
\dot \Pi &=& -\frac{1}{\tau_{_\Pi}}\Bigl[ \Pi +\zeta \theta +
\Pi\zeta \gamma_{_\Pi}\Bigr] 
= -\frac{1{+}\gamma_{_\Pi}\zeta}{\tau_{_\Pi}}
\left[ \Pi + \frac{\zeta}{1{+}\gamma_{_\Pi}\zeta}\,\theta\right]
= -\frac{1}{\tau'_{_\Pi}}\Bigl[ \Pi +\zeta'\, \theta\Bigr]
\end{eqnarray}
(where $\gamma_{_\Pi}{\,\equiv\,}T \partial_\mu\left(\frac{\tau_{_\Pi} 
u^\mu}{2\zeta T}\right)$), and similarly for the shear pressure tensor.
One sees that the second-order term in the first square bracket modifies 
both the kinetic relaxation time and the viscosity, by an amount 
${\sim}\gamma_\Pi$ that involves the macroscopic expansion rate 
$\partial_\mu u^\mu$. 
In regions of rapid hydrodynamic expansion and/or large shear flow,
this effectively lowers both the Navier-Stokes limits of the dissipative
flows and the relaxation times for approaching them, thereby effectively
limiting excursions of the dissipative flows away from their Navier-Stokes
limits \cite{Song:2008si,Denicol:2008ua}.
Numerical studies \cite{Song:2008si} show that this reduces the sensitivity 
of final physical observables to the choice of $\tau_\pi,\,\tau_\Pi$.
Equations (\ref{eq33},\ref{eq34}), through the introduction of non-zero 
microscopic relaxation times $\tau_\pi,\,\tau_\Pi$, thus resolve the 
issues with acausal signal propagation and numerical instability of 
the relativistic Navier-Stokes equations (at least for modes with 
macroscopic wave lengths $\lambda>c\tau_\pi$) while largely preserving 
their physics content.
In the second order Israel-Stewart formalism, one solves the dissipative 
hydrodynamic equations (\ref{eq22},\ref{eq23}) simultaneously with kinetic 
relaxation equations of the type (\ref{eq33},\ref{eq34}) for the 
irreversible flows. 
The second-order terms displayed on the right hand sides of 
Eqs.~(\ref{eq33},\ref{eq34}) do not exhaust all terms that one 
could write down based on symmetries and tensor structure 
alone \cite{Baier:2007ix}. 
Indeed, microscopic derivations of the dissipative corrections to the 
ideal-fluid decomposition (\ref{eq9}) of the energy-momentum tensor, 
starting from Boltzmann kinetic theory for the distribution function 
$f(x,p)$ and expanding it around the local equilibrium form (\ref{eq4}), 
produce many more second-order terms than obtained from the macroscopic 
approach described here \cite{York:2008rr,Baier:2007ix,Betz:2008me}.
In Boltzmann theory the coefficients of all second-order terms 
are found to be proportional to (powers of) the microscopic relaxation 
time $\tau_\pi$ \cite{York:2008rr}.  
While this is an active area of research, it is expected that within 
the range of applicability of Israel-Stewart theory the exact values of 
these coefficients are practically irrelevant, i.e. that physical
observables show little sensitivity to the value of $\tau_\pi$ and
to the choice of second-order terms (other than those that can be 
derived macroscopically) that are included.   

%%%%%%%%%%%%%%%%%%%%%%%%%%%%%%%%%%%%%%%%%%%%%%%%%%%%%%%%%%%%%%%%%%%%%%%%%%%%%
\section{The beginning and end of the hydrodynamic stage in heavy-ion 
collisions}
\label{sec:init_and_fin}
%%%%%%%%%%%%%%%%%%%%%%%%%%%%%%%%%%%%%%%%%%%%%%%%%%%%%%%%%%%%%%%%%%%%%%%%%%%%%
%
Hydrodynamics does not rest on the availability of an underlying kinetic 
theory in terms of colliding particles, but it does require the system to 
be close to local thermal equilibrium (a concept that can be formulated 
even for strongly coupled quantum systems that are too hot and dense to 
allow for a particle description because large scattering rates never let 
any of the particles go on-shell). 
Hydrodynamics can never be expected to describe the earliest stage of the 
collision, just after nuclear impact, during which a fraction of the 
energy stored in the initial coherent motion along the beam direction 
is redirected into the transverse directions and randomized.
The results of this initial thermalization process enter the hydrodynamic 
description through {\em initial conditions} for $T^{\mu\nu}(x)$, i.e. 
for the macroscopic density and (dissipative) flow distributions, 
implemented at a suitable starting time $\tau_0$ for the hydrodynamic 
evolution.
If a microscopic description of the early pre-equilibrium stage based on
first principles is available, these initial conditions can be calculated
from the pre-equilibrium energy-momentum tensor by matching it to the
form (\ref{eq17}) (with $W^\mu\equiv0$) through Landau matching 
conditions, as described in Sec.~\ref{sec3}.
Presently there is no sufficiently mature pre-equilibrium description
available, so initial conditions for the ideal fluid components of
$T^{\mu\nu}$ are adjusted to experimental data for final observables
in {\em central} collisions and then extrapolated to {\em non-central} 
collisions using geometric considerations. 
Central collisions provide more data than necessary for adjusting the
hydrodynamic initial conditions:
As we will see in Sec.~\ref{sec:initialization}, a complete initialization 
requires the total charged hadron multiplicity density at midrapidity
$(dN_\mathrm{ch}/dy)(y{=}0)$, its dependence on centrality, and the 
shapes of the transverse momentum spectra of two hadron species with 
very different masses. 
The spectra of all other hadron species from central collisions, as 
well as all spectra (including their anisotropies) from non-central 
collisions can thus be considered as tests for the validity of the 
hydrodynamic model.
As the fluid evolves hydrodynamically, there is continuous competition
between the local rate of expansion, which drives the system away from
equilibrium, and microscopic relaxation processes attempting to restore 
local equilibrium.   
For longitudinally boost-invariant initial conditions that best reflect
our present understanding of the microscopic initial particle production
processes at high collision energy, the expansion rate 
$\tau_\mathrm{exp}^{-1}=\partial_\mu u^\mu$ is $\sim 1/\tau$, where
$\tau=\sqrt{t^2{-}z^2}$ is the longitudinal proper time after
nuclear impact. 
It is huge at very early times but decreases rapidly.
On the other hand, all of the local scattering rates (elastic and 
inelastic) are proportional to the local temperature $T(x)$, 
$\tau_\mathrm{scatt}^{-1}\sim T$, $T$ being the only dimensionful 
quantity in a thermalized system of (approximately) massless quarks 
and gluons.
For boost-invariant longitudinal expansion temperature decreases with time 
as $T\sim\tau^{-1/3}$, i.e. more slowly than the expansion rate. 
Hence, the ratio $\tau_{\rm scatt}/\tau_\mathrm{exp}$ initially decreases 
with time, improving the conditions for local thermalization.
As time proceeds, transverse flow is generated and the initially entirely 
longitudinal expansion eventually turns 3-dimensional. 
For 3-d expansion, the temperature decreases like $1/\tau$ (due to
rela\-tivistic effects even somewhat faster), thus the scattering rate 
now decreases in lockstep with the expansion rate.
If by then the system has not reached local thermal equilibrium, it
never will.
Below the quark-hadron phase transition the conditions for local 
thermalization deteriorate quickly since now, due to finite hadron
masses, the density falls exponentially with temperature while, as a 
result of color confinement, the hadronic scattering cross sections 
saturate and become temperature independent. 
Once the mean collision time becomes larger than the local 
``Hubble time'' $\tau_\mathrm{exp}=1/\partial{\cdot}u$, the system
quickly falls out of equilibrium \cite{BGZ78,HLR87,LRH88,HS98}, turning 
into a gas of free-streaming hadrons soon afterwards.
This ``decoupling process'' defines the end of the hydrodynamic evolution.
In the next two subsections we discuss beginning and end of
the hydrodynamic stage in more detail.
%

%%%%%%%%%%%%%%%%%%%%%%%%%%%%%%%%%%%%%%%%%%%%%%%%%%%%%%%%%%%%%%%%%%%%%%%%
%   SUBSECTION: INITIALIZATION
%   last updated: 1/25/09 
%%%%%%%%%%%%%%%%%%%%%%%%%%%%%%%%%%%%%%%%%%%%%%%%%%%%%%%%%%%%%%%%%%%%%%%%
%%%%%%%%%%%%%%%%%%%%%%%%%%%%%%%%%%%%%%%%%%%%%%%%%%%%%%%%%%%%%%%%%%%%%%%%
\suse{Initialization}
\label{sec:initialization}
%%%%%%%%%%%%%%%%%%%%%%%%%%%%%%%%%%%%%%%%%%%%%%%%%%%%%%%%%%%%%%%%%%%%%%%%
%
Lacking a microscopic theory of the early pre-equilibrium evolution,
initial profiles for hydrodynamics are usually parametrized 
geometrically, with normalization parameters adjusted to final 
observables in central heavy-ion collisions.
Ideal fluid simulations for heavy-ion collisions at RHIC 
energies have been performed in 2+1 dimensions \cite{KSH99,Teaney:1999gr,%
KSH00,KHHH01,TLS01b,HKHRV01,KHHET01,TLS01,TLS02,PFKthesis02,HK02,HK02HBTosci,%
KR03,Kolb:2003gq,Kolb:2003dz,Chojnacki:2006tv,Broniowski:2008vp} and in 
3+1 dimensions \cite{Morita:1999vj,NHM00,Hirano01v2etaRHIC,Hirano:2001yi,%
Morita:2002av,HT02,Nonaka:2004pg,Hirano:2005xf,Nonaka:2006yn}.
(The first number indicates the number of spatial dimensions, the +1 
stands for time.)
Viscous hydrodynamic simulations \cite{Heinz:2005bw,Chaudhuri:2005ea,%
Baier:2006um,Chaudhuri:2006jd,Baier:2006gy,Romatschke:2007jx,%
Chaudhuri:2007zm,Romatschke:2007mq,Song:2007fn,Song:2007ux,Chaudhuri:2008sj,%
Luzum:2008cw,Denicol:2008rj,Song:2008si,Heinz:2008qm,Denicol:2008ha,%
Denicol:2008ua,Song:2008hj} have up to now been restricted to at most 
2+1 dimensions.
The (2+1)-d simulations assume longitudinal boost-invariance, 
i.e. initial density profiles that do not depend on space-time rapidity 
$\eta_s=\frac{1}{2}\ln[(z{+}t)/(z{-}t)]$ (where $z$ is the beam 
direction), whereas the (3+1)-d simulations make no such assumption.
Both types of simulations assume an initial longitudinal expansion 
velocity profile with boost-invariance, $y_L=\eta_s$, where 
$y_L=\frac{1}{2}\ln[(1{+}v_z)/(1{-}v_z)]$ is the fluid rapidity in 
beam direction. 
(All velocities $v$ are in units of $c$.)
This corresponds to an initial longitudinal flow velocity profile 
$v_z=z/t$, as suggested by an initial particle production process
that, at infinite collision energy, is independent of longitudinal 
reference frame and depends only on the longitudinal proper time 
$\tau$ (i.e. the time in the particles'\ longitudinal rest 
frame \cite{Bjorken83}.).
In the (2+1)-d simulations, the identity $y_L=\eta_s$ holds for
all times $\tau$, due to boost-invariant initial densities and the
resulting absence of longitudinal pressure gradients 
$\partial_{\eta_s}p$ \cite{Bjorken83}.
In the (3+1)-d simulations, non-vanishing longitudinal pressure
gradients $\partial_{\eta_s}p\ne 0$ lead to longitudinal acceleration
of the fluid, i.e. the longitudinal density profiles $e(\tau,x,y,\eta_s)$
etc. broaden with time.
Due to the logarithmic nature of the rapidity variable, at large
values of $\eta_s$ even small shifts in rapidity require large
changes in longitudinal momentum, so these rapidity-broadening effects
are limited and, at RHIC and LHC energies, typically well below one
unit of rapidity.
The (3+1)-d simulations require input for the initial space-time rapidity 
profiles of the energy density $e$ and baryon density $n$. 
They are adjusted to the final rapidity distributions of pions and 
protons in momentum space. 
Due to the limited rapidity evolution just mentioned and the assumed 
initial identity of $\eta_s$ and $y_L$, the initial space-time rapidity
distributions of $e$ and $n$ look very similar to the final momentum-space 
rapidity distributions of pions ($\pi^\pm$) and net protons ($p$-$\bar p$), 
respectively.
The initial space-time rapidity density profiles can be taken 
independent of transverse position $\bm{r_\perp}$ relative to the 
beam axis \cite{Nonaka:2006yn} or, more realistically, 
$\bm{r_\perp}$-dependent \cite{Hirano:2005xf}.
This makes little difference for the final charged hadron rapidity
distributions, but matters for a correct description of the rapidity 
distributions of net baryons and elliptic flow as a function of
collision centrality.  
For computing the initial transverse distributions of energy and 
baryon density, two leading models are on the market: The Glauber 
model \cite{Glauber59,BBC76}, and the Kharzeev-Levin-Nardi (KLN) model
\cite{Kharzeev:2001gp,Hirano:2004en,Kharzeev:2004if,Drescher:2006pi,%
Lappi:2006xc,Drescher:2006ca} based on the Color Glass Condensate (CGC) 
theory. 
Since these define (in a sense detailed below) the outer limits of viable 
initializations, both will be briefly outlined in the following.
Although not discussed here in more detail, also other initializations, 
such as the pQCD + final state saturation model (EKRT) \cite{EKR97,%
Eskola:1999fc}, have been applied for hydrodynamics at RHIC and LHC 
\cite{Eskola:2001bf,Eskola:2002wx,Eskola:2005ue}.
The initial transverse collective flow velocity is typically assumed 
as zero.
This makes sense if the hydrodynamic stage starts early, at times
$\tau_0 \ll 1$\,fm.
For later starting times, some pre-equilibrium transverse flow should
be allowed for and has been introduced in some simulations to
improve the agreement with experimental data \cite{PFKthesis02,KR03,%
Vredevoogd:2008id,Pratt:2008qv,Broniowski:2008qk}.
However, lacking guidance from {\em ab initio} pre-equilibrium 
calculations, it is difficult to accurately determine the initial
transverse flow phenomenologically.
%

\begin{figure} 
\centerline{
\epsfig{file=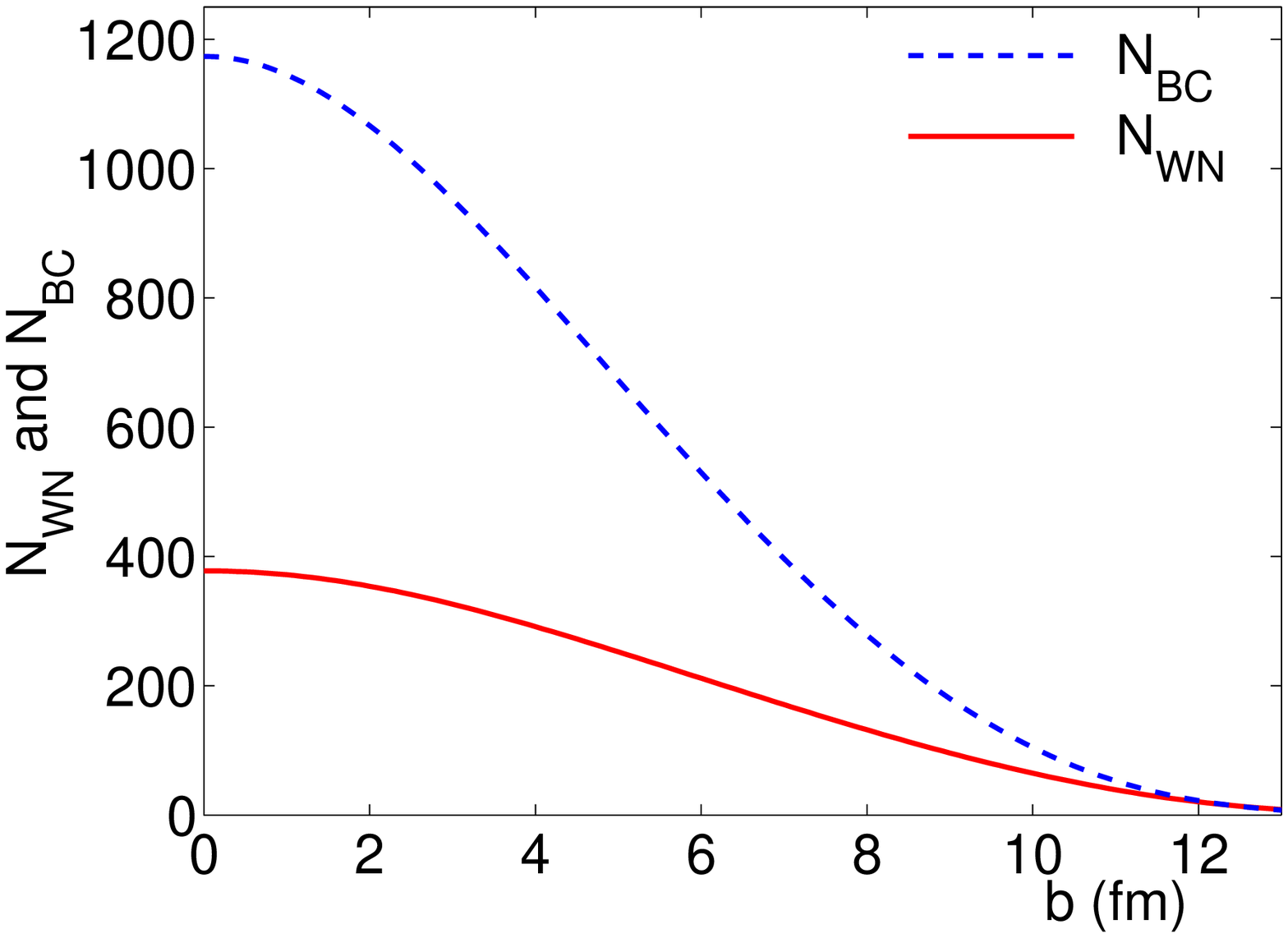,width=6cm}%\hfill
\hspace*{10mm}
\epsfig{file=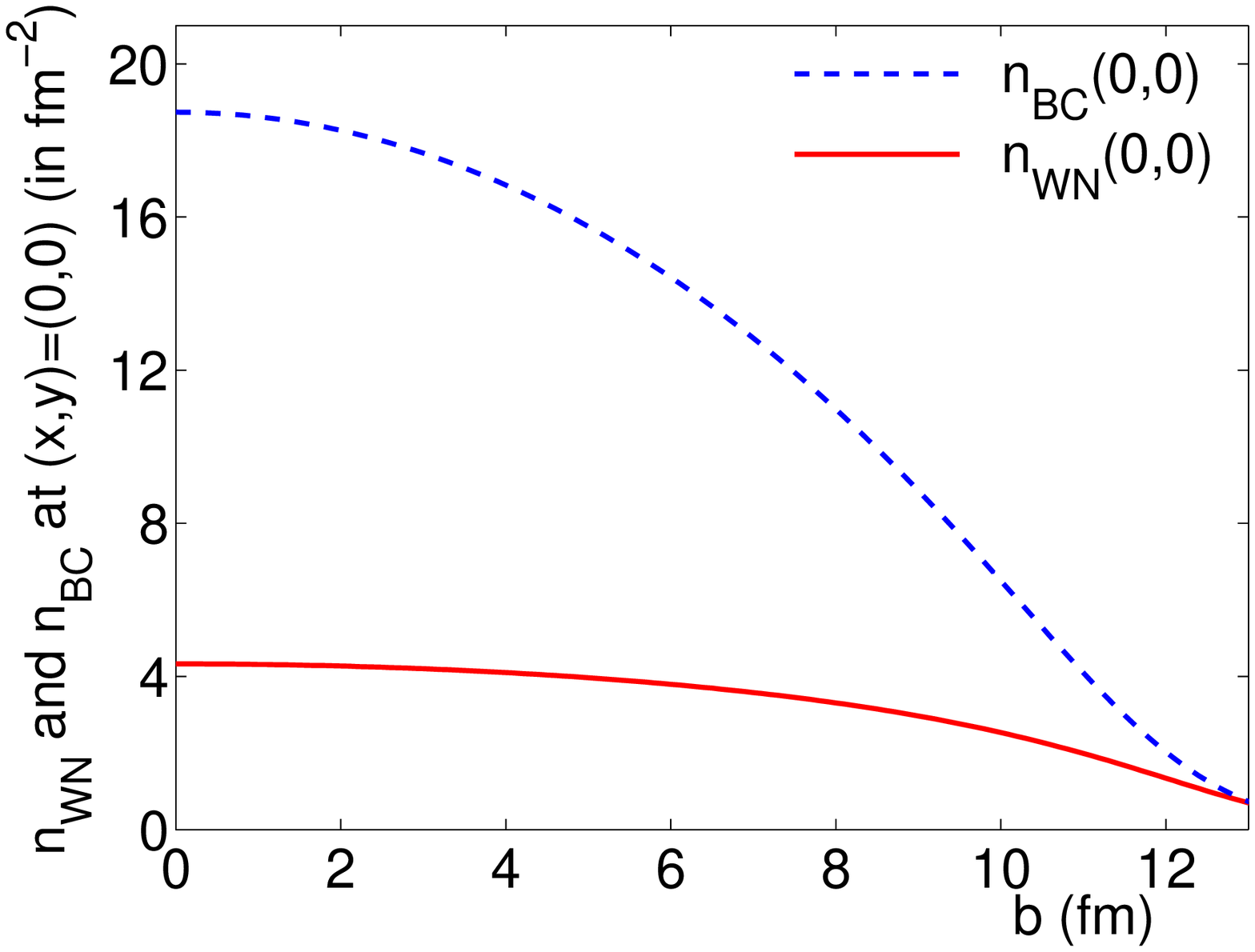,width=6cm}
}  
\vspace*{-10mm}
\caption{{\sl Left:} Number of wounded nucleons and binary collisions as a
         function of impact parameter, for Au+Au collisions 
         $\sqrt{s}=130\,A$\,GeV ($\sigma_0=40$\,mb). 
         {\sl Right:} Density of wounded nucleons and binary collisions in 
         the center of the collision as a function of impact parameter. 
}
\label{fig:NWNNBCnWNnBC} 
\end{figure}

%%%%%%%%%%%%%%%%%%%%%%%%%%%%%%%%%%%%%%%%%%%%%%%%%%%%%%%%%%%%%%%%%%%%%%%%%
\sususe{Glauber model}
\label{Glauber}
%%%%%%%%%%%%%%%%%%%%%%%%%%%%%%%%%%%%%%%%%%%%%%%%%%%%%%%%%%%%%%%%%%%%%%%%%
%
The microscopic processes that generate the initial entropy are still 
poorly understood.
Before the advent of the Color Glass Condensate theory (c.f. 
Ref. \cite{McLerran:2008pe} and references therein) which describes the 
initial transverse distribution as a dense gluon
system characterized by an $\bm{x}_\perp$-dependent saturation momentum
$Q_s(\bm{x}_\perp,\eta_s)$ (see Sec.~\ref{CGC}), the only available
model was the Glauber model which assumes that initial entropy production 
is controlled by some combination of wounded nucleon and binary 
nucleon-nucleon collision distributions \cite{Glauber59}:
\bea{alpha}
%\label{alpha}
  s(\bm{x}_\perp,\tau_0;b)= \kappa_s\bigl(\alpha n_\WN(\bm{x}_\perp;b)
  +(1{-}\alpha)n_\BC(\bm{x}_\perp;b)\bigr)
\end{eqnarray}
One assumes that ``soft'' processes scale with the number of wounded
nucleons per unit transverse area while ``hard'' processes scale with
the areal density of binary collisions.
The soft fraction $\alpha$ and the overall normalization are adjusted 
such \cite{KHHET01,KN01} that the experimentally observed rapidity density 
of charged hadrons at the end of the collision \cite{PHOBOS02dN130,%
PHOBOS02dN200} and its dependence on the collision centrality 
\cite{PHENIX01dNcent,PHOBOS02dNcent} are reproduced \cite{KHHET01,%
Kolb:2003dz,Hirano:2005xf,Kuhlman:2005ts}.
To compute these distributions in the transverse plane one starts from 
Saxon-Woods profiles describing the density 
distributions of the colliding nuclei with mass numbers $A$ and $B$,
respectively,
\beq{equ:WoodsSaxon}
\rho_A(r) = \frac{\rho_0}{e^{(r-R_A)/\xi}+1}\, ,
\end{equation}
with nuclear radius $R_A{\,=\,}(1.12\,A^{1/3}{-}0.86\,A^{-1/3})$\,fm 
and surface diffuseness $\xi{\,=\,}0.54$\,fm \cite{BM69}.
The nuclear thickness function is defined as the optical path-length 
through the nucleus along the beam direction:
\beq{equ:thickness}
T_A(x,y)=\int_{-\infty}^{\infty} dz \, \rho_A(x,y,z).
\end{equation}
The coordinates $x,y$ parametrize the transverse plane, with $x$
pointing in the direction of the impact parameter $\bb$ (such that
$(x,z)$ span the reaction plane) and $y$ perpendicular to the 
reaction plane.
For a non-central collision with impact parameter $b$, the density of 
binary nucleon-nucleon collisions $n_\BC$ at a point $(x,y)$ in the 
transverse plane is proportional to the product of the two nuclear 
thickness functions, transversally displaced by $b$:
\beq{equ:nBC}
n_\BC(x,y;b)=\sigma_0 \, T_A(x+b/2,y)\,T_B(x-b/2,y).
\end{equation}
$\sigma_0$ is the total inelastic nucleon-nucleon cross section; it 
enters here only as a multiplicative factor which is later absorbed 
in the proportionality constant between $n_\BC(x,y;b)$ and the ``hard'' 
component of the initial entropy deposition \cite{KHHET01}.
Integration over the transverse plane (the $(x,y)$-plane)
yields the total number of binary collisions
\beq{equ:NBC}
N_\BC(b)=\int dx\,dy \; n_\BC(x,y;\,b). 
\end{equation}
Its impact parameter dependence, as well as that of the maximum density 
of binary collisions in the center of the reaction zone, $n_\BC(0,0;b)$,
are shown as the dashed lines in Fig.~\ref{fig:NWNNBCnWNnBC}. 

The ``soft'' part of the initial entropy deposition is assumed to scale
with the density of ``wounded nucleons'' \cite{BBC76},
defined as those nucleons in the projectile and target which 
{\em participate} in the particle production process by suffering
at least one collision with a nucleon from the other nucleus.
The Glauber model gives the following transverse density distribution 
of wounded nucleons \cite{BBC76}:
%
%\begin{eqnarray}
%\label{equ:nWN}
\bea{equ:nWN}
 n_\WN(x,y;b) 
       = T_A(x+b/2,y) 
         \left( 1- \left( 1 - \frac{\sigma_0 T_B(x-b/2,y)}{B}\right)^B\right)
 \nonumber \\
         + \;
         T_B(x-b/2,y) 
         \left( 1- \left( 1 - \frac{\sigma_0 T_A(x+b/2,y)}{A}\right)^A\right).
\end{eqnarray}
Here the value $\sigma_0$ of the total inelastic nucleon-nucleon
cross section plays a more important role since it influences the 
shape of the transverse density distribution $n_\WN(x,y;b)$, and
its dependence \cite{PDG02} on the collision energy $\sqrt{s}$ must 
be taken into account.
The total number of wounded nucleons is obtained by integrating
Eq.~(\ref{equ:nWN}) over the transverse plane.
Its impact parameter dependence, as well as that of the maximum density 
of wounded nucleons in the center of the reaction zone, $n_\WN(0,0;b)$,
are shown as the solid lines in Fig.~\ref{fig:NWNNBCnWNnBC}.
%

%
%%%%%%%%%%%%%%%%%%%%%%%% Fig. 1 %%%%%%%%%%%%%%%%%%%%%%%%%%%%%%%%%%%%%%%%%%%
%%%%%%%%%%%%%%%%%%%%%%%%%%%%%%%%%%%%%%%%%%%%%%%%%%%%%%%%%%%%%%%%%%%%%%%%%%%
%

%
Hydrodynamic calculations with soft fraction $\alpha=0.75-0.85$,
i.e. with initial conditions that ascribe between 75 and 85\% of the 
initial entropy production to ``soft'' processes (scaling with 
$n_\WN(x,y;b)$) and 15--25\% to ``hard'' processes (scaling with 
$n_\BC(x,y;b)$), were found \cite{KHHET01,KN01,Hirano:2005xf,Kuhlman:2005ts} 
to give reasonable descriptions of the measured \cite{PHENIX01dNcent,%
PHOBOS02dNcent} centrality dependence of charged particles produced per 
participating (``wounded'') nucleon.
For simplicity and lack of other information, the initial tranverse
distributions of baryon and entropy density are often assumed to have 
the same shape (i.e. the entropy per baryon is constant in the transverse
plane), but other calculations simply set the net baryon density proportional 
to the density of wounded nucleons.
Their relative normalization is controlled by the net proton to pion
ratio at midrapidity.
At midrapidity, the net baryon density is small, and the mentioned 
differences in the initial transverse baryon density profile do not 
matter.
For (3+1)-d calculations, phenomenology requires that the entropy per
baryon decreases at forward rapidities; the $\eta_s$-dependence of
$s/n$ is thus another parameter in such simulations that needs adjusting.  
Entropy conservation in ideal fluid dynamics allows to fix the 
normalizations of the initial entropy density profile from measurements 
of the total charged hadron multiplicity $dN_\mathrm{ch}/dy$ (which is 
a measure of the total final entropy per unit rapidity) in central 
collisions \cite{KHHET01}.
It is natural to assume that the entropy produced per wounded nucleon or
per binary nucleon-nucleon collision depends only on collision energy but
not on collision geometry.
In this case, once their normalization has been fixed in central collisions,
normalization and shape of the initial density distributions in peripheral
collisions are predicted without additional parameters.
%

%%%%%%%%%%%%%%%%%%%%%%%%%%%%%%%%%%%%%%%%%%%%%%%%%%%%%%%%%%%%%%%%%%%%%%%%%
\sususe{Color Glass Condensate theory and KLN model}
\label{CGC}
%%%%%%%%%%%%%%%%%%%%%%%%%%%%%%%%%%%%%%%%%%%%%%%%%%%%%%%%%%%%%%%%%%%%%%%%%
%
The second type of initial conditions described here is based on the CGC 
model \cite{McLerran:2008pe}.
For simplicity, I describe the original Kharzeev-Levin-Nardi (KLN) 
approach \cite{Kharzeev:2001gp,Hirano:2004en,Hirano:2005xf} even though 
a somewhat improved version has recently been developed
\cite{Drescher:2006pi,Lappi:2006xc,Drescher:2006ca}.
In this approach, the energy distribution of produced gluons with 
rapidity $y$ is given by the $k_T$-factorization formula \cite{GLR83}
\bea{eq:ktfac}
   \frac{dE_T}{d^2x_{\perp}dy}&=&
   \frac{4\pi^2N_c}{N_c^2-1} \int\frac{d^2\pt}{\pt}
   \int^{\pt} d^2\kt \,\alpha_s(Q^2)\,   
    \phi_A(x_1,\kt^2;\bm{x}_\perp)\, \phi_B(x_2,(\pt{-}\kt)^2;\bm{x}_\perp), 
\end{eqnarray}
where $x_{1,2}{\eq}\pt\exp(\pm y)/\sqrt{s}$ are the longitudinal 
momentum fractions of the fusing gluons from nucleus $A$ and $B$, 
and $\pt$ is the transverse momentum of the produced gluon. 
For the unintegrated gluon distribution function one uses
\begin{equation}
\label{eq:uninteg}
  \phi_A(x,k^2_T;\bm{x}_\perp)
  =\left\{\begin{array}{l}
   \frac{\kappa C_F}{2\pi^3\alpha_s(Q^2_s)}\frac{Q_s^2}{Q_s^2+\Lambda^2}, 
   \,                      \quad  k_T\,\leq\,Q_s, \\
   \frac{\kappa C_F}{2\pi^3\alpha_s(Q^2_s)}\, \frac{Q^2_s}{k^2_T+\Lambda^2},
              \quad k_T\,>\,Q_s,
\end{array}
\right.
\end{equation}
where $C_F\eq\frac{N_c^2{-}1}{2N_c}$, $Q_s$ denotes the gluon saturation 
momentum, and $\Lambda=0.2$\,GeV is a soft regulator. 
The overall normalization $\kappa$ is determined by fitting the 
multiplicity of charged hadrons at midrapidity at $\sqrt{s_{NN}}=200$\,GeV 
for the most central collisions. 
The saturation momentum $Q_s$ of nucleus $A$ in $A{+}B$ collisions, needed 
in the function $\phi_A$, is obtained by solving the following implicit 
equation at fixed momentum fraction $x$ and transverse position 
$\bm{x}_{\perp}$:
\begin{equation}
\label{eq:saturation}
   Q^2_s(x, \bm{x}_{\perp}) = \frac{2\pi^2}{C_F}
   \,\alpha_s(Q^2_s)\,x G(x,Q^2_s)\,
   \frac{dN^A_{\mathrm{part}}}{d^2 x_\perp}.
\end{equation}
Here $dN^A_{\mathrm{part}}/d^2x_\perp\equiv n_\WN^A(\bm{x}_\perp)$ is
the transverse density of wounded nucleons in nucleus $A$, given by
the first term in Eq.~(\ref{equ:nWN}).
An analogous equation holds for the saturation momentum of nucleus $B$
in $\phi_B$. 
For the gluon distribution function $G$ inside a nucleon one takes the 
simple ansatz \cite{Kharzeev:2001gp}
\begin{equation}
\label{eq:xG}
  xG(x,Q^2) = K\ln\left( \frac{Q_s^2 + \Lambda^2}
               {\Lambda_{\mathrm{QCD}}^2}\right)
	       x^{-\lambda} (1-x)^4
\end{equation}
with $\Lambda\eq\Lambda_{\mathrm{QCD}}\eq0.2$\,GeV. 
Choosing $K\eq0.7$ and $\lambda\eq0.2$ ensures that the average saturation 
momentum in the transverse plane yields 
$\langle Q_s^2(x{=}0.01)\rangle{\,\sim\,}2.0$\,GeV$^2/c^2$
in central 200\,$A$\,GeV Au+Au collisions at RHIC. 
For the running coupling constant $\alpha_s$ in (\ref{eq:saturation}) 
one uses the standard perturbative one-loop formula with an additional 
cut-off in the infra-red region of small $Q_s$ (i.e. near the surface of 
the nuclear overlap region where the produced gluon density is low), by 
limiting the coupling constant to $\alpha_s{\,\leq\,}0.5$.
From Eq.~(\ref{eq:ktfac}) one obtains the energy density distribution at 
time $\tau_0$ as $e(\tau_0,\bm{x}_\perp,\eta_s){\eq}dE_T/(\tau_0 
d\eta_s d^2x_\perp)$, where $y$ is identified with $\eta_s$.
The KLN model predicts a centrality dependence of the produced charged
hadron multiplicity per wounded nucleon that agrees with RHIC
measurements \cite{Kharzeev:2001gp,Drescher:2006ca}.
A similar dependence can be obtained in the Glauber model by judicious
choice of the ``soft'' fraction $\alpha$ (see Eq.~(\ref{alpha})).
The main prediction of the CGC approach is the near independence of 
$\alpha$ of the collision energy, which is so far confirmed by 
experiment. 
%

%%%%%%%%%%%%%%%%%%%%%%%%%%%%%%%%%%%%%%%%%%%%%%%%%%%%%%%%%%%%%%%%%%%%%%%%%
\sususe{Non-central collisions and initial fireball eccentricity}
\label{ecc}
%%%%%%%%%%%%%%%%%%%%%%%%%%%%%%%%%%%%%%%%%%%%%%%%%%%%%%%%%%%%%%%%%%%%%%%%%
%
  
%
A key feature of non-central collisions between large nuclei is that
they produce deformed fireballs. 
This breaks the azimuthal symmetry inherent in central collisions
between spherical nuclei. 
In a strongly interacting fireball, the initial geometric anisotropy of 
the reaction zone gets transferred onto the final momentum spectra and
thus becomes experimentally accessible. 
As we will see, this provides a window into the very early collision 
stages that central collisions between spherical nuclei do not provide. 
Full-overlap collisions between {\em deformed} nuclei, such as
U, allow to explore the same physics with better resolution and
higher initial energy densities \cite{Heinz:2004ir}, but this 
requires careful event selection \cite{Kuhlman:2005ts}. 
%

%
%%%%%%%%%%%%%%%%%%% Fig. 2 %%%%%%%%%%%%%%%%%%%%%%%%%%%%%%%%%%%%%%%%%%%%
\begin{figure} 
\centerline{\epsfig{file=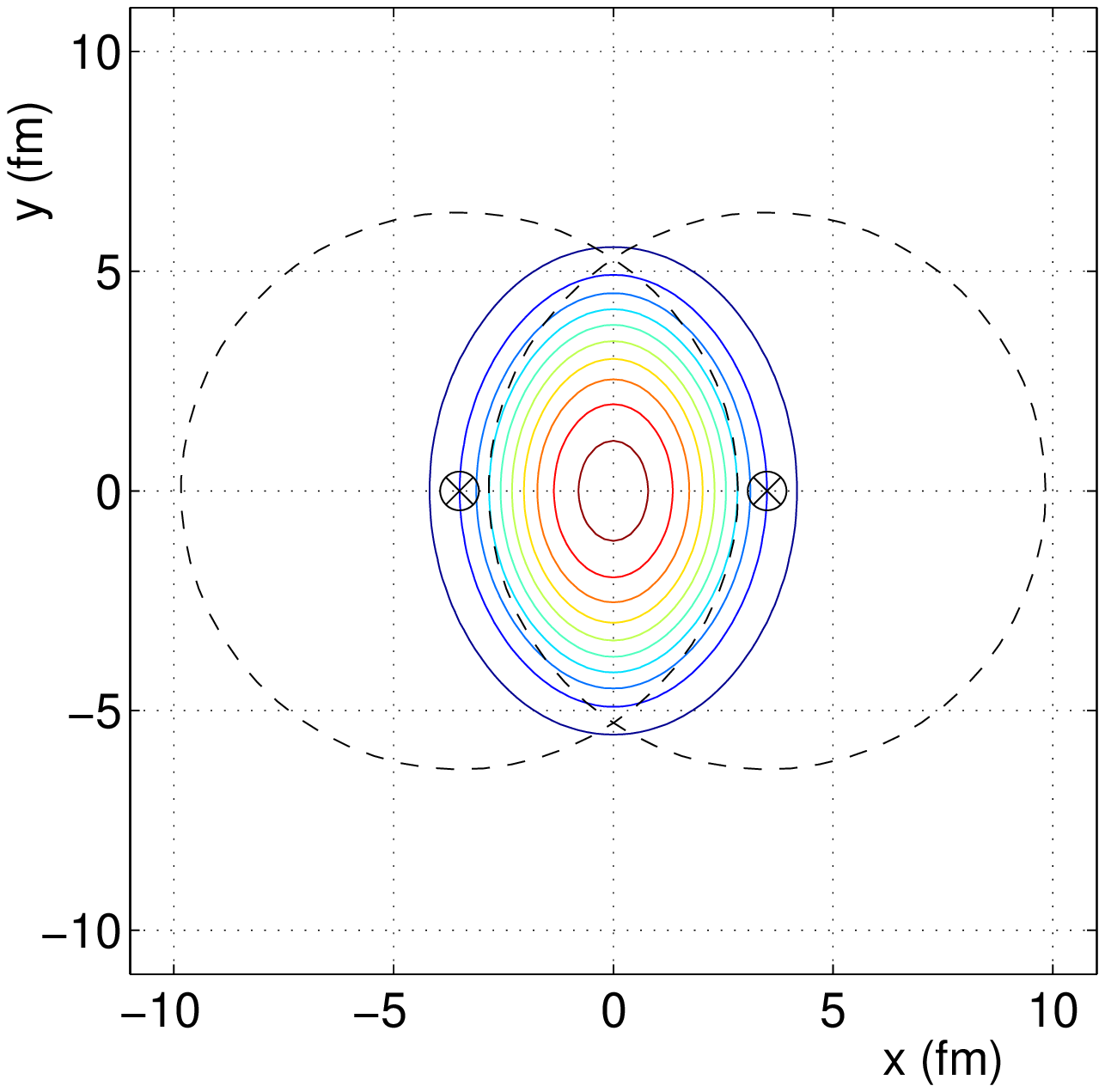,width=5cm}
\hspace*{10mm}
            \epsfig{file=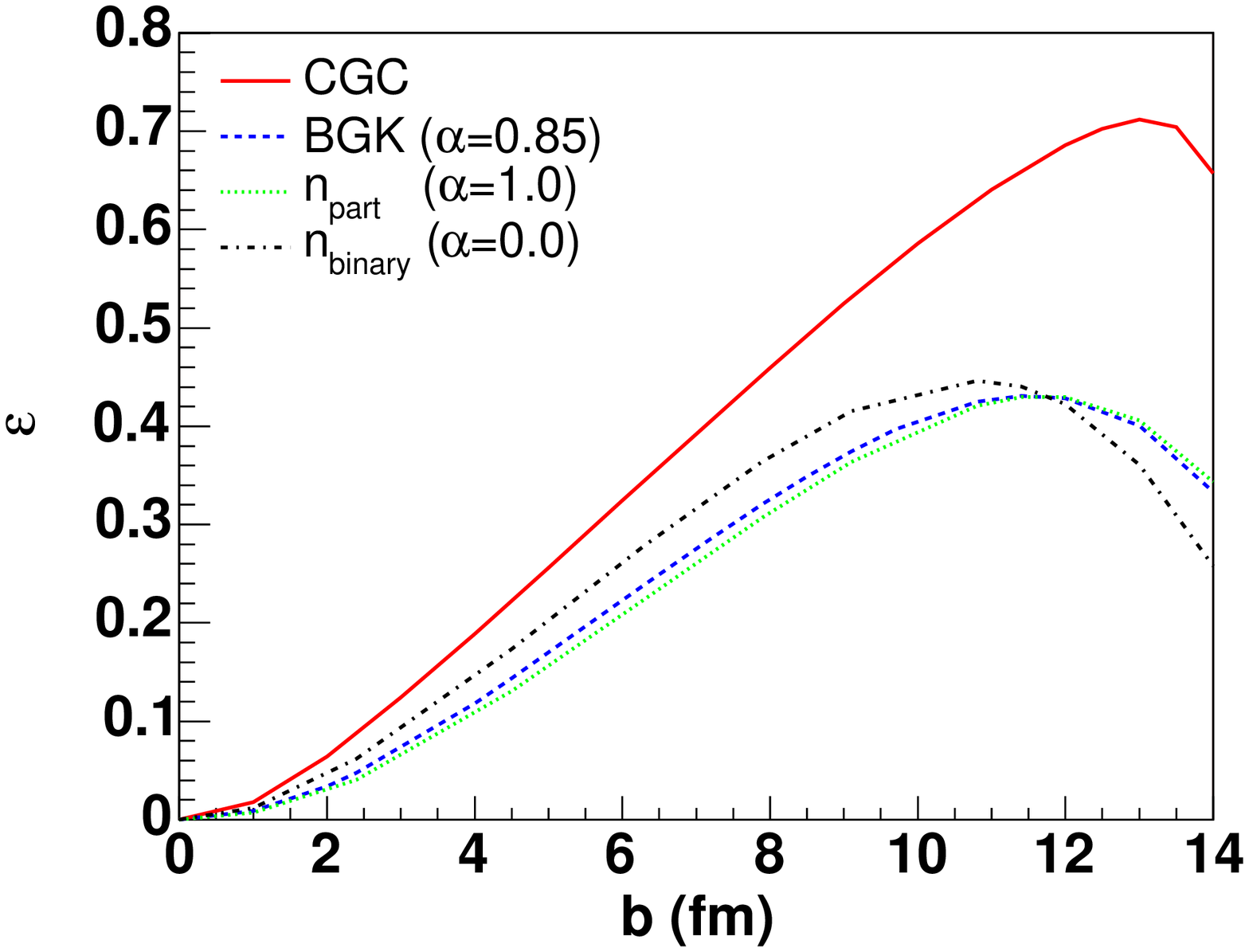,width=8cm}}  
\caption{{\sl Left:} Density of binary collisions in the transverse plane
         for a Au+Au collision with impact parameter $b=7$~fm.
         Shown are contours of constant density together with the 
         projection of the initial nuclei (dashed lines). 
         {\sl Right:} Spatial eccentricity $\epsilon$ as a function of the 
         impact parameter \protect\cite{Hirano:2005xf}, calculated with 
         Eq.~(\ref{equ:epsilonx}) using the initial energy density as 
         weight function, for four different models as described in the text.}
\label{fig:anisotropies} 
\end{figure} 
%%%%%%%%%%%%%%%%%%%%%%%%%%%%%%%%%%%%%%%%%%%%%%%%%%%%%%%%%%%%%%%%%%%%%%%
%

%
The left panel of Fig.~\ref{fig:anisotropies} shows the distribution 
of binary collisions in the transverse plane for Au+Au collisions at
impact parameter $b=7$~fm.
Shown are lines of constant density at 5, 15, 25,~\dots\% 
of the maximum value. 
The dashed lines indicate the Woods-Saxon circumferences of the two 
colliding nuclei, displaced by $\pm b/2$ from the origin.
The clearly visible geometric deformation of the overlap region can be 
quantified by the {\em spatial eccentricity}
\beq{equ:epsilonx}
\epsilon_x(b) = \frac{\la y^2 - x^2 \ra}{\la y^2+x^2 \ra} \, ,
\end{equation}
where the average is to be taken with the energy density as weight
function \cite{KSH00}.
The initial energy density is obtained from the initial entropy density
through the equation of state (EOS, see Sec.~\ref{sec:nucleareos}).
The right panel of Fig.~\ref{fig:anisotropies} shows the initial
spatial eccentricity for three models where the initial {\em entropy} density
is taken proportional to the density of wounded nucleons ($n_\mathrm{part}$,
green dotted line), of binary nucleon-nucleon collisions ($n_\mathrm{binary}$,
black dash-dotted line), and of a superposition of these two with 85\% weight
for the ``soft'' component (BGK, blue dashed line).
These are compared with a fourth model (solid red line) that uses directly 
the initial {\em energy} density (\ref{eq:ktfac}) of gluons from the KNL model.
One sees that, at any given impact parameter, the KLN model (``CGC'') 
predicts almost 50\% larger spatial eccentricities than the standard 
Glauber initialization (``BGK'') \cite{Hirano:2005xf}.
A recently improved version of the model called fKLN \cite{Drescher:2006pi,%
Drescher:2006ca} produces somewhat smaller eccentricities but even those 
exceed the Glauber model values by 25--30\%.
%

%%%%%%%%%%%%%%%%%%%%%%%%%%%%%%%%%%%%%%%%%%%%%%%%%%%%%%%%%%%%%%%%%%%%%%%%
%   SECTION: DECOUPLING AND FREEZE-OUT
%   last updated: 1/26/09 UH
%%%%%%%%%%%%%%%%%%%%%%%%%%%%%%%%%%%%%%%%%%%%%%%%%%%%%%%%%%%%%%%%%%%%%%%%

%%%%%%%%%%%%%%%%%%%%%%%%%%%%%%%%%%%%%%%%%%%%%%%%%%%%%%%%%%%%%%%%%%%%%%%%
\suse{Decoupling and freeze-out}
\label{sec:breakdown}
%%%%%%%%%%%%%%%%%%%%%%%%%%%%%%%%%%%%%%%%%%%%%%%%%%%%%%%%%%%%%%%%%%%%%%%%
\sususe{Two-stage decoupling}
\label{twostage}
%%%%%%%%%%%%%%%%%%%%%%%%%%%%%%%%%%%%%%%%%%%%%%%%%%%%%%%%%%%%%%%%%%%%%%%%

%
As explained in Sec.~\ref{sec:init_and_fin}, the hydrodynamic 
description begins to break down again once the transverse expansion
becomes so rapid and the matter density so dilute that local thermal 
equilibrium can no longer be maintained. 
Detailed studies \cite{SH94,HLS89a} comparing local mean free paths with 
the overall size of the expanding fireball and the local Hubble radius 
(inverse expansion rate) have shown that bulk freeze-out happens
{\em dynamically}, i.e. it is driven by the expansion of the fireball
and not by its finite size.
This is similar to the decoupling of the primordial nuclear abundances 
and the cosmic microwave background in the early universe which, too, 
were controlled by the cosmological expansion rate.  
The similarities between the ``Little Bangs'' created in heavy-ion 
collisions and the Big Bang that created our universe do not end here.
Similar to the Big Bang, local thermodynamic equilibrium breaks in two 
stages:
In the early universe, {\em primordial nucleosynthesis} signals the
end of inelastic nuclear reactions that can change its chemical 
composition; it takes hundreds of millions of years after this point 
to restart nuclear reactions in the cores of stars formed by 
gravitational collapse of density inhomogeneities.
In heavy-ion collisions, an analogous process of {\em chemical 
decoupling} happens once inelastic reaction rates among hadrons
become too low to maintain chemical equilibrium among the various 
hadron species.
At RHIC energies, chemical decoupling is observed to happen at
a temperature of about 160\,MeV and appears to be driven by 
the hadronization process at the quark-hadron phase transition.
The 2.7\,K thermal background radiation in our universe reflects
its {\em thermal decoupling} at $T\sim 3000$\,K, cosmologically
redshifted by about a factor 1000.
At this temperature ions and electrons combined into neutral atoms and
the cosmological photons stopped rescattering, thus freezing in their
Bose-Einstein thermal energy distribution.
The analogous process in the ``Little Bang'' is called {\em kinetic}
or {\em thermal freeze-out} and happens when the matter is so dilute
that even elastic collisions cease among hadrons, thereby freezing 
in their momentum distributions. 
The kinetic freeze-out temperature in heavy-ion collisions is about
100\,MeV. 
In the Big Bang, chemical and thermal freeze-out are separated
by about 400,000 years. 
Since the fireballs created in the Little Bangs expand about $10^{18}$ 
times faster than the early universe did at similar temperatures, the 
time separation between chemical and thermal decoupling shrinks to a 
few fm/$c$ in heavy-ion collisions.
That the two decoupling processes do not happen simultaneously but
hierarchically is easily seen from the kinetic decoupling 
criterium:\cite{BGZ78,HLR87,LRH88,SH94,HS98,HLS89a,Heinz:2006ur,Eskola:2007zc,%
Dusling:2007gi}
\begin{equation}
\label{focrit}
  \tau_\mathrm{exp}(x) \equiv \frac{1}{\partial\cdot u(x)} = 
  \xi\, \tau^{(i)}_\mathrm{scatt}(x) \equiv \xi\, \frac{1}{\sum_j
  \langle\sigma_{ij}v_{ij}\rangle\rho_j(x)},
\end{equation}
where $\xi$ is an (unknown) parameter of order 1. 
Local equilibrium requires the mean free time $\tau_\mathrm{scatt}$
between scatterings to be much shorter than the local ``Hubble time''
$\tau_\mathrm{exp}$ describing the fireball expansion.
Equilibrium breaks when the two time scales become of the same order.
The scattering rate involves the product of the scattering cross 
section with the density of scatterers. 
Since chemical transformations require inelastic processes which
constitute only a small fraction of the total cross section whereas
momenta get changed by almost all types of collisions, thermal 
equilibration is driven by much larger cross sections and happens 
considerably faster than chemical equilibration.
Correspondingly, Eq.~(\ref{focrit}) tells us that, in a medium with
given hydrodynamical expansion rate, chemical freeze-out happens at
higher particle densities (and thus higher temperatures) than
thermal freeze-out.
Furthermore, the equation predicts that in general different particle
species freeze out at different temperatures, since scattering cross 
sections are species-specific.
%

%
%%%%%%%%%%%%%%%%%%% Fig. 3 %%%%%%%%%%%%%%%%%%%%%%%%%%%%%%%%%%%%%%%%%%%
\begin{figure}[b]
\begin{center}
\begin{minipage}[h]{10cm} 
       \epsfig{file=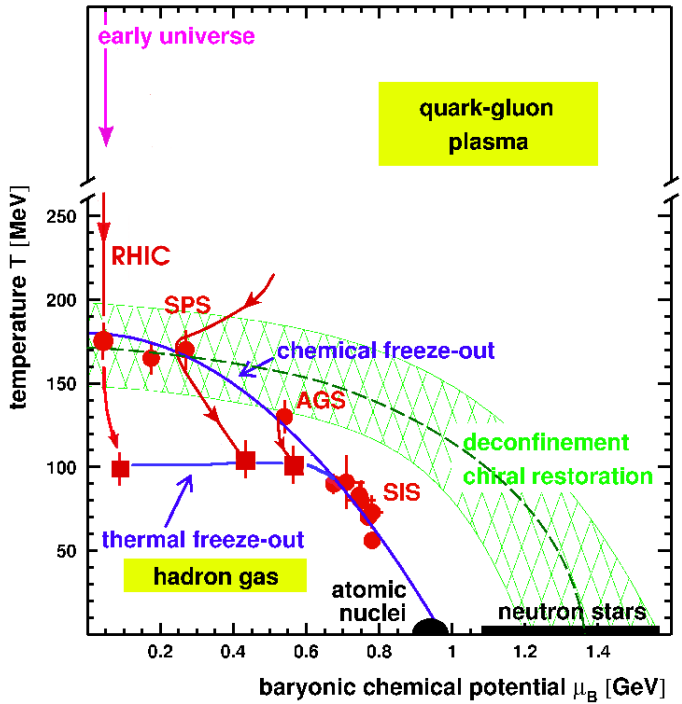,width=9cm}
\end{minipage}
\hspace*{5mm}
\begin{minipage}[h]{4.5cm}
\caption{\protect
         Chemical and thermal freeze-out points extracted from heavy-ion
         collisions at the GSI SIS, BNL AGS, CERN SPS and RHIC. The shaded
         area indicates the likely location of the quark-hadron phase 
         transition as extracted from lattice QCD and theoretical
         models. An updated version adding many more chemical freeze-out 
         points can be found in Ref.~\protect\cite{Cleymans:2007kj}.}
\label{F3} 
\end{minipage}
\end{center}
\end{figure} 
%\vspace*{-5mm}
%%%%%%%%%%%%%%%%%%%%%%%%%%%%%%%%%%%%%%%%%%%%%%%%%%%%%%%%%%%%%%%%%%%%%%%
%
%
Equation~(\ref{focrit}) is a {\em local} criterium.
The set of points $(\bm{x},\tau_f(\bm{x}))$ satisfying Eq.~(\ref{focrit}) 
defines the {\em freeze-out hypersurface} $\Sigma_\mathrm{f}$. 
It is a 3-dimensional surface imbedded in 4-dimensional space-time. 
The shapes of these freeze-out surfaces depend on the hydrodynamic
expansion rate $\partial{\cdot}u(x)$, and their computation thus
requires a dynamical simulation.
Since the matter near the transverse edge of the fireball is dilute 
and thus freezes out early, the freeze-out surface typically closes
on the initialization surface where the hydrodynamic evolution is started.
(It may even close above the initialization surface, i.e. at times
$\tau>\tau_0$, if hydrodynamics is initialized too early \cite{Dusling:2007gi};
since the expansion rate diverges like $1/\tau$ at early times, hydrodynamics
cannot be started until the longitudinal expansion rate has dropped
enough to allow for local thermal equilibrium.)
Numerical studies \cite{SH94,HLS89a,Heinz:2006ur,Eskola:2007zc,Dusling:2007gi} 
show that, except near the transverse edge of the fireball where the 
expansion rate changes rapidly with position, the kinetic freeze-out 
surfaces defined by Eq.~(\ref{focrit}) can be well approximated by surfaces 
of constant temperature. 
Making use of the fact that at RHIC energies pions form the most abundant
species and their kinetic decoupling thus controls thermal freeze-out of
all other hadrons, one can approximate the thermal decoupling of all
hadron species by a single surface of temperature $T_\mathrm{dec}$ 
corresponding to pion freeze-out. 
Its value can be determined phenomenologically from so-called blast-wave
model fits \cite{SSHPRC93,SSH93,Broniowski:2001we,Broniowski:2001uk,%
Retiere:2003kf,Biedron:2006vf} to experimental hadron spectra.
These models characterize the decoupling medium by an average freeze-out
temperature and an average transverse flow velocity.
Similarly, one can fit the observed final hadron abundance ratios with
a thermal model and extract from this the chemical decoupling temperature
$T_\mathrm{chem}$. 
The result of such an exercise \cite{FMHS99,BMMRS01,Cleymans:2007kj}, 
performed on a huge set of heavy-ion collision data from SIS to RHIC 
energies, is shown in Fig.~\ref{F3}.
The figure demonstrates a clear separation of chemical from thermal
decoupling for collision energies above $\sqrt{s_{NN}}\sim 5$\,GeV.
%

%
%%%%%%%%%%%%%%%%%%%%%%%% Fig. 4 %%%%%%%%%%%%%%%%%%%%%%%%%%%%%%%%%%%%%%%%
\begin{figure}[ht]
\centerline{\epsfig{file=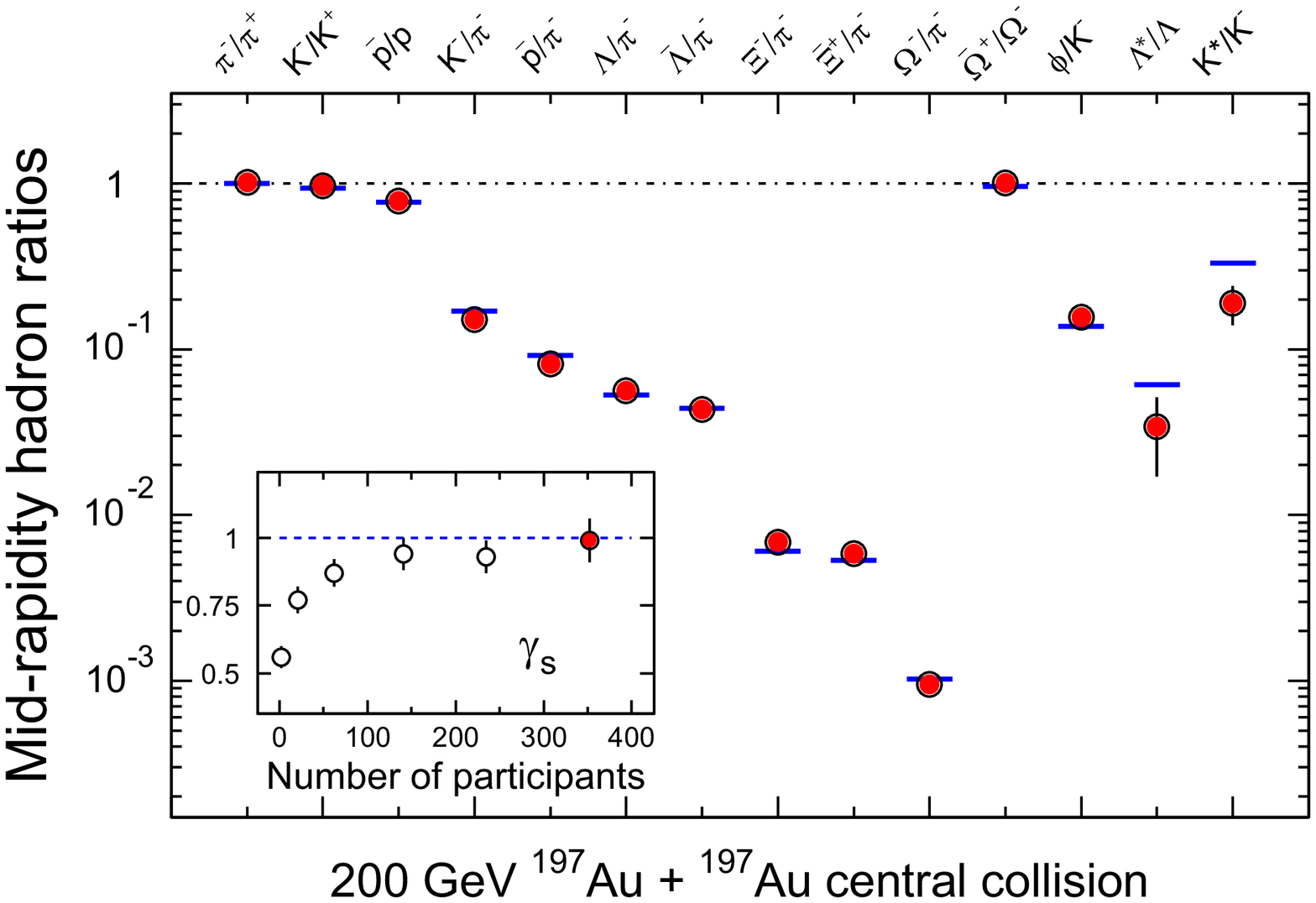,height=5.2cm}
            \epsfig{file=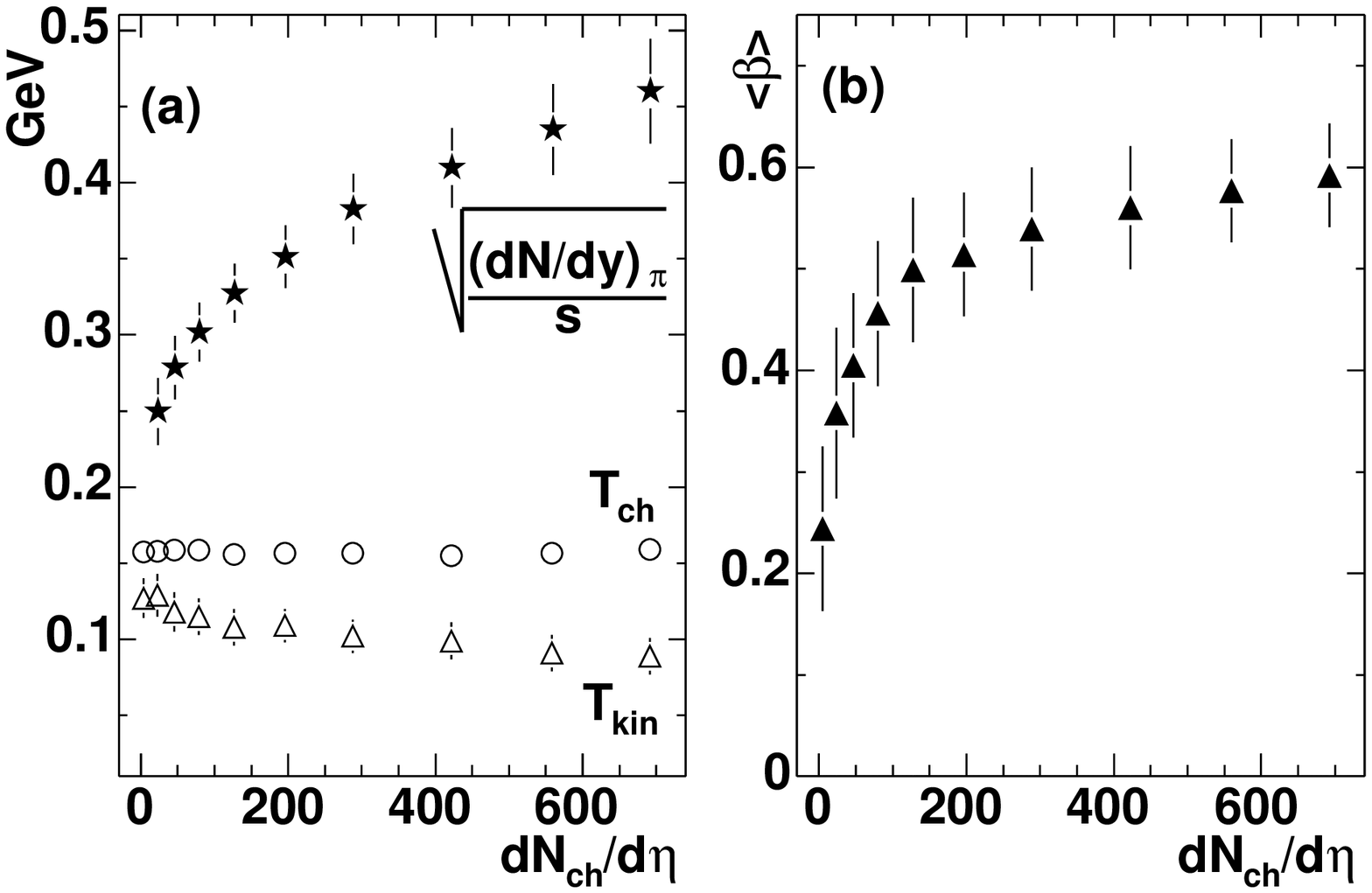,height=4.8cm}}
\caption{{\sl Left:} Abundance ratios of stable hadrons from central
         $200\,A$\,GeV Au+Au collisions at RHIC \protect\cite{Adams:2005dq}. 
         The blue lines show predictions from a thermal model fit 
         with $T_\mathrm{chem}=163\pm4$\,MeV, $\mu_B=24\pm4$\,MeV, and a 
         strangeness saturation factor $\gamma_s=0.99\pm0.07$ 
         \protect\cite{Adams:2005dq}. The inset shows the centrality 
         dependence of $\gamma_s$. 
         {\sl Right:} Centrality dependence (with centrality measured by 
         charged hadron rapidity density $dN_\mathrm{ch}/d\eta$) of (a) 
         the thermal freeze-out temperature 
         $T_\mathrm{kin}{\equiv}T_\mathrm{dec}$ (open triangles), the
         chemical freeze-out temperature $T_\mathrm{chem}$ (open circles),
         and the square root of the transverse areal density of pions
         $(dN_\pi/d\eta)/S$ (solid stars), and (b) the average transverse 
         flow velocity $\langle\beta\rangle{\equiv}\langle v_\perp\rangle$
         (solid triangles), for the same collision 
         system \protect\cite{STAR_Tdec}.}
\label{F4}
\end{figure}
%%%%%%%%%%%%%%%%%%%%%%%%%%%%%%%%%%%%%%%%%%%%%%%%%%%%%%%%%%%%%%%%%%%%%%%%%
%
%
The kinetic freeze-out criterium (\ref{focrit}) predicts a dependence
of the (average) freeze-out temperature on the (average) hydrodynamic 
expansion rate.  
The latter changes with system size and collision centrality. 
The right panel of Fig.~\ref{F4} shows that the thermal decoupling 
temperature $T_\mathrm{kin}\equiv T_\mathrm{dec}$ in Au+Au collisions
at RHIC indeed depends on centrality.
This dependence is consistent with hydrodynamic predictions and
Eq.~(\ref{focrit}) \cite{Heinz:2006ur}:
Larger collision systems created in more central collisions cool down
further and develop larger radial flow $\langle v_\perp\rangle$ than
the smaller fireballs formed in peripheral collisions.
In contrast, the chemical decoupling temperature shows no sensitivity
whatsoever to collision centrality and the accompanying change in
expansion rate.
(The excellent quality of the chemical fits is shown in the left panel 
of Fig.~\ref{F4}.)
The baryon chemical potential $\mu_B$ and the strangeness saturation 
factor $\gamma_s$ (which indicates to what extent strange hadrons are
suppressed relative to non-strange hadrons) decrease somewhat in peripheral
collisions, but $T_\mathrm{chem}$ is completely independent of 
centrality \cite{Adams:2005dq,Manninen:2008mg}.
Chemical freeze-out at RHIC can therefore not be driven by a local 
competition between inelastic hadron scattering and hydrodynamic 
expansion, as described by Eq.~(\ref{focrit}) \cite{Heinz:2006ur}.
The observed universality of the measured chemical freeze-out temperature
and the proximity of the value extracted from experiment to the critical
temperature $T_c$ of the quark-hadron phase transition predicted by
lattice QCD \cite{Karsch:2003jg,Aoki:2006br,Cheng:2006qk} can only be 
understood if one assumes that the phase transition itself controls the 
chemical freeze-out process. 
At $T_c$, hadrons are created from quarks and gluons in a state of
maximum entropy, with thermal abundances reflecting a temperature 
$T_\mathrm{chem}\approx T_c$ that characterizes the critical energy 
density for hadronization \cite{Becattini:1997rv}.
At that point, the fireball is already expanding so rapidly and the
hadron resonance gas is so dilute that inelastic hadronic reactions
can no longer change its chemical composition.
The chemical temperature is thus frozen at $T_c$, allowing us to
measure the quark-hadron phase transition temperature directly
through hadron abundances. 
Between chemical decoupling at $T_c$ and thermal decoupling at 
$T_\mathrm{dec}$, hadrons continue to rescatter quasi-elastically
through a rich spectrum of hadronic scattering resonances with large
cross sections.
Since the resonances typically decay into the same particles from 
which they were created (up to quark exchange), this does not
affect the chemical composition, but it changes the hadrons'\ momenta.
For a while they can thus maintain approximate thermal equilibrium even 
though chemical equilibrium is broken.
As long as thermal equilibration can be locally maintained, hydrodynamics
continues to be applicable.
The equation of state through which pressure gradients are evaluated 
must, however, correctly reflect the non-equilibrium chemical composition 
of the HRG below $T_\mathrm{chem}$ \cite{Bebie:1991ij,HT02,Teaney02,Rapp02,%
KR03}.
The latter is also essential for the computation of elliptic flow
since the distribution of the total momentum-anisotropy of the 
energy-momentum tensor over the various hadron species depends on
their relative abundance, i.e. on the (non-equilibrium) chemical 
composition at thermal freeze-out \cite{HT02,KR03,Hirano:2005wx,%
Huovinen:2007xh}.
%

%%%%%%%%%%%%%%%%%%%%%%%%%%%%%%%%%%%%%%%%%%%%%%%%%%%%%%%%%%%%%%%%%%%%%%%%
\sususe{Final hadron momentum spectra}
\label{spectra}
%%%%%%%%%%%%%%%%%%%%%%%%%%%%%%%%%%%%%%%%%%%%%%%%%%%%%%%%%%%%%%%%%%%%%%%%
%
The breakdown of local equilibrium ends the hydrodynamic stage of 
a heavy-ion collision. 
A relatively easy way to implement this into hydrodynamics is through
the Cooper-Frye prescription \cite{CF74} which postulates a sudden
transition from a thermalized fluid to free-streaming particles on a 
decoupling surface of, say, temperature $T_\mathrm{dec}$.
In this subsection we describe how this procedure allows to compute
final hadron momentum spectra, multiplicities and elliptic flow,
both in ideal and viscous fluid dynamics.
The idealization of a sudden freeze-out has, of course, limitations.
Even if one correctly accounts for the non-equilibrium chemical
composition in the hadronic phase below $T_c$ through appropriate
chemical potentials $\mu_i(T)$, it is not a priori clear that
a sudden transition can capture all phenomenologically important  
aspects of the freeze-out process.
Real-life freeze-out happens gradually, is particle specific, and
should thus be described in a microscopic kinetic approach.
Quantitative model predictions for hadron spectra from heavy-ion 
collisions will thus eventually require matching the hydrodynamic
evolution to a hadronic rescattering cascade that describes the final
expansion stage \cite{BDBBZSG99,BD00,SBD01,TLS01b,TLS01,TLS02,Hirano:2005xf,%
Nonaka:2006yn,Petersen:2008dd}.
In this case, the Cooper-Frye prescription discussed here is used
at a suitable switching temperature $T_\mathrm{dec}<T_\mathrm{sw}<T_c$ 
to generate thermally distributed hadrons in an expanding ensemble, 
which are then used as discrete input into a hadronic cascade that
follows their further evolution until all collisions have ceased.
To accumulate enough statistics for the final hadron spectra, the
hadronic cascade must be run many times with initial conditions 
sampled by a Monte-Carlo simulation of the Cooper-Frye spectra 
at $T_\mathrm{sw}$.   
This is numerically expensive, and therefore not many such 
calculations from a hydro+cascade hybrid approach are presently
available \cite{BDBBZSG99,BD00,SBD01,TLS01b,TLS01,TLS02,Hirano:2005xf,%
Nonaka:2006yn,Petersen:2008dd}.
All of these use ideal fluid dynamics to generate the input for the 
hadron cascade. 
No systematic studies exist that show the existence of a 
window of switching temperatures that produces final results 
independent of $T_\mathrm{sw}$.
One may expect that, if it exists, that window will be larger
when viscous hydrodynamics (with viscosities matched to those
of the hadronic cascade) is used to initialize the late kinetic
stage. 
The Cooper-Frye formalism is based on the following expression 
for the final momentum spectrum \cite{CF74}:
\beq{equ:CooperFrye}
     E \frac{dN_i}{d^3p} 
   = \frac{dN_i}{dy \pT d\pT d\phi_p}
   = \frac{g_i}{(2\pi)^3}
     \int_\Sigma p{\cdot}d^3\sigma(x)\,f_i\bigl(x,p\bigr) \,.
\end{equation}
Here $d^3\sigma_\mu(x)$ is the outward normal vector on the freeze-out
surface $\Sigma(x)$ such that $p^\mu d^3\sigma_\mu\,f_i$ is the
local flux through this surface of particles of species $i$ with 
momentum $p$. 
In ideal fluid dynamics, the phase-space distribution $f_i$ in this 
formula is the local equilibrium distribution {\em just before} 
decoupling, 
\beq{equ:distribution}
   f_{i,\mathrm{eq}}(x,p) = 
   \frac{1}{\exp[(p{\cdot}u(x)-\mu_i(x))/T(x)]\pm 1} \,,
\end{equation}
boosted with the local flow velocity $u^\mu(x)$ to the global reference 
frame by the substitution $E\mapsto p\cdot u(x)$. 
$\mu_i(x)$ and $T(x)$ are the chemical potential of particle 
species $i$ and the local temperature along $\Sigma$, respectively.
The temperature and chemical potentials on $\Sigma$ are computed from 
the hydrodynamic output for the energy density $e$, net baryon density 
$n$ and pressure $p$ with the help of the equation of state 
\cite{SHKRPV97}.
The quantum statistical correction $\pm1$ in the denominator matters
only for pions where Bose corrections can reach 10--20\% (depending
on the pion chemical potential at freeze-out). 
For all other hadron spectra the Boltzmann approximation is sufficiently
accurate.
In viscous hydrodynamics, the distribution function along the 
decoupling hypersurface is in general slightly out of equilibrium,
by an amount proportional to the dissipative flows $\Pi(x)$, $q^\mu$
and $\pi^{\mu\nu}(x)$ on $\Sigma$. For vanishing bulk viscosity and
heat conduction one finds \cite{Israel:1979wp,Teaney:2003kp}
\begin{eqnarray}
\label{viscf}
  f(x,p) &=& f_\mathrm{eq}(x,p) \left[
  1 + \bigl(1{\mp}f_\mathrm{eq}(x,p)\bigr)\frac{c_2}{2} 
      \frac{p^\mu p^\nu}{T^2(x)}\frac{\pi_{\mu\nu}(x)}{e(x){+}p(x)}
      \right]
\nonumber\\
  &\approx& f_\mathrm{eq}(x,p) \left[
  1 + \frac{1}{2} \frac{p^\mu p^\nu}{T^2(x)}
      \frac{\pi_{\mu\nu}(x)}{e(x){+}p(x)}
      \right]
  \equiv f_\mathrm{eq}(x,p) + \delta f(x,p).
\end{eqnarray}
Here $c_2{\,=\,}1$ in Boltzmann approximation; for massless bosons, 
$c_2{\,=\,}1.04$.
For massive bosons, $c_2$ is a temperature dependent function 
that interpolates between these limits \cite{Teaney:2003kp}. 
Replacing in Eq.~(\ref{viscf}) the factor $1{\mp}f_\mathrm{eq}$ 
by 1 is an excellent approximation even for pions since it deviates from 1 
only at small momenta where the non-equilibrium correction is suppressed 
by two powers of $p$. 
The reader should note that shear viscous pressure effects modify 
the {\em shape} of the {\em local} momentum distribution by an amount 
that increases quadratically with $p$. Even for very small shear viscous 
pressure at freeze-out, the non-equilibrium correction $\delta f$ of the 
local distribution function eventually becomes big and comparable with 
the equilibrium contribution if $p$ gets sufficiently large. At this 
point, the near-equilibrium expansion breaks down, and the spectrum 
calculated from (\ref{viscf}) can no longer be trusted. This emphasizes 
the nature of (viscous) hydrodynamics as an effective theory that 
applies at large distances (low momenta) but breaks down at short 
distances.
To apply Cooper-Frye freeze-out, one first lets the hydrodynamic code 
run up to large times, assuming hydrodynamics to be valid everywhere.
One then determines the space-time hypersurface $\Sigma(x)$ by 
identifying which fluid cells satisfy the freeze-out criterium.
Back-reaction effects arising from the (in principle) non-hydrodynamic 
behaviour of the matter outside the decoupling surface on the hydrodynamic 
evolution inside the thermalized space-time region are ignored.
The Cooper-Frye formalism is used to calculate the momentum distributions 
of all directly emitted hadrons, stable and unstable.
Unstable resonances are then allowed to decay if they do so via strong
or electromagnetic interactions, accounting for the appropriate branching 
ratio of different decay channels \cite{PDG02}. 
Weakly decaying particles are considered as stable because they are
usually reconstructed in the experiments.
The stable decay products are added to the thermal momentum spectra of 
the directly emitted stable hadrons to give the total measured particle 
spectra \cite{SKH90,Wiedemann:1996ig,Broniowski:2003ax,Wheaton:2004qb,%
Kisiel:2005hn}.
%

%%%%%%%%%%%%%%%%%%%%%%%%%%%%%%%%%%%%%%%%%%%%%%%%%%%%%%%%%%%%%%%%%%%%%%%%
%   SECTION: THE NUCLEAR EQUATION OF STATE
%   last updated: 1/24/09
%%%%%%%%%%%%%%%%%%%%%%%%%%%%%%%%%%%%%%%%%%%%%%%%%%%%%%%%%%%%%%%%%%%%%%%%

%%%%%%%%%%%%%%%%%%%%%%%%%%%%%%%%%%%%%%%%%%%%%%%%%%%%%%%%%%%%%%%%%%%%%%%%
\section{The nuclear equation of state}
\label{sec:nucleareos}
%%%%%%%%%%%%%%%%%%%%%%%%%%%%%%%%%%%%%%%%%%%%%%%%%%%%%%%%%%%%%%%%%%%%%%%%
%
As emphasized in Sec.~\ref{sec:ideal}, the hydrodynamic equations
require the input of an equation of state (EOS) $p(e,n)$ for closure,
and this EOS, through the speed of sound $c_s^2(T)=\frac{\partial p}
{\partial e}$, defines the ``pushing power'' of the medium, i.e.
how strongly the matter accelerates in reaction to pressure gradients.
%

%
%%%%%%%%%%%%%%%%%%%%%%% Fig. 5 %%%%%%%%%%%%%%%%%%%%%%%%%%%%%%%%%%%%%%%%%%
\begin{figure}
\centerline{\epsfig{file=Fig5a.eps,width=6.8cm}\hspace*{5mm}
            \epsfig{file=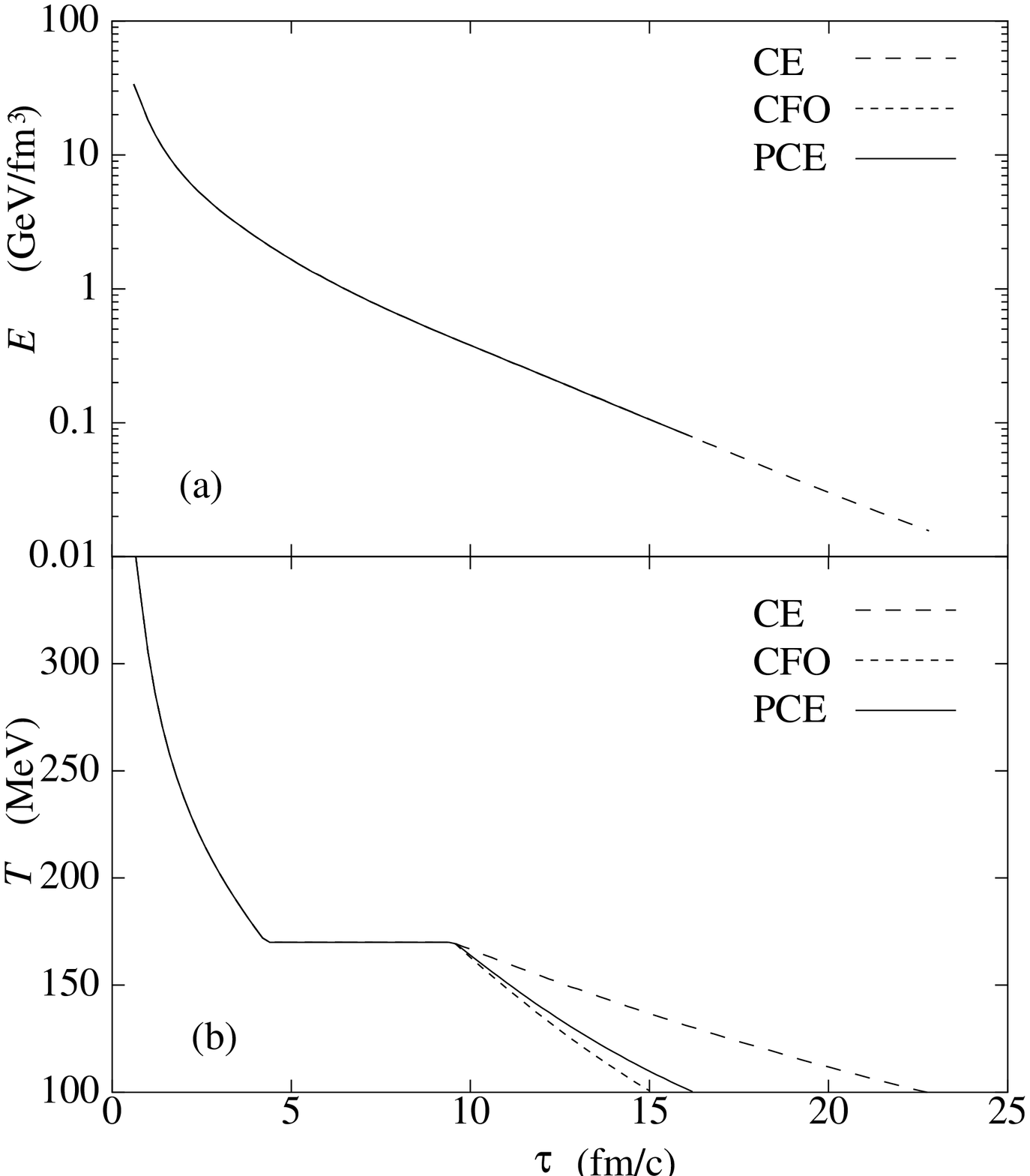,width=7.2cm,height=9.3cm}}
\caption{{\sl Left: }The equation of state for baryon-free QCD matter.
         The upper plot shows the pressure $p$ as a function of energy 
         density $e$ and (in the inset) the squared speed of sound 
         $c_s^2=\frac{\partial p}{\partial e}$ as a function of 
         temperature $T$. The lower panel shows $c_s^2$ as a function 
         of energy density $e$ \protect\cite{Song:2008si}. The solid red
         line (SM-EOS~Q) is a slightly smoothed version of EOS~Q (green 
         dashed line).
         {\sl Right:} Energy density (top) and temperature (bottom) of the
         central cell as a function of longitudinal proper time from a 
         (2+1)-d ideal fluid dynamical simulation with of Au+Au collisions 
         at RHIC \protect\cite{HT02}, for an EOS with a first-order 
         quark-hadron transition at $T_c=170$\,MeV and three choices of 
         the chemical composition of the HRG below $T_c$: CE (dashed)
         assumes full hadronic chemical equilibrium at all temperatures 
         (this case corresponds to the green dashed lines in the left panel); 
         CFO (dotted) assumes chemical freeze-out of {\em all} hadronic 
         species (stable and unstable) at $T_c$; PCE (solid) makes the
         realistic assumption that unstable resonances continue to 
         re-equilibrate in the HRG phase via resonance scattering, but
         that the final yields of all stable decay products remain
         unchanged below $T_c$. The $p(e)$ curves for all three choices
         are almost identical \protect\cite{HT02}, resulting in identical
         time evolutions of the energy density $e(\tau)$.}
\label{F5}
\end{figure}
%%%%%%%%%%%%%%%%%%%%%%%%%%%%%%%%%%%%%%%%%%%%%%%%%%%%%%%%%%%%%%%%%%%%%%%%%
%

%
At RHIC energies the net baryon density $n$ is very small at midrapidity, 
and the dependence of the EOS $p(e,n)$ on $n$ is weak.
For hydrodynamic purposes we can thus use the EOS at $n=0$ with excellent 
accuracy near midrapidity.
To obtain correctly normalized hadron spectra at freeze-out it is,
however, important that the used EOS incorporates all relevant 
hadronic species with the correct chemical composition.
For this, the $n$-dependence of the EOS matters.
A simple and in the past very popular procedure to construct an EOS for 
QCD matter (known as EOS~Q \cite{KSH99,KSH00}) is to match a non-interacting 
massless quark-gluon gas (shown as EOS~I in Fig.~\ref{F5}) with 
adjustable bag constant $B$ to a non-interacting, chemically equilibrated 
hadron resonance gas that includes all known hadron resonances with their 
measured masses up to a certain mass cutoff (typically between 1.6 and 
2\,GeV) \cite{LRH88b,SHKRPV97}.  
Adjusting $B$ to obtain $T_c=165$\,MeV in accordance with lattice 
QCD data, this construction results in a first order phase transition 
with a mixed quark-hadron phase for energy densities $0.45 < e < 
1.6$\,GeV/fm$^3$.
The squared speed of sound for EOS~Q is $c_s^2=\frac{1}{3}$ above $T_c$, 
$c_s^2\approx 0.15$ between $T_\mathrm{dec}$ and $T_c$ \cite{KSH99}, 
and $c_s^2=0$ for all energy density values in the mixed phase (green 
dashed lines in Fig.~\ref{F5}).
Some viscous hydrodynamic calculations require a slightly smoothed
version of this EOS, called SM-EOS~Q (solid red line in the left
panel of Fig.~\ref{F5}) for numerical stability.
Modern lattice QCD data \cite{Karsch:2003jg,Aoki:2006br,Cheng:2006qk,%
Aoki:2006we} show that this modelling is unrealistic in two aspects: 
Lattice QCD shows a continuous cross-over phase transition without
phase coexistence, instead of a first-order discontinuity at $T_c$.
So the speed of sound, while becoming small and developing a minimum
(``softest point'') near $T_c$ (dash-dotted curve in the left panel of 
Fig.~\ref{F5}), never drops to zero as assumed in EOS~Q.  
Above $T_c$, lattice QCD data show clear deviations from an ideal
gas of massless quarks and gluons which, in the temperature range
explored by heavy-ion collisions at RHIC, reduce the squared speed
of sound by a significant (and temperature-dependent) fraction,
making the EOS softer than EOS~Q.
More and more hydrodynamic simulations are therefore now being 
performed with equations of state that are better matched to
lattice QCD data (such as EOS~L shown in Fig.~\ref{F5}).
Another unrealistic aspect of EOS~Q that turns out to be more 
difficult to fix is the assumption of chemical equilibrium below 
$T_c$.
As already discussed, the experimental data indicate chemical 
freeze-out near $T_c$ \cite{BMMRS01}. 
This requires the introduction of non-equilibrium chemical potentials 
for the stable hadron species in the hadron resonance gas (HRG) 
phase \cite{Bebie:1991ij,Teaney02,HT02,KR03} that must be anchored at
$T_c$, using the correct non-zero baryon chemical potential at $T_c$.
A proper matching to the QGP phase must thus be done at all relevant 
non-zero values of net baryon density $n$.     
Lattice QCD data at non-zero $n$ have recently become available (see 
Refs.~\cite{Fodor:2001pe,deForcrand:2006pv,Karsch:2007dp,Endrodi:2009sd} 
and references therein), and successful quasiparticle parametrizations
of lattice QCD data that allow to extrapolate data at $n=0$ to nonzero
net baryon densities are also available \cite{Thaler:2003uz,%
Bluhm:2004xn,Bluhm:2007nu}.
However, a proper smooth matching of these data to a chemically 
non-equilibrated HRG has not yet been performed.
Existing equations of state that properly describe the non-equilibrium
chemical composition of the HRG below $T_c$ match to an ideal massless
quark-gluon gas through a first-order transition \cite{HT02,KR03}. 
As shown in the right panel of Fig.~\ref{F5}, the hydrodynamical 
evolution with such an EOS is virtually indistinguishable from EOS~Q,
since the non-equilibrium chemical potentials have only very
small effects on the EOS $p(e)$ in the HRG phase below $T_c$.
What does change, however, is the relationship between energy density
and temperature in the HRG. 
In the chemically frozen HRG baryons and antibaryons as well as pions
are not allowed to annihilate, which is ensured by giving the
non-equilibrium chemical potentials that grow as $T$ 
decreases \cite{Bebie:1991ij,Teaney02,HT02,KR03}. 
This stores more of the energy density in particle rest masses, 
reducing the thermal energy and temperature corresponding to
a given energy density.
(Surprisingly, the pressure $p(e)$ does not decrease.) 
This has obvious consequences for the final hadron spectra: at the
same decoupling energy density $e_\mathrm{dec}$, the hydrodynamic
flow is the same as with EOS~Q but the temperature $T_\mathrm{dec}$
is smaller, so the transverse momentum spectra are steeper.
As we will see, this also results in a significant reshuffling of the
momentum anisotropy in non-central collisions which strongly modifies
the elliptic flow coefficients \cite{HT02,KR03,Hirano:2005wx,Huovinen:2007xh}. 
A given set of experimental spectra thus requires a retuning of
hydrodynamic initial conditions, to ensure that more radial flow is 
generated to compensate for the lower decoupling temperature.
The predicted elliptic flow will then be different for the chemically 
non-equilibrated hadron gas than for the unrealistic EOS~Q, with
important consequences for the interpretation of the data as to how
much room they offer for non-zero viscosity of the expanding QCD 
matter.
%

%%%%%%%%%%%%%%%%%%%%%%%%%%%%%%%%%%%%%%%%%%%%%%%%%%%%%%%%%%%%%%%%%%%%%%%
% expansion.tex
% last edited on 1/27/09
%%%%%%%%%%%%%%%%%%%%%%%%%%%%%%%%%%%%%%%%%%%%%%%%%%%%%%%%%%%%%%%%%%%%%%%

%%%%%%%%%%%%%%%%%%%%%%%%%%%%%%%%%%%%%%%%%%%%%%%%%%%%%%%%%%%%%%%%%%%%
\section{Phenomenology of the transverse expansion}
\label{sec:expansion}
%%%%%%%%%%%%%%%%%%%%%%%%%%%%%%%%%%%%%%%%%%%%%%%%%%%%%%%%%%%%%%%%%%%%

In this section we study the transverse fireball expansion at 
midrapidity as it follows from the hydrodynamic equations of 
motion (Section~\ref{sec:hydrodynamics}) with the equation of state 
described in Section \ref{sec:nucleareos} and the initial conditions 
from Section \ref{sec:initialization}.
We analyze results from (2+1)-d simulations with longitudinal 
boost-invariance for which both ideal and viscous fluid dynamical 
codes are available.
Results from (3+1)-d ideal fluid dynamics (with the initial conditions
discussed in Sec.~\ref{sec:initialization}) largely agree at midrapidity 
with those from (2+1)-d ideal fluid simulations. 
In Section~\ref{sec:evolutioncentral} we begin by discussing azimuthally
symmetric radial expansion in central collisions ($b{=}0$). 
Both Au+Au and Cu+Cu collisions have been simulated but, as it happens,
a detailed comparison between ideal and viscous dynamics with and 
without transverse expansion has only been done for Cu+Cu collisions.
We therefore use these for illustration.
The collision energy is reflected in the initial entropy or energy
density which is adjusted to the final charged hadron multiplicity
as discussed in Sec.~\ref{sec:initialization}.
One usually quotes the peak energy density $e_0$ or peak entropy density
$s_0$ in the fireball center for $b{=}0$ collisions for reference.
These values then determine the shapes and normalization of the initial 
density profiles at all impact parameters. 
Unless stated otherwise, the simulations were started at $\tau_0=0.6$\,fm/$c$,
and thermal freeze-out was implemented on a hypersurface of constant
energy density $e_\mathrm{dec}=0.075$\,GeV/fm$^3$.
These choices will be motivated in Sec.~\ref{sec:momspacobservables}.
In Section~\ref{sec:evolutionnoncentral} we address non-central 
collisions and discuss the special opportunities provided by the 
breaking of azimuthal symmetry in this case.
We discuss how the initial spatial deformation transforms rapidly 
into a momentum space anisotropy which ultimately manifests itself through
a dependence of the emitted hadron spectra and their momentum correlations
on the azimuthal emission angle relative to the reaction plane (``elliptic
flow'').
%

%%%%%%%%%%%%%%%%%%%%%%%%%%%%%%%%%%%%%%%%%%%%%%%%%%%%%%%%%%%%%%%%%%%
% SECTION: RADIAL EXPANSION IN CENTRAL COLLISIONS
% last updated 1/27/09
%%%%%%%%%%%%%%%%%%%%%%%%%%%%%%%%%%%%%%%%%%%%%%%%%%%%%%%%%%%%%%%%%%%%
\suse{Radial expansion in central collisions}
\label{sec:evolutioncentral}
%%%%%%%%%%%%%%%%%%%%%%%%%%%%%%%%%%%%%%%%%%%%%%%%%%%%%%%%%%%%%%%%%%%%
%
Even though for boost-invariant longitudinal expansion there is no
longitudinal acceleration in $\eta_s$ direction, the thermodynamic 
pressure performs longitudinal work on the fluid at the expense
of thermal energy.
For a fluid that has initially no transverse expansion but features
boost-invariant longitudinal flow, the velocity shear tensor 
$\sigma^{\mu\nu}$ has non-zero diagonal elements that induce
a negative shear pressure component in the longitudinal direction and
equal positive pressure components of half the size in the two transverse
directions \cite{Danielewicz:1984ww,Heinz:2005bw}.
As a result, the fluid does less longitudinal work than in the ideal case,
while transverse pressure gradients are increased and transverse expansion 
is accelerated.
%

%
%%%%%%%%%%%%%%%%%%%%%%% Fig. 6 %%%%%%%%%%%%%%%%%%%%%%%%%%%%%%%%%%%%%%%%%%%
\begin{figure} 
\centerline{\epsfig{file=Fig6a.eps,width=7cm} \hfill
            \epsfig{file=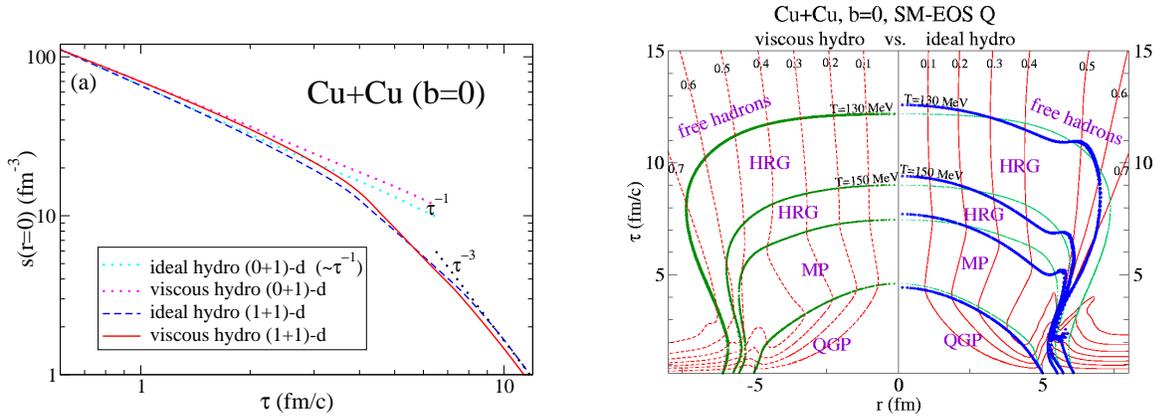,width=7cm}} 
\caption{{\sl Left:} Time evolution of the entropy density in the central
         cell of the expanding fireball. Shown as an
         example are central Cu+Cu collisions with an initial peak energy 
         density $e_0\equiv e(\bm{x}_\perp{=}0,\tau_0{=}0.6\,\mathrm{fm}/c)
         = 30$\,GeV/fm$^3$, with SM-EOS~Q. The two dotted lines
         show ideal (light blue) and viscous (magenta) boost-invariant
         longitudinal expansion without transverse expansion. The lower
         set of lines include tranverse radial expansion for ideal (dashed)
         and viscous (solid red) fluids, again assuming longitudinal
         boost invariance. The viscous simulations use $\eta/s=1/4\pi$ 
         for the specific shear viscosity and zero bulk viscosity.
         {\sl Right:} Surfaces of constant temperature $T$ and constant 
         radial flow velocity $v_\perp$ for viscous (left half) and ideal 
         fluid dynamics (right half) in 2+1 dimensions, for the same 
         collision system and EOS. MP indicates the mixed phase between 
         QGP and HRG \protect\cite{Song:2007ux}.       
\label{F6}
} 
\end{figure} 
%%%%%%%%%%%%%%%%%%%%%%%%%%%%%%%%%%%%%%%%%%%%%%%%%%%%%%%%%%%%%%%%%%%%%%%
%

%
This is documented in the left panel of Fig.~\ref{F6}: 
In the absence of transverse flow (which can be simulated by providing
initial conditions that are independent of transverse position 
$\bm{x}_\perp$), longitudinally boost-invariant ideal fluid dynamics
causes the entropy density to decrease like $1/\tau$ whereas in the 
viscous case it decreases more slowly (dotted lines).
In the QGP phase $s\sim T^3$, so this implies a reduced cooling rate
in the viscous case.
As the volume increases linearly with $\tau$ in this 1-dimensional 
situation, total entropy is conserved in the ideal fluid but increases
with time in the viscous fluid.
Transverse expansion leads to additional cooling, but at early times
the dominant viscous effect is still a reduced cooling rate due to
reduced longitudinal work (dashed and solid lines in the left panel
of Fig.~\ref{F6}).
At later times, however, the additional radial flow caused by the
positive viscous contribution to the transverse pressure gradients
increases the cooling rate so much that, at least in the fireball    
center, the entropy density decreases more rapidly in the viscous 
than in the ideal fluid.
At long times, the expansion becomes fully 3-dimensional, so the volume
increases approximately like $\tau^4$ and the entropy density in both 
ideal and viscous hydro decreases like $\tau^{-3}$.
Entropy production ceases at late times because all viscous pressure
components become very small \cite{Song:2007ux}.
The right panel in Fig.~\ref{F6} gives a picture of the time evolution
of the fireball in the transverse plane. 
It shows surfaces of constant temperature and lines of constant radial
flow velocity, for viscous hydrodynamics (green dots) on the left (and
mirrored on the right) and ideal hydrodynamics (blue dots) on the right
side of the plot.
The most prominent feature of the viscous hypersurfaces is their utter
smoothness:
For the ideal fluid, the isothermal hypersurfaces feature prominents 
structures arising from the first-order phase transition which are 
completely smeared out in viscous hydrodynamics.
The reason are large velocity gradients near the QGP-MP and MP-HRG 
interfaces, caused by the sudden change of the speed of sound at these
interfaces.
These velocity gradients contribute to the shear flow tensor and generate
shear viscous pressure gradients which suppress large velocity gradients
and, at the same time, mask the discontinuities in the EOS, turning 
the first-order phase transition effectively into a smooth 
crossover \cite{Song:2007ux}.
The additional radial flow generated in viscous fluid dynamics by the 
positive transverse components of the shear pressure tensor generates, 
for identical initial conditions, flatter transverse momentum spectra 
than for ideal fluids \cite{Teaney:2003kp,Baier:2006um,Baier:2006gy,%
Song:2007fn,Song:2007ux}.
This requires a retuning of initial conditions if one attempts to 
describe a given set of experimental spectra \cite{Romatschke:2007jx}.
This will be discussed in more detail in Section~\ref{sec:observables}.
%

%
%%%%%%%%%%%%%%%%%%%%%%  Fig. 7 %%%%%%%%%%%%%%%%%%%%%%%%%%%%%%%%%%%%%%%%%%%%%%
\begin{figure} 
\centerline{\hspace*{3mm}
            \epsfig{file=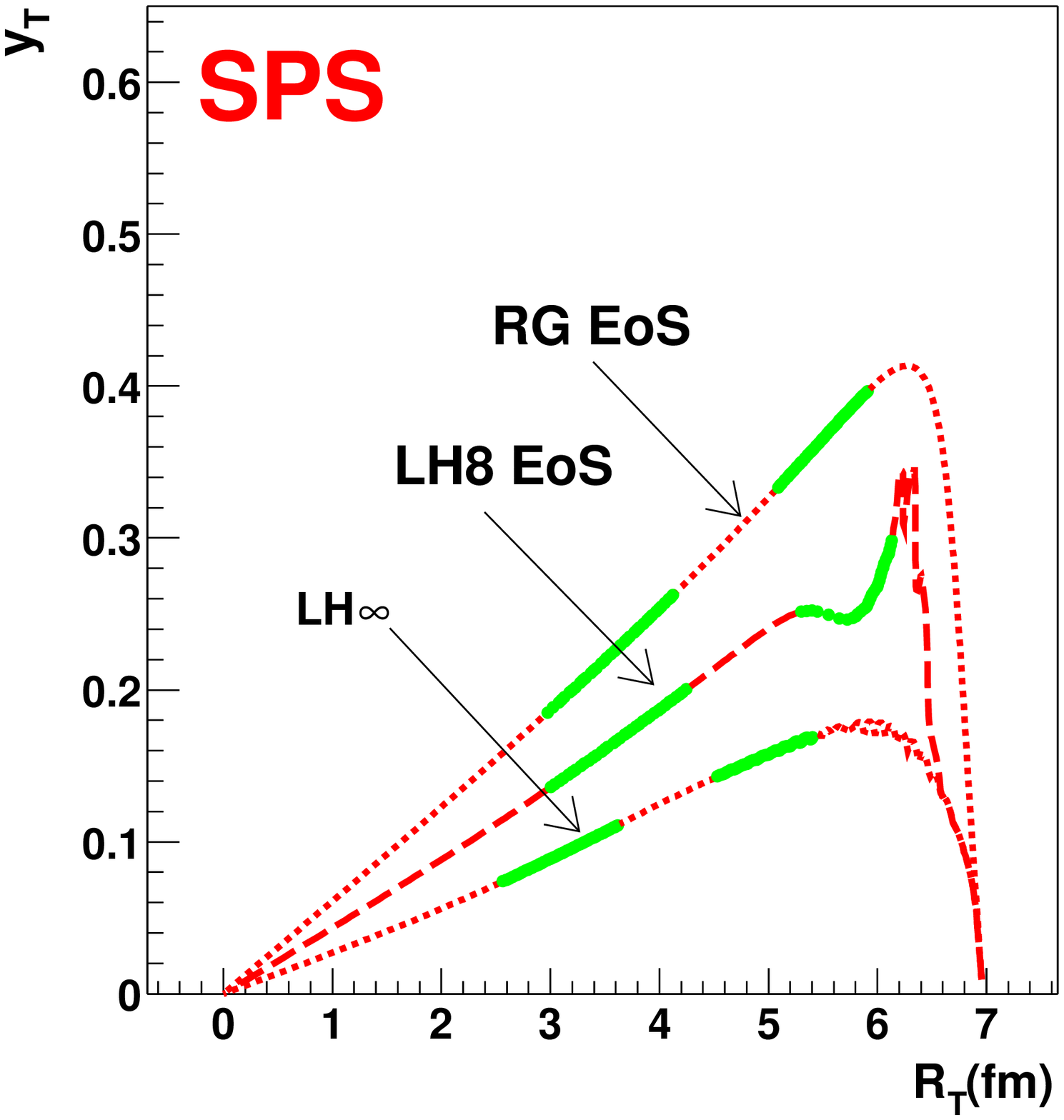,width=6cm} 
            \hspace*{-5mm}
            \epsfig{file=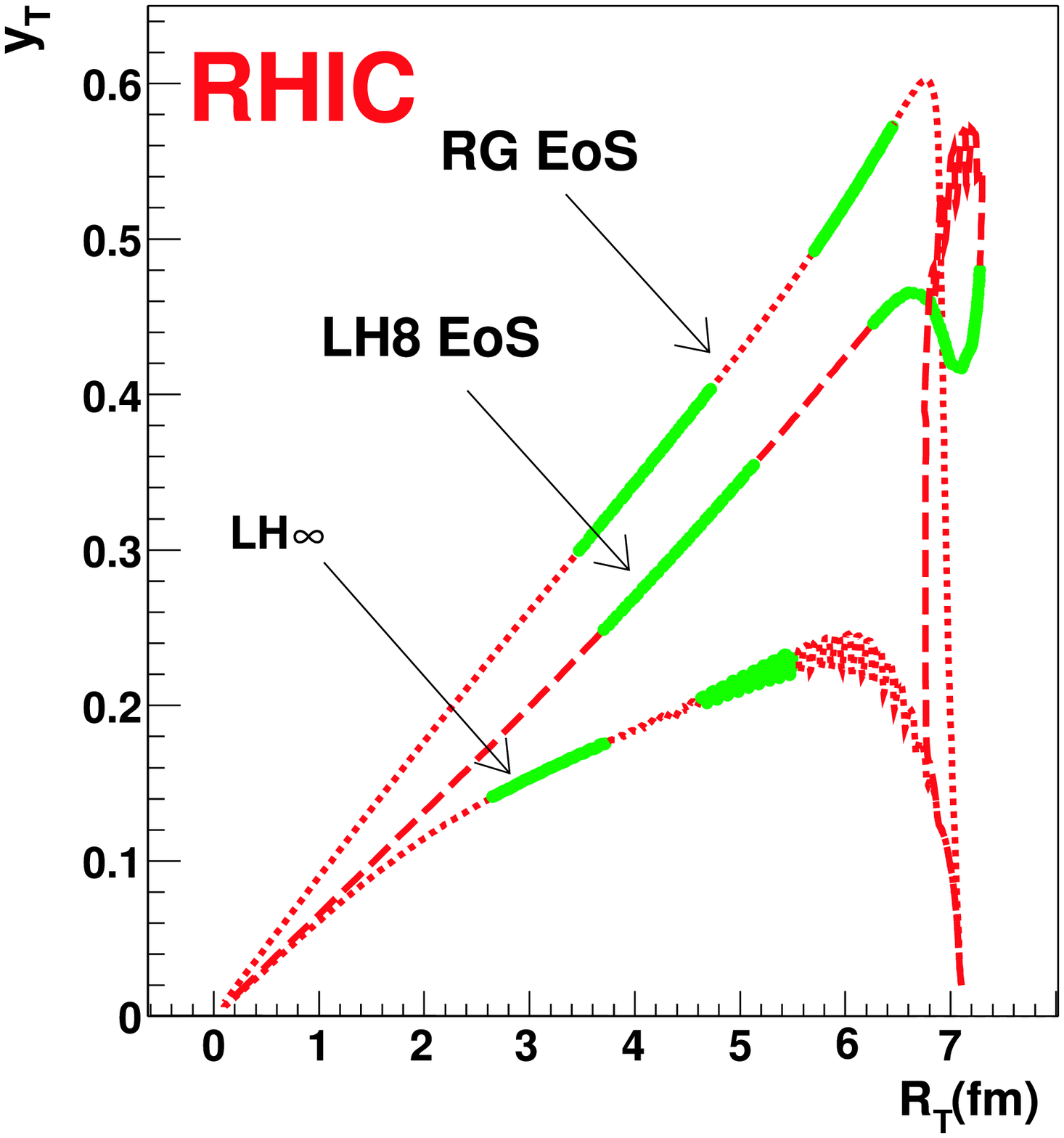,width=6.15cm,height=5.56cm}}  
\caption{The transverse flow rapidity $y_\perp=\frac{1}{2}\ln[(1{+}v_\perp)/
         (1{-}v_\perp)]$ as a function of 
         radial distance $r$ along a surface of constant energy
         density $e\eq0.45$\,GeV/fm$^3$, for Pb+Pb collisions
         at the SPS (left) and for Au+Au collisions at RHIC 
         (right) \protect\cite{TLS01}.
         Three different equations of state have been explored
         in this figure \protect\cite{TLS01}, with LH8 corresponding
         most closely to EOS~Q shown in Figure~\protect\ref{F5}. The 
         dashed and solid line segments subdivide the surface into 5 
         pieces through each of which flow 20\% of the entropy.
\label{F7} 
} 
\end{figure} 
%%%%%%%%%%%%%%%%%%%%%%%%%%%%%%%%%%%%%%%%%%%%%%%%%%%%%%%%%%%%%%%%%%%%%%%
%

%
The hydrodynamically generated radial flow rapidity profiles in 
heavy-ion collisions are typically linear, with a slope that initially 
increases with time but eventually saturates \cite{Kolb:2003dz}.
(More precisely it is the flow rapidity $y_\perp=\frac{1}{2}
\ln[(1{+}v_\perp)/(1{-}v_\perp)]$ --which is not constrained
by the speed of light-- that is proportional to $r$.)
If the fireball matter passes through a phase transition, the
transition generates non-monotonic structures in the radial flow 
profiles at early times \cite{KSH00} which eventually disappear
at late times.
The developing flow pattern thus approaches a Hubble form where
fluid cells recede from the fireball center with flow rapidities
that increase linearly with distance -- not only in longitudinal 
direction where this is imposed by the assumed boost-invariance, but
also in the transverse directions.
Contrary to our cosmos, however, this expansion is not isotropic: it
features different ``Hubble constants'' in longitudinal and 
transverse directions, and the latter depend on collision energy
and the EOS. 

Figure~\ref{F7} shows that the transverse flow rapidity profile is 
approximately linear not only at fixed proper time, but also along
the decoupling surface.
It compares the radial flow rapidity profile $y_\perp(r)$ for Pb+Pb 
or Au+Au collisions at SPS and RHIC energies for three different 
equations of state \cite{TLS01}, with LH8 corresponding most closely 
to EOS~Q.
Figure~\ref{F7} provides welcome support for the phenomenologically 
very successful blast-wave parametrization \cite{SSHPRC93,SSH93,%
Broniowski:2001we,Broniowski:2001uk,Retiere:2003kf,Biedron:2006vf}
which is usually employed with a linear transverse velocity or rapidity 
profile, for reasons of simplicity.
(Note that for the range of velocities covered in the figure the 
difference between rapidity $y_\perp$ and velocity 
$v_\perp=\tanh y_\perp$ can be neglected.)
As discussed in Section~\ref{twostage}, particle freeze-out is 
controlled by the competition between the macroscopic expansion 
time scale \cite{SH94,HS98} $\tau_{\rm exp}\eq(\partial{\cdot}u)^{-1}$ 
and the microscopic scattering time scale 
$\tau_{\rm scatt}^i\eq1/\sum_j \la \sigma_{ij}v_{ij}\ra\rho_j$.
Figure~\ref{F7} shows that the expansion rate $\partial{\cdot}u$ 
changes significantly between SPS and RHIC:
For boost-invariant longitudinal flow and a linear transverse flow
rapidity profile $y_\perp\eq\xi r$ the expansion rate is calculated 
as\cite{Kolb:2003gq}
\begin{equation}
\label{rate}
  \partial\cdot u =\frac{\cosh(\xi r)}{\tau} + \xi \left(
  \cosh(\xi r) + \frac{\sinh(\xi r)}{\xi r}\right) \approx
  \frac{1}{\tau} + 2\xi,
\end{equation}
where the approximation \cite{Tomasik:2002qt} holds in the region 
$\xi r{\,\ll\,}1$. 
Equation~(\ref{rate}) gives $\tau(\partial{\cdot}u)\eq1+2\xi\tau$.
From Figure~\ref{F7} we read off $\xi\approx0.07$ at RHIC energies,
but at SPS energies $\xi$ is about 30\% smaller.
At freeze-out ($\tau_\dec{\,\simeq\,}15{-}17$\,fm/$c$ \cite{KSH00,TLS01})
the expansion rate at RHIC is thus about 25\% larger than at the SPS
$\bigl((\partial{\cdot}u)_\dec{\,\approx\,}0.21$\,fm$^{-1}$ for Au+Au at 
$\sqrt{s}\eq130\,A$\,GeV vs. 
$(\partial{\cdot}u)_\dec{\,\approx\,}0.16$\,fm$^{-1}$ 
for Pb+Pb at $\sqrt{s}\eq17\,A$\,GeV$\bigr)$.
The corresponding ``Hubble times'' at freeze-out are 
$\tau_{\rm exp}^\dec({\rm RHIC}){\,\approx\,}4.8$\,fm/$c$ and 
$\tau_{\rm exp}^\dec({\rm SPS}){\,\approx\,}6.1$\,fm/$c$.
Barring a dramatic change in the scattering mean free times between SPS 
and RHIC energies that could result from different chemical compositions, 
one is led to the conclusion that at RHIC freeze-out should happen at 
somewhat higher decoupling temperatures than at the SPS.
Note, however, that Fig.~\ref{F3} does not support this conclusion.
%

%%%%%%%%%%%%%%%%%%%%%%%%%%%%%%%%%%%%%%%%%%%%%%%%%%%%%%%%%%
% Section: ANISOTROPIC FLOW IN NON-CENTRAL COLLISIONS
% last updated 1/27/09
%%%%%%%%%%%%%%%%%%%%%%%%%%%%%%%%%%%%%%%%%%%%%%%%%%%%%%%%%%
\suse{Anisotropic flow in non-central collisions}
\label{sec:evolutionnoncentral}
%%%%%%%%%%%%%%%%%%%%%%%%%%%%%%%%%%%%%%%%%%%%%%%%%%%%%%%%%%%%%%%%%%%%

\begin{figure}[ht] 
\centerline{\epsfig{file=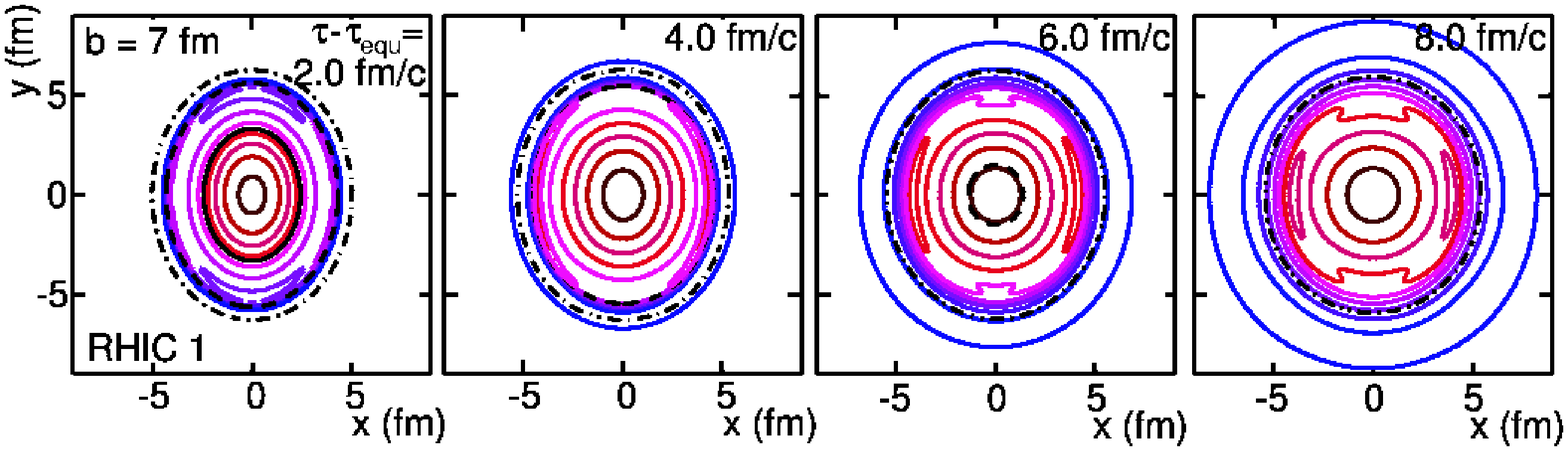,width=\textwidth}}  
\caption{Contours of constant energy density in the transverse plane
         at different times (2, 4, 6 and 8~fm/$c$ after equilibration) 
         for a Au+Au collision at $\scm=130$~GeV and impact parameter 
         $b=7$~fm \protect\cite{KSH00,PFKthesis02}. Contours indicate 
         5,\,15,\,\dots,\,95~\% of the maximum energy density. Additionally, 
         the black solid, dashed and dashed-dotted lines indicate   
         the transition to the mixed-phase, to the resonance gas phase
         and to the decoupled stage, where applicable.          
\label{fig:evolutionsnapshots} 
} 
%\vspace*{-3mm}
\end{figure}

In Section \ref{sec:initialization} we have already addressed some of 
the great opportunities offered by non-central collisions.
The most important ones are related to the broken azimuthal symmetry,
introduced through the spatial deformation of the nuclear overlap zone 
at non-zero impact parameter (see Figure~\ref{fig:anisotropies}).
If the system evolves hydrodynamically, driven by its internal pressure 
gradients, it will expand more strongly in its short direction (i.e. 
into the direction of the impact parameter) than perpendicular to the
reaction plane where the pressure gradient is smaller \cite{Ollitrault92}. 
This is shown in Figure~\ref{fig:evolutionsnapshots} where contours of 
constant energy density are plotted at times 2, 4, 6 and 8 fm/$c$ after 
thermalization. 
The figure illustrates qualitatively that, as the system evolves,
it becomes less and less deformed.
In addition, some interesting fine structure develops at later
times: 
After about 6~fm/$c$ the energy density distribution along the 
$x$-axis becomes non-monotonous, forming two fragments of a shell
that enclose a little 'nut' in the center \cite{Teaney:1999gr}. 
Unfortunmately, when plotting a cross section of the profiles shown in 
Figure~\ref{fig:evolutionsnapshots} one realizes that this effect is 
rather subtle, and it was also found to be fragile, showing a strong
sensitivity to details of the initial density profile \cite{KSH00}
and to even small amounts of viscosity (see Fig.~\ref{F6}).
%

%
%%%%%%%%%%%%%%%%%%%%%%% Fig. 8 %%%%%%%%%%%%%%%%%%%%%%%%%%%%%%%%%%%%%%%%%%%%%
%%%%%%%%%%%%%%%%%%%%%%%%%%%%%%%%%%%%%%%%%%%%%%%%%%%%%%%%%%%%%%%%%%%%%%%
%

%
A more quantitative characterization of the contour plots in 
Figure~\ref{fig:evolutionsnapshots} and their evolution with time
is provided by defining the {\em spatial eccentricity}
\beq{equ:epsilonxdef}
   \epsilon_x(\tau) = \frac{\la y^2-x^2 \ra}{\la y^2 + x^2 \ra},
\end{equation}
where the brackets indicate an average over the transverse plane 
with the local energy density $e(x,y;\tau)$ as weight function,
and the {\em momentum anisotropy}
\beq{equ:epsilonpdef}
\epsilon_p(\tau) = \frac{\int dx dy \, (T^{xx}-T^{yy})}
                        {\int dx dy \, (T^{xx}+T^{yy})}\;.
\end{equation}
Note that with these sign conventions, the spatial eccentricity is 
positive for out-of-plane elongation (as is the case initially)
whereas the momentum anisotropy is positive if the preferred flow 
direction is {\em into} the reaction plane.

%
%%%%%%%%%%%%%%%%%%%%%%%%% Fig. 9 %%%%%%%%%%%%%%%%%%%%%%%%%%%%%%%%%%%%%%
\begin{figure}[ht]
\begin{center}
\begin{minipage}[h]{8.5cm} 
      \epsfig{file=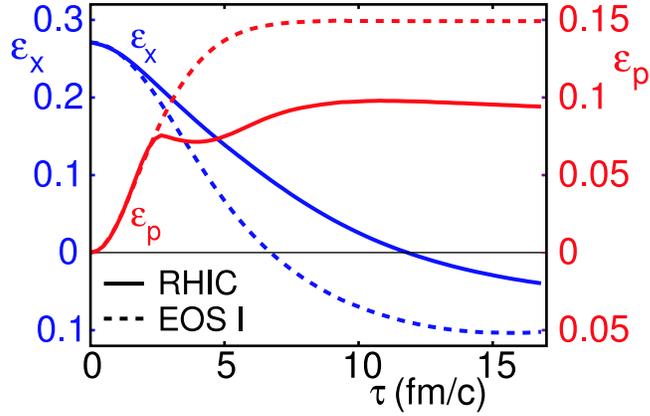,width=8.5cm}
\end{minipage}
\hspace*{5mm}
\begin{minipage}[h]{5cm} 
\caption{\protect Time evolution in ideal fluid dynamics of the spatial 
         eccentricity $\epsilon_x$ and the momentum anisotropy 
         $\epsilon_p$ for Au+Au collisions at RHIC with 
         $b\eq7$\,fm \protect\cite{HK02HBTosci}. 
\vspace*{1cm}
\label{fig:anisoovertau} 
} 
\end{minipage}
\end{center}
\end{figure}
%%%%%%%%%%%%%%%%%%%%%%%%%%%%%%%%%%%%%%%%%%%%%%%%%%%%%%%%%%%%%%%%%%%%%%%
%

\begin{figure}[ht] 
\begin{minipage}[h]{7.5cm}
    \epsfig{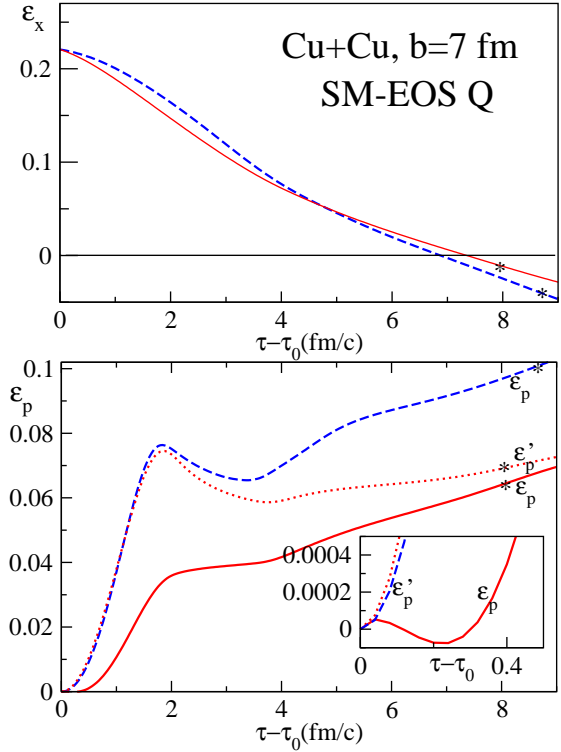} 
\caption{\protect
Time evolution for the spatial eccentricity $\epsilon_x$, momentum 
anisotropy $\epsilon_p$ and total momentum anisotropy $\epsilon_p'$ 
(see text for definitions), calculated for $b\eq7$\,fm Cu+Cu collisions
with SM-EOS~Q. Dashed lines are for ideal hydrodynamics while the solid 
and dotted lines show results from viscous hydrodynamics. Stars indicate
completion of freeze-out. See text for discussion.
\label{F10}
}
\end{minipage}
\hspace*{5mm}
\begin{minipage}[h]{7cm}
The dotted line in the bottom panel of Fig.~\ref{F10} shows that
for the viscous expansion the flow anisotropy is less than for the
ideal fluid, especially at later times. 
This causes the spatial eccentricity of the viscous fireball to 
decrease more slowly than that of the ideal fluid towards the end
of the expansion phase (solid line in the top panel).
It is instructive to compare the behaviour of the flow-induced
ideal-fluid contribution to the momentum anisotropy, $\epsilon'_p$, 
with that of the total momentum anisotropy $\epsilon_p$. 
At early times they are very different, with $\epsilon_p$ being much 
smaller than $\epsilon'_p$ and even turning slightly {\em negative} at 
very early times (see insets in the lower panel of 
Fig.~\ref{fig:anisoovertau}). 
This reflects very large {\em negative} contributions to the anisotropy 
of the total energy momentum tensor from the shear viscous pressure 
whose gradients along the out-of-plane direction $y$ strongly exceed 
those within the reaction plane along the $x$ direction. 
At early times this effect almost compensates for the larger in-plane 
gradient of the thermal pressure. 
The {\em negative} viscous pressure gradient anisotropy is responsible 
for reducing the growth of flow anisotropies, thereby causing the 
flow-induced momentum anisotropy $\epsilon'_p$ to significantly
lag behind its ideal fluid value at later times. 
The
\end{minipage} 
\end{figure}

Figure \ref{fig:anisoovertau} shows the time evolution of the spatial  
and momentum anisotropies for Au+Au collisions at impact parameter 
$b\eq7$\,fm, for RHIC initial conditions with a realistic equation of
state (EOS~Q, solid lines) and for a much higher initial energy density
(initial temperature at the fireball center =\,2\,GeV) with a massless 
ideal gas equation of state (EOS~I, dashed lines) \cite{HK02HBTosci}. 
The initial spatial asymmetry at this impact parameter is 
$\epsilon_x(\tau_{\rm equ})\eq0.27$, and obviously 
$\epsilon_p(\tau_{\rm equ})\eq0$ since the fluid is initially at 
rest in the transverse plane.
The spatial eccentricity is seen to disappear before the fireball 
matter freezes out, in particular for the case with the very high
initial temperature (dashed lines) where the source is seen to
switch orientation after about 6\,fm/$c$ and becomes in-plane-elongated
at late times \cite{HK02HBTosci}.
One also sees that the momentum anisotropy $\epsilon_p$ saturates 
at about the same time when the spatial eccentricity $\epsilon_x$ 
vanishes.
All of the momentum anisotropy is built up during the first 6\,fm/$c$.
Near a phase transition (in particular a first order transition) 
the equation of state becomes very soft, and this inhibits the 
generation of transverse flow.
This also affects the generation of transverse flow {\em anisotropies} 
as seen from the solid curves in Figure~\ref{fig:anisoovertau}:
The rapid initial rise of $\epsilon_p$ suddenly stops as a significant
fraction of the fireball matter enters the mixed phase. 
It then even decreases somewhat as the system expands radially without 
further acceleration, thereby becoming more isotropic in both coordinate
and momentum space. 
Only after the phase transition is complete and pressure gradients 
reappear, the system reacts to the remaining spatial eccentricity
by a slight further increase of the momentum anisotropy.
The softness of the equation of state near the phase transition 
thus focusses the generation of anisotropic flow to even earlier
times, when the system is still entirely partonic and has not even
begun to hadronize.
At RHIC energies this means that almost all of the finally observed
elliptic flow is created during the first 3-4 fm/$c$ of the collision
and reflects the hard QGP equation of state of an ideal gas of
massless particles ($c_s^2\eq\frac{1}{3}$) \cite{KSH00}. 
Microscopic kinetic studies of the evolution of elliptic flow lead to 
similar estimates for this time scale \cite{Sorge97,Sorge99,ZGK99,MG02}.
The anisotropic flow effects seen in non-central collisions turn out 
to be very sensitive to viscosity.
All examples shown in this subsection so far assumed a perfect fluid.
In Fig.~\ref{F10} we show the evolution of the spatial eccentricity 
(\ref{equ:epsilonxdef}) (top panel) and of the momentum anisotropies 
$\epsilon_p$ and $\epsilon'_p$ (bottom panel). 
Here $\epsilon_p$ is the total momentum anisotropy as defined in 
Eq.~(\ref{equ:epsilonpdef}), using the complete energy momentum tensor. 
$\epsilon'_p =
{\langle T^{xx}_\mathrm{eq}{-}T^{yy}_\mathrm{eq}\rangle}\big/
{\langle T^{xx}_\mathrm{eq}{+}T^{yy}_\mathrm{eq}\rangle}$
is a variant of the momentum anisotropy that includes only the
ideal fluid part $T^{\mu\nu}_\mathrm{eq}$ and thus measures 
only the anisotropy of the transverse momentum density arising 
from anisotropies in the collective flow pattern. 
It ignores contributions to the anisotropy arising from the viscous 
pressure components which reflect momentum anisotropies of the 
phase-space distribution in the local fluid rest frame, caused by 
anisotropic deviations $\delta f$ of that distribution from local 
equilibrium. 
The top panel of Figure~\ref{F10} shows that the viscous fireball 
loses its spatial deformation initially faster than if it were a 
perfect fluid.
This results mostly from the faster buildup of radial flow due to
initially large viscous tranverse pressure gradients -- the fact
that these gradients are themselves anisotropic plays only a minor 
role here. Early pressure gradient {\em anisotropies} manifest themselves 
in the initial growth rate of the flow-induced momentum anisotropy 
$\epsilon'_p$ which is seen to slightly exceed that observed in the 
ideal fluid at times up to about 1\,fm/$c$ after the beginning of the 
transverse expansion (bottom panel in Fig.~\ref{F10}).
%

%
%%%%%%%%%%%%%%%%%%%%%%%%% Fig. 10 %%%%%%%%%%%%%%%%%%%%%%%%%%%%%%%%%%%%%%%%%%
%%%%%%%%%%%%%%%%%%%%%%%%%%%%%%%%%%%%%%%%%%%%%%%%%%%%%%%%%%%%%%%%%%%%%%%
%
\vspace*{-5mm}
\noindent
negative viscous pressure anisotropies responsible for the difference 
between $\epsilon_p$ and $\epsilon'_p$ disappear at later times, since 
all viscous pressure components then become very small \cite{Song:2007ux}.
%

%
%%%%%%%%%%%%%%%%%%%%%%%%% Fig. 11 %%%%%%%%%%%%%%%%%%%%%%%%%%%%%%%%%%%%%%
\begin{figure}[htb]
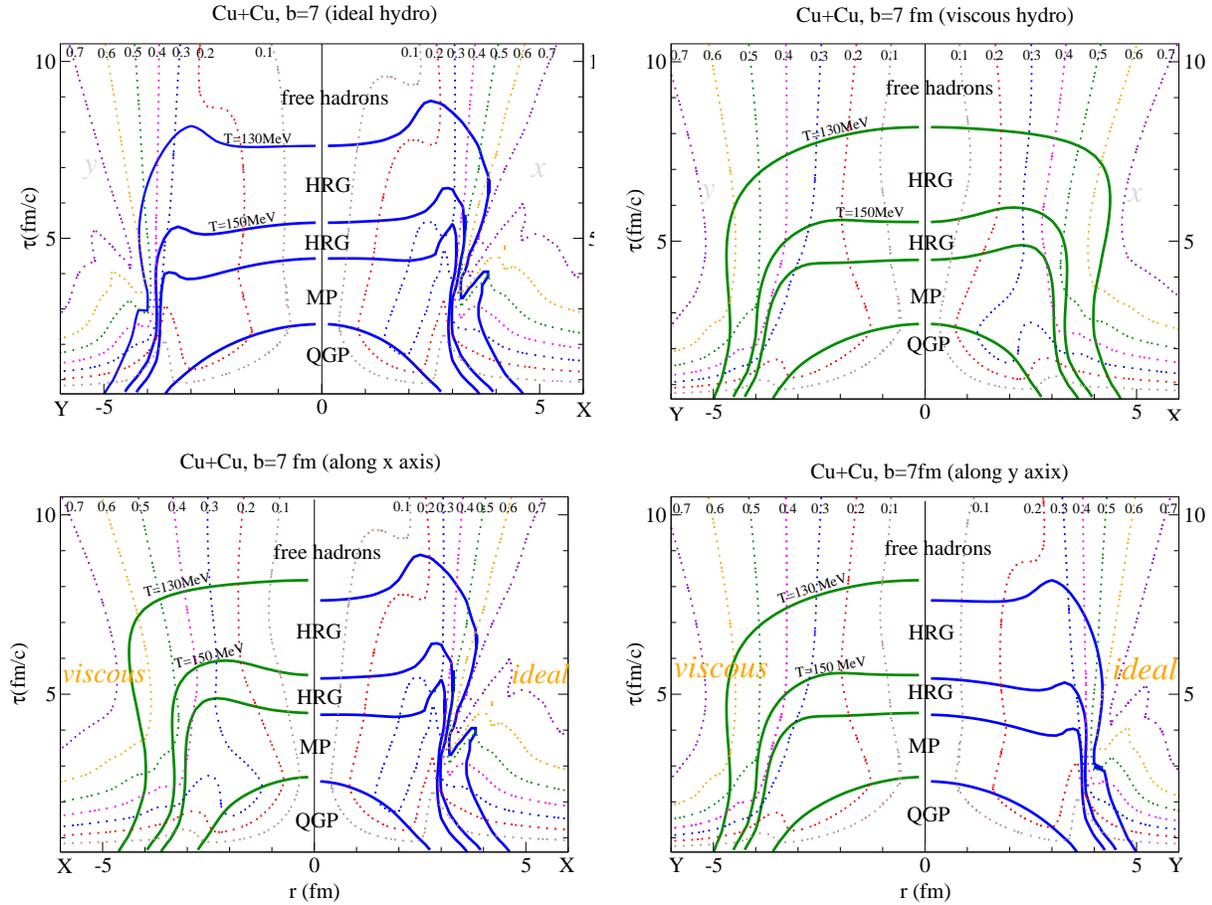
 
\centerline{\epsfig{file=Fig11a.eps,width=8cm}
            \epsfig{file=Fig11b.eps,width=8cm}}  
\centerline{\epsfig{file=Fig11c.eps,width=8cm}
            \epsfig{file=Fig11d.eps,width=8cm}}
\caption{Surfaces of constant temperature $T$ and constant transverse 
         flow velocity $v_\perp$ for semi-peripheral Cu+Cu collisions 
         at $b\eq7$\,fm, evolved with SM-EOS~Q \protect\cite{Song:2007ux}.
         In the {\em top row} we contrast ideal (left panel) and viscous 
         (right panel) fluid dynamics, with a cut along the $x$ axis 
         (in the reaction plane) shown in the right half while the left 
         half shows a cut along the $y$ axis (perpendicular to the reaction 
         plane). In the {\em bottom row} we compare ideal and viscous 
         evolution in the same panel, with cuts along the $x$ ($y$) 
         direction shown in the left (right) panel. See Fig.~\ref{F6} for 
         comparison with central Cu+Cu collisions.
\label{F11} 
} 
%\vspace*{-2mm}
\end{figure} 
%%%%%%%%%%%%%%%%%%%%%%%%%%%%%%%%%%%%%%%%%%%%%%%%%%%%%%%%%%%%%%%%%%%%%%%
%

%
The net result of this interplay is a total momentum anisotropy (i.e. 
a source of elliptic flow $v_2$) that for a ``minimally'' viscous fluid 
with $\frac{\eta}{s}\eq\frac{1}{4\pi}$ is 40-50\% lower in Cu+Cu collisions
than for an ideal fluid.
Initially this is due to strong momentum anisotropies in the local rest 
frame, with momenta pointing preferentially out-of-plane, induced by 
deviations from local equilibrium and associated with large shear 
viscous pressure.
Towards the end momentum isotropy in the local fluid rest frame is 
approximately restored, but at the expense of a reduced anisotropy
of the collective flow.
Figure~\ref{F11} shows isothermal hypersurfaces and contours of 
constant transverse flow velocity for non-central Cu+Cu collisions, 
computed in ideal and viscous hydrodynamics. 
We again see the smoothing effects of shear viscosity which
smears out all structures related to the assumed first-order
phase transition, and the viscous slowdown of the cooling process
(which now persists until freeze-out because in peripheral Cu+Cu 
collisions enough transverse flow is never generated to overcome
the effects of reduced longitudinal cooling).
%

%
%%%%%%%%%%%%%%%%%%%%%%%%%%%% Fig. 12 %%%%%%%%%%%%%%%%%%%%%%%%%%%%%%%%%%%%%%%%
\begin{figure}[b] 
\begin{minipage}[h]{7cm}
    \epsfig{file=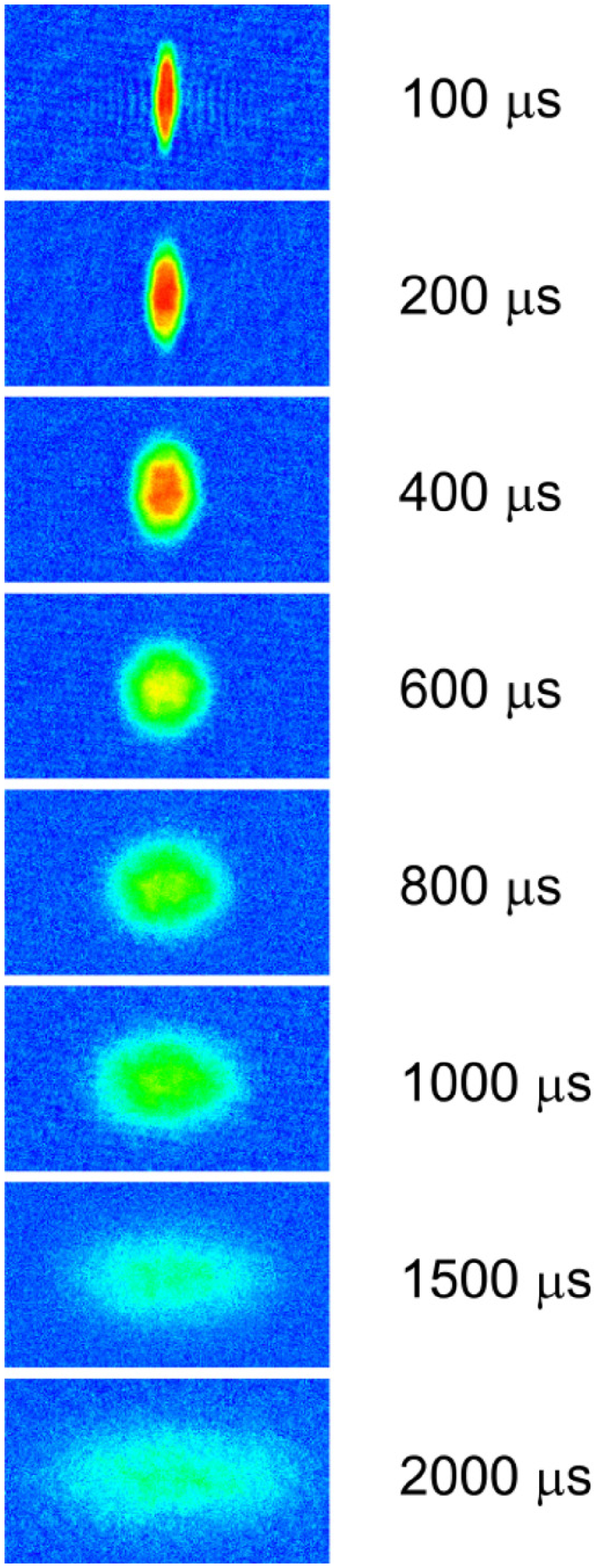,width=7cm} 
\end{minipage}
\hspace*{-5mm}
\begin{minipage}[h]{7cm}
    \epsfig{file=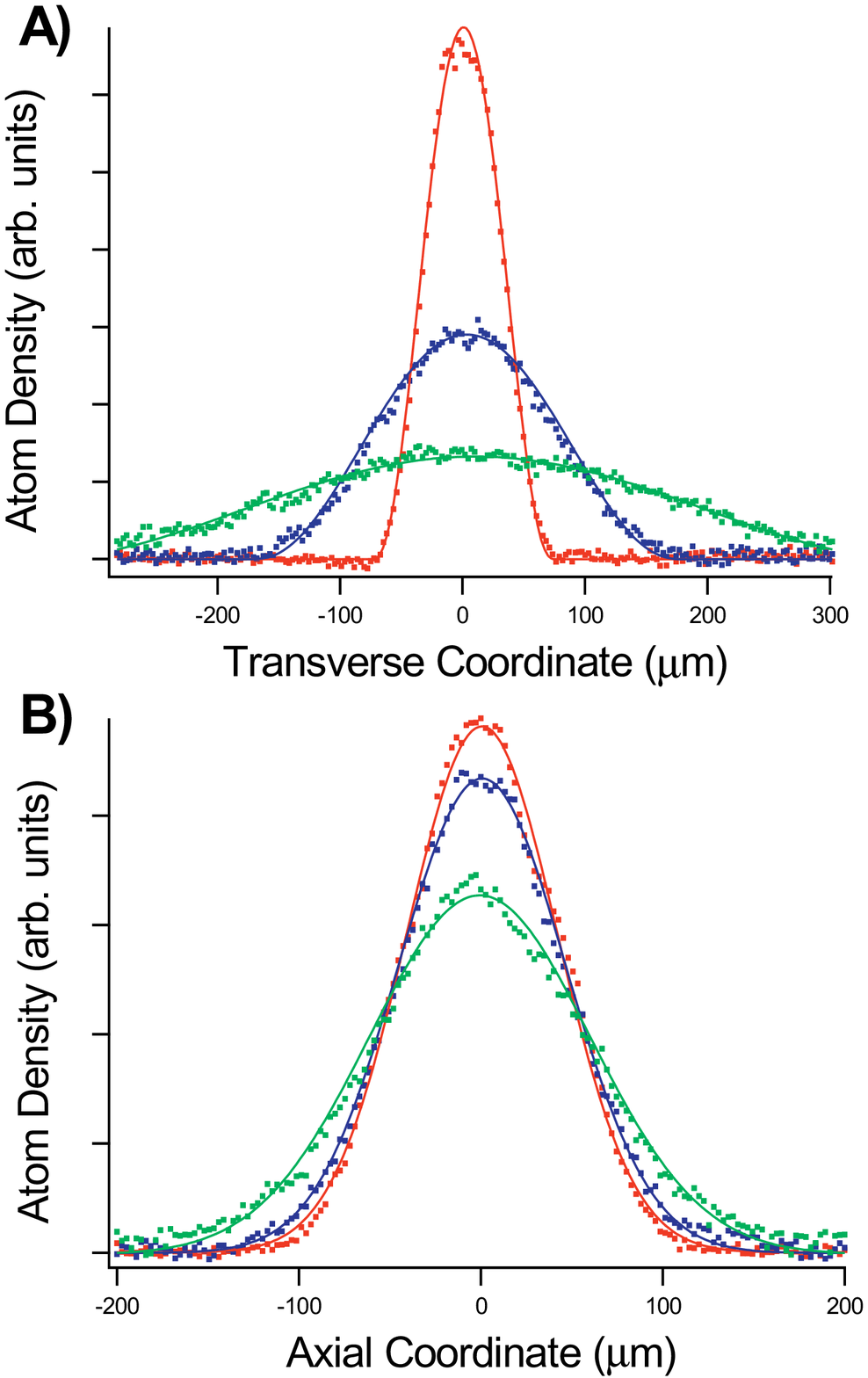,width=7cm,height=10cm}\\[4ex]
\hspace*{5mm}
\begin{minipage}[h]{7cm}
\caption{\protect
{\sl Left:} False color absorption images of a strongly interacting
degenerate Fermi gas of ultracold $^6$Li atoms as a function of time after
release from a laser trap. 
{\sl Right:} Atomic density distributions in the
initially shorter (top) and longer (bottom) directions at times
0.4\,ms (red, narrowest), 1.0\,ms (blue) and 2.0\,ms (green, widest)
after release from the trap. Reprinted with permission from O'Hara 
{\it et al.} \protect\cite{OHGGT02}
\copyright\, 2002 AAAS.
\label{fig:atomtrap}
}
\end{minipage} 
\end{minipage}
\vspace*{-3mm}
\end{figure} 
%%%%%%%%%%%%%%%%%%%%%%%%%%%%%%%%%%%%%%%%%%%%%%%%%%%%%%%%%%%%%%%%%%%%%%%
%

%
We close this Section with a beautiful example of elliptic flow from 
outside the field of heavy-ion physics where the hydrodynamically
predicted spatial expansion pattern shown in 
Figure~\ref{fig:evolutionsnapshots} has for the first time been 
directly observed experimentally \cite{OHGGT02}:
Figure~\ref{fig:atomtrap} shows absorption images of an ensemble 
of about 200,000 $^6$Li atoms which were captured and cooled to ultralow
temperatures in a CO$_2$ laser trap and then suddenly released by
turning off the laser. 
The trap is highly anisotropic, creating a pencil-like initial 
spatial distribution with an aspect ratio of about 29 between the 
length and diameter of the pencil.
The interaction strength among the fermionic atoms can be tuned with 
an external magnetic field by exploiting a Feshbach resonance. 
The pictures shown in Figure~\ref{fig:atomtrap} correspond to the case 
of very strong interactions.
The right panels in Figure~\ref{fig:atomtrap} show that the fermion gas
expands in the initially short (``transverse'') direction much more 
rapidly than along the axis of the pencil. 
As argued in the paper \cite{OHGGT02}, the measured expansion rates
in either direction are consistent with hydrodynamic calculations
\cite{MPS02}.
At late times the gas evolves into a pancake oriented perpendicular to the 
pencil axis. 
The aspect ratio passes through 1 (i.e. $\epsilon_x\eq0$) about
600\,$\mu$s after release and continues to follow the hydrodynamic
predictions to about 800\,$\mu$s after release. 
At later times it continues to grow, but more slowly than predicted by 
hydrodynamics, perhaps indicating a gradual breakdown of local thermal 
equilibrium due to increasing dilution.
It is important to note that this phenomenon is {\em only} observed
if the atoms interact strongly through the Feshbach resonance; off-resonance
their interaction is weak, and instead of the pattern shown in 
Fig.~\ref{fig:atomtrap} one observes ballistic expansion in
all directions, with the cloud becoming spherical at late times.
This shows that hydrodynamic behaviour, manifesting itself here in
elliptic flow, requires a strongly coupled fluid.
%

%%%%%%%%%%%%%%%%%%%%%%%%%%%%%%%%%%%%%%%%%%%%%%%%%%%%%%%%%%%%%%%%%%%%%%%
% observables.tex 
% last edited by UH on 1/27/09
%%%%%%%%%%%%%%%%%%%%%%%%%%%%%%%%%%%%%%%%%%%%%%%%%%%%%%%%%%%%%%%%%%%%%%%

%%%%%%%%%%%%%%%%%%%%%%%%%%%%%%%%%%%%%%%%%%%%%%%%%%%%%%%%%%%%%%%%%%%%%%%%%
\section{Comparison with experiment}
\label{sec:observables}
%%%%%%%%%%%%%%%%%%%%%%%%%%%%%%%%%%%%%%%%%%%%%%%%%%%%%%%%%%%%%%%%%%%%%%%%%

In heavy-ion collisions, the small size and short lifetime prohibit a 
similar direct observation of the spatial evolution of the fireball.
Only the momenta of the emitted particles are directly experimentally
accessible, and spatial information must be extracted somewhat indirectly
using {\em momentum correlations}.
We here discuss the single-particle hadron momentum spectra measured
at RHIC.
These test the space-time integrated aspects of collective flow in
the dynamical models, through their manifestation in the final momentum
distributions, in particular through their dependence on the hadron rest
masses.
A comprehensive review of two-particle correlations and their 
implications for the space-time structure of the collision fireball
is presented in the accompanying article by Lisa and Pratt 
\cite{Lisa:2008gf}.
This section consists of three parts.
In the first two we discuss 1) the rapidity and azimuthally integrated 
transverse momentum spectra and 2) the momentum-space anisotropies, in 
particular of the elliptic flow of various hadron species.
In the third part we focus on signs of failure of the ideal fluid 
dynamical approach and point to evidence for viscous effects.
We will concentrate on available comparisons with hydrodynamic 
model calculations.
Many more experimental details and data on momentum anisotropies
can be found in the accompanying review by Voloshin, Poskanzer and
Snellings \cite{Voloshin:2008dg}.
%
 
%%%%%%%%%%%%%%%%%%%%%%%%%%%%%%%%%%%%%%%%%%%%%%%%%%%%%%%%%%%%%%%%%%%%%%%
% section: MOMENTUM SPACE OBSERVABLES
% last updated 1/27/09
%%%%%%%%%%%%%%%%%%%%%%%%%%%%%%%%%%%%%%%%%%%%%%%%%%%%%%%%%%%%%%%%%%%%%%%
\suse{Azimuthally integrated momentum spectra}
\label{sec:momspacobservables}
%%%%%%%%%%%%%%%%%%%%%%%%%%%%%%%%%%%%%%%%%%%%%%%%%%%%%%%%%%%%%%%%%%%%%%%

The primary single-particle observables in heavy-ion collisions are 
the triple-differential momentum distributions of identified hadrons 
$i$ as a function of collision centrality (impact parameter $b$):
\beq{equ:fourierexpansion}
    \frac{dN_i}{\pT d\pT \, dy \, d\phi_p}(b) 
  = \frac{1}{2 \pi}\frac{dN_i}{\pt d\pt \, dy}(b) 
    \Bigl( 1 + 2 \sum_{n=1}^\infty\,v_n^i(\pt,y;b) \cos(n\phi_p) \Bigr) \,.
\end{equation}
We have expanded the dependence on the azimuthal emission angle $\phi_p$
relative to the reaction plane into a Fourier series \cite{VZ96}. 
Due to reflection symmetry with respect to the reaction plane, only
cosine terms appear in the expansion. 
At midrapidity $y\eq\ln[(E{+}p_z)/(E{-}p_z)]\eq0$ all odd harmonics 
(in particular the {\em directed flow} coefficient $v_1^i$) vanish
in symmetric collisions.
We begin by studying in Sec.~\ref{rapidity} the $\pt$- and 
$\phi_p$-integrated rapidity distributions $dN_i/dy$. 
Section~\ref{ptdist} will focus on the $\phi_p$-integrated 
transverse momentum distributions at midrapidity.
In Sec.~\ref{sec:ellipticflow} finally, we discuss the azimuthal 
momentum anisotropies, in particular the elliptic flow coefficient
$v_2(\pt,y;b)$.
%

%%%%%%%%%%%%%%%%%%%%%%%%%%%%%%%%%%%%%%%%%%%%%%%%%%%%%%%%%%%%%%%%%%%%%%%%%
\sususe{Rapidity distributions}
\label{rapidity}
%%%%%%%%%%%%%%%%%%%%%%%%%%%%%%%%%%%%%%%%%%%%%%%%%%%%%%%%%%%%%%%%%%%%%%%%%
%
The final hadron rapidity distributions reflect their longitudinal
collective dynamics at freeze-out.
At high collision energies, the theoretically best justified initial
particle production models implement longitudinal boost-invariance
which identifies the particles'\ initial rapidity $y$ with the
space-time rapidity $\eta_s$ of their production point.
Even after the particles begin to interact with each other, this imparts 
on the ensemble of produced particle a collective longitudinal expansion 
characterized by the identity $y_L=\eta_s$ (where $y_L$ is the average 
flow rapidity of the particles in a cell located at space-time
rapidity $\eta_s$). 
At high energies, this initial longitudinal collectivity completely 
dominates the final motion in beam direction.
This is true even though the {\em density} of produced particles is
not boost-invariant (i.e. it depends on rapidity). 
Hydrodynamic deceleration or acceleration effects due to longitudinal
density and pressure gradients are weak to negligible, such that
the final rapidity distribution of the particles closely resembles
their initial space-time rapidity distribution.
Since collective flow affects hadrons of different masses in 
characteristic ways, and these masses appear only after hadronization 
but play no role in the initial particle production at the partonic level, 
one might hope that by comparing rapidity distributions of different 
hadron species one could explore the validity of the hydrodynamic picture.
However, such mass-dependent flow effects are concentrated at low 
momenta, i.e. one would have to search for them near midrapidity ($y{=}0$)
where all rapidity distributions are flat and thus have essentially
the same shape \cite{Arsene:2004fa}.
In any case, no hydrodynamic model comparisons with rapidity spectra of 
identified hadrons have been done so far.
%
 
%
%%%%%%%%%%%%%%%%%%%%%%% Fig. 13 %%%%%%%%%%%%%%%%%%%%%%%%%%%%%%%%%%%%%%%%%
\begin{figure}[ht]
\centerline{
            \epsfig{file=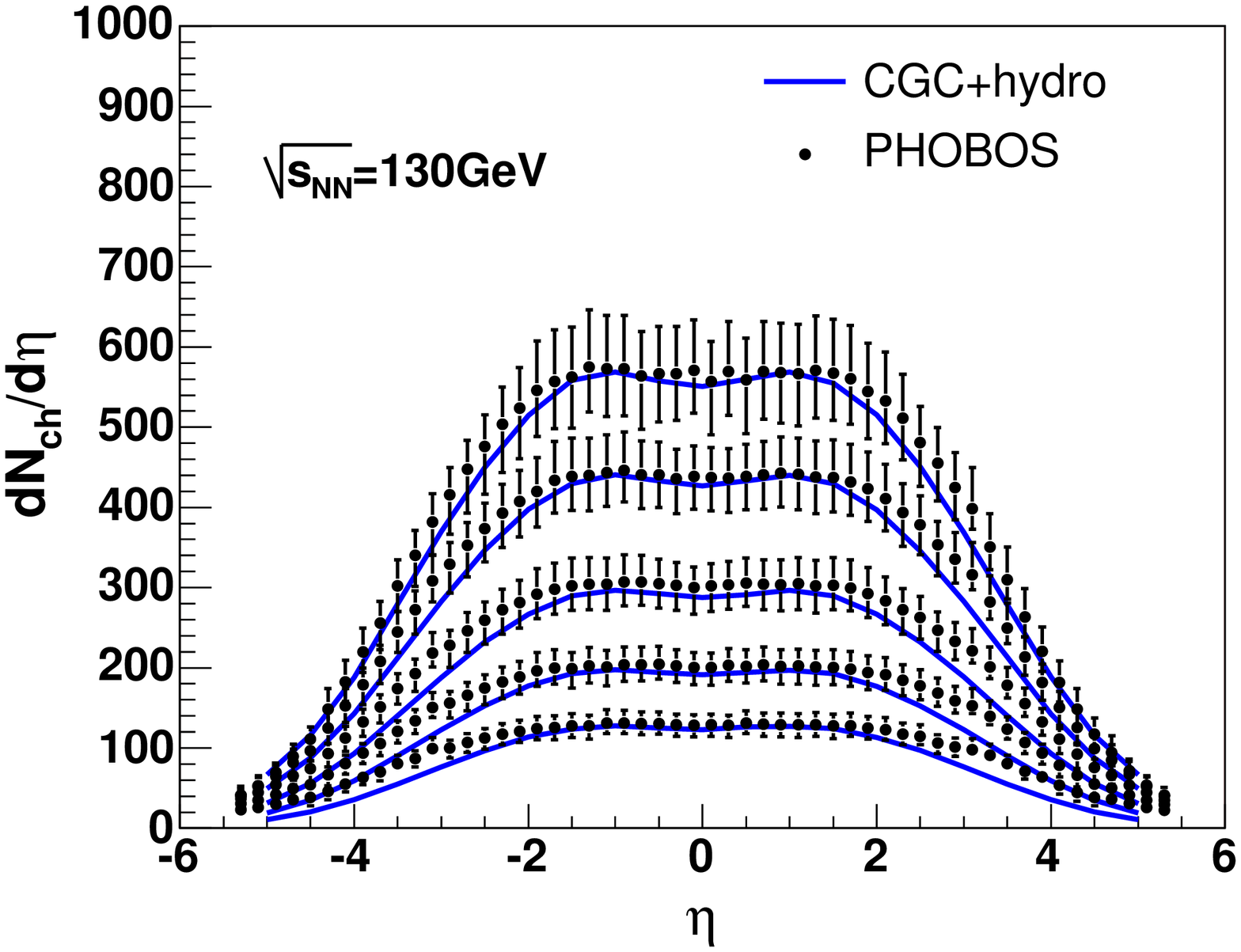,width=7.5cm}
\hspace*{5mm}
            \epsfig{file=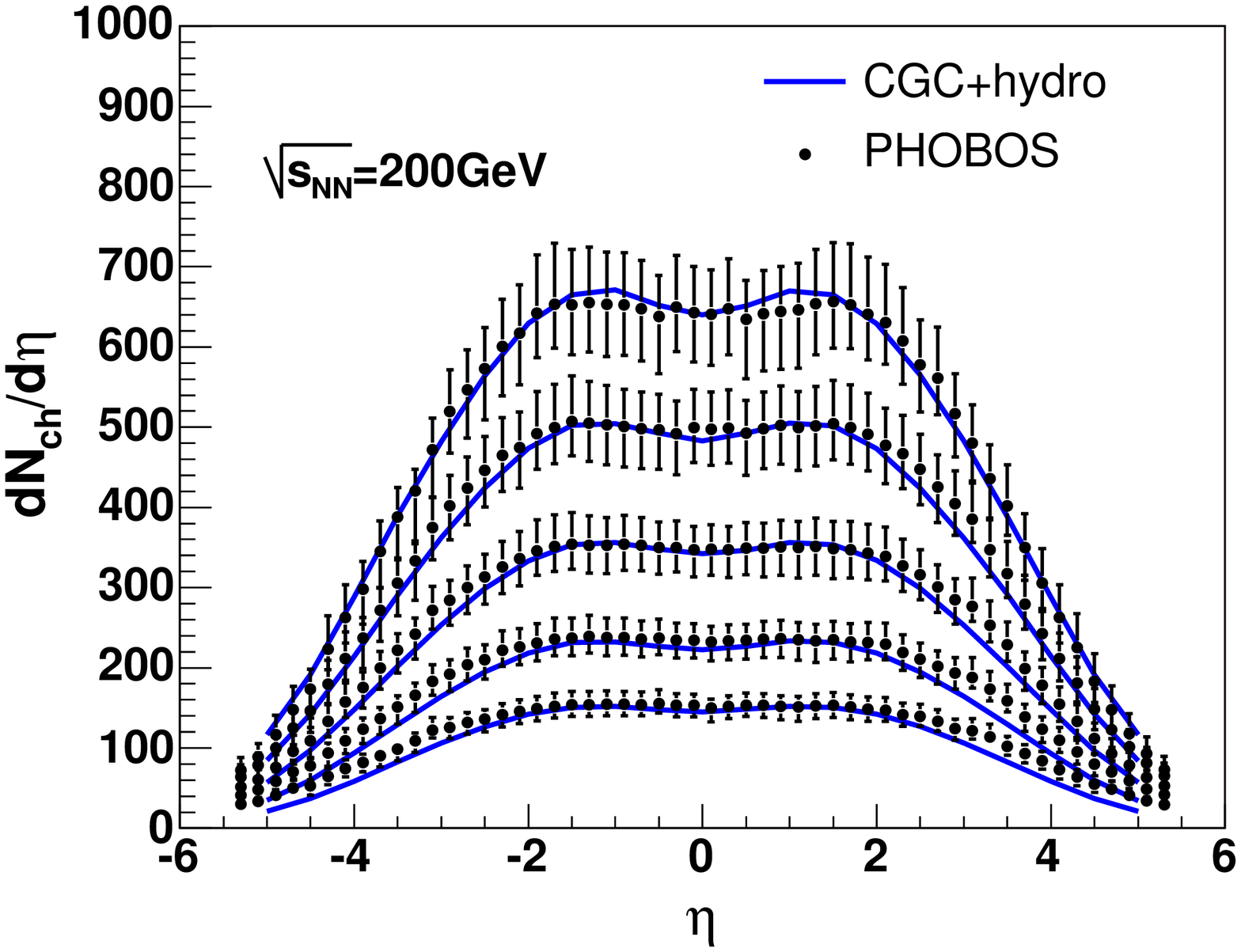,width=7.5cm}
            }  
\caption{Charged hadron pseudorapidity ($\eta{=}\frac{1}{2}\ln
         [(p{+}p_z)/(p{-}p_z)]$) distributions from
         Au+Au collisions at $\scm{=}130$\,GeV (left) and
         $\scm{=}200$\,GeV (right) for 5 different centrality
         bins. The data are from the PHOBOS 
         experiment \protect\cite{Back:2002wb}.
         The lines show (3+1)-d ideal fluid dynamical simulations
         with initial distributions derived from the CGC model
         (see Eq.~(\ref{eq:ktfac}) \protect\cite{Hirano:2004en}.
\label{F13} 
} 
\vspace*{-3mm}
\end{figure} 
%%%%%%%%%%%%%%%%%%%%%%%%%%%%%%%%%%%%%%%%%%%%%%%%%%%%%%%%%%%%%%%%%%%%%%%
%

%
We conclude that the measured charged hadron rapidity distributions 
test our ability to predict the initial particle distribution in rapidity
space, but not much else. 
Of course, one can always parametrize the initial space-time rapidity
distribution such that the final charged hadron distribution matches 
experiment, and this was indeed done in many (3+1)-d hydrodynamical 
studies \cite{Morita:1999vj,Hirano:2001rg,Hirano:2001yi,Morita:2002av,%
Nonaka:2006yn}.
Such a procedure provides, however, little predictive power.
To reproduce the collision energy dependence one must adjust
parameters, and the impact parameter dependence is largely dictated
by the overlap geometry in the transverse plane. 
Still, such studies is have provided one important insight
\cite{Nonaka:2006yn}:
For the rapidity distributions, it is irrelevant whether one ends
the hydrodynamic evolution suddenly via Cooper-Frye freeze-out or
allows decoupling to happen gradually by matching the hydrodynamic
model to a hadronic cascade for the late hadronic stage.
The resulting charged hadron rapidity distributions are identical.
There exists one model that claims to {\em predict} the rapidity,
beam energy and centrality dependence of the initial particle 
production: the Color Glass Condensate theory \cite{Kharzeev:2001gp}.
Figure~\ref{F13} shows that this claim is well supported by experiment.
This is a non-trivial success of the CGC model.
The Glauber model and its generalizations to non-zero 
rapidity \cite{Sollfrank:1998js,Heinz:2004et,Adil:2005qn,Hirano:2005xf}
cannot predict the collision energy dependence of the
rapidity distribution nor of the hard fraction $(1{-}\alpha)$ in 
Eq.~(\ref{alpha}) that controls the non-linearity of the charged
hadron multiplicity as a function of participant number $N_\mathrm{part}$.
The CGC model does so successfully.
%

%%%%%%%%%%%%%%%%%%%%%%%%%%%%%%%%%%%%%%%%%%%%%%%%%%%%%%%%%%%%%%%%%%%%%%%%%
\sususe{Transverse momentum and transverse mass distributions at
midrapidity}
\label{ptdist}
%%%%%%%%%%%%%%%%%%%%%%%%%%%%%%%%%%%%%%%%%%%%%%%%%%%%%%%%%%%%%%%%%%%%%%%%%

%
The parameters of the hydrodynamic model are 
fixed by reproducing the measured centrality dependence of the total 
charged multiplicity $dN_\mathrm{ch}/dy$ as well as the shape of the 
pion and proton $\pt$-spectra in central collisions at midrapidity.
The shapes of other hadron spectra, their centrality dependence and 
the dependence of the elliptic flow coefficient $v_2^i$ on $\pt$, 
centrality and hadron species $i$ are then all parameter free 
predictions of the model \cite{HKHRV01}. 
These predictions will be compared with experiment and used to test 
the hydrodynamic approach and to extract physical information from its
successes and failures.
The free parameters of the hydrodynamic model are the starting 
(thermalization) time $\tau_\equ$, the entropy and net baryon
density in the center of the reaction zone at this time, and the 
freeze-out energy density $e_\dec$. 
The corresponding quantities at other fireball points at $\tau_\equ$ 
are then determined by the Glauber profiles discussed in 
Sec.~\ref{sec:initialization}.
The ratio of net baryon to entropy density is fixed by the measured
proton/pion ratio.
Since the measured chemical composition of the final state at RHIC
was found \cite{BMMRS01} to accurately reflect a hadron resonance gas 
in chemical equilibrium at the hadronization phase transition, we
require the hydrodynamic model to reproduce this $p/\pi$ ratio on a 
hypersurface of temperature $\Tcrit$.
In ideal fluid dynamics, the final total charged multiplicity 
$dN_{\rm ch}/dy$ fixes by entropy conservation the initial product 
$(s\cdot\tau)_\equ$ \cite{KSH99,Bjorken83,Ollitrault92}. 
The value of $\tau_\equ$ controls how much transverse flow can be 
generated until freeze-out.
Since the thermal motion and radial flow affect light and heavy particles 
differently at low $\pt$ \cite{SSHPRC93,LHS90}, a simultaneous fit of the 
final pion and proton spectra separates the radial flow from the thermal
component.
The final flow strength then ``fixes''%
\footnote{It should be noted that this determination of $\tau_\equ$
is not very precise since earlier starting times also lead to earlier 
freeze-out, limiting the buildup of radial flow. One really obtains
only an upper limit for $\tau_\equ$ in this way. In viscous hydrodynamics 
one must additionally reduce the product $(s\cdot\tau)_\equ$ when 
shortening $\tau_\equ$, to account for entropy viscous 
production \cite{Luzum:2008cw}. The consequences of this for the final 
shape of the $\pt$-spectra have not yet been fully explored.}  
$\tau_\equ$ whereas the freeze-out temperature determines the energy 
density $e_\dec$ at decoupling.
%

%
%%%%%%%%%%%%%%%%%%%%% Fig. 14 %%%%%%%%%%%%%%%%%%%%%%%%%%%%%%%%%%%%%%%%%%%%
\begin{figure}[ht]
\begin{center}
\epsfig{file=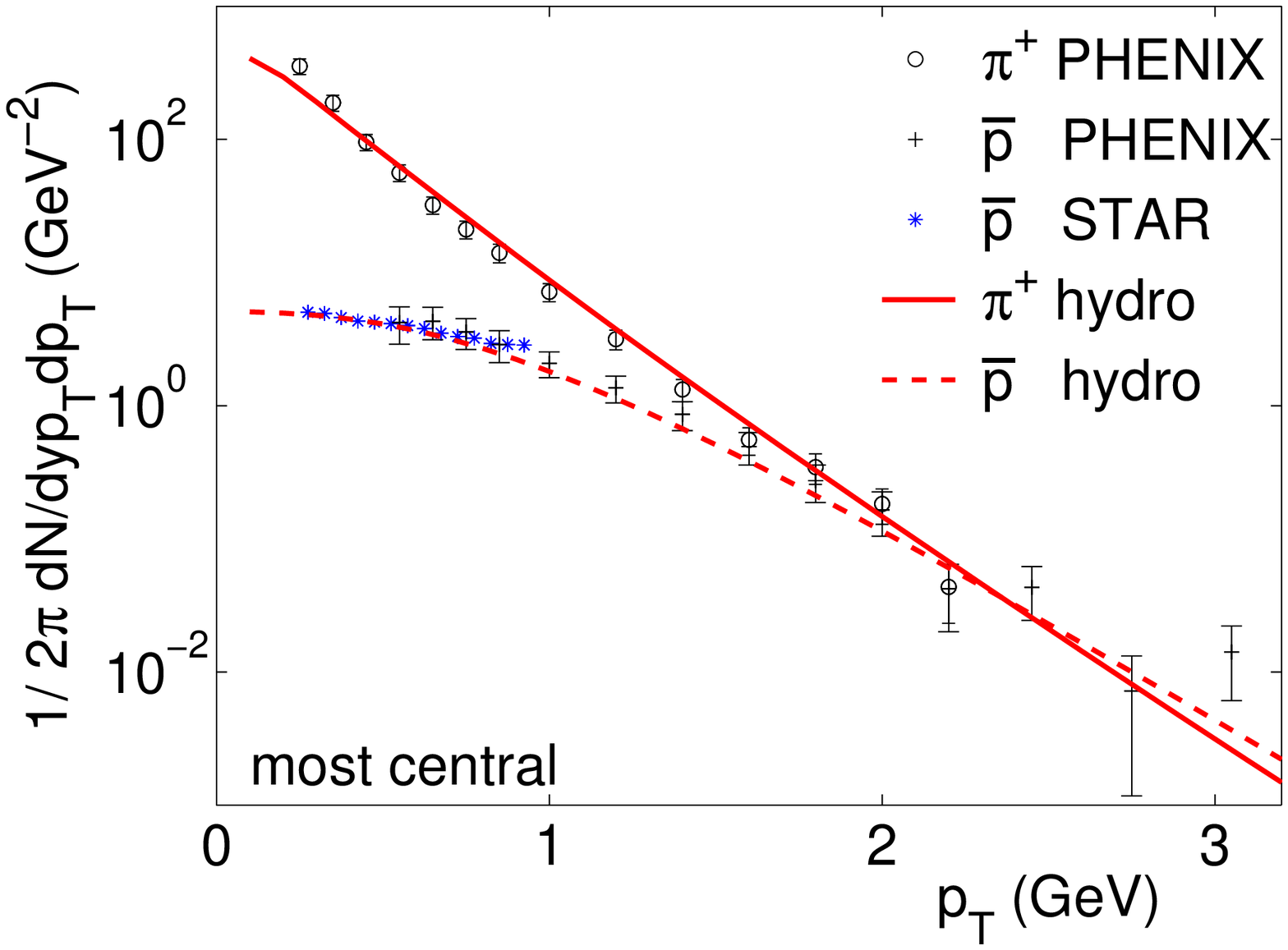,width=56mm,height=50mm}
\epsfig{file=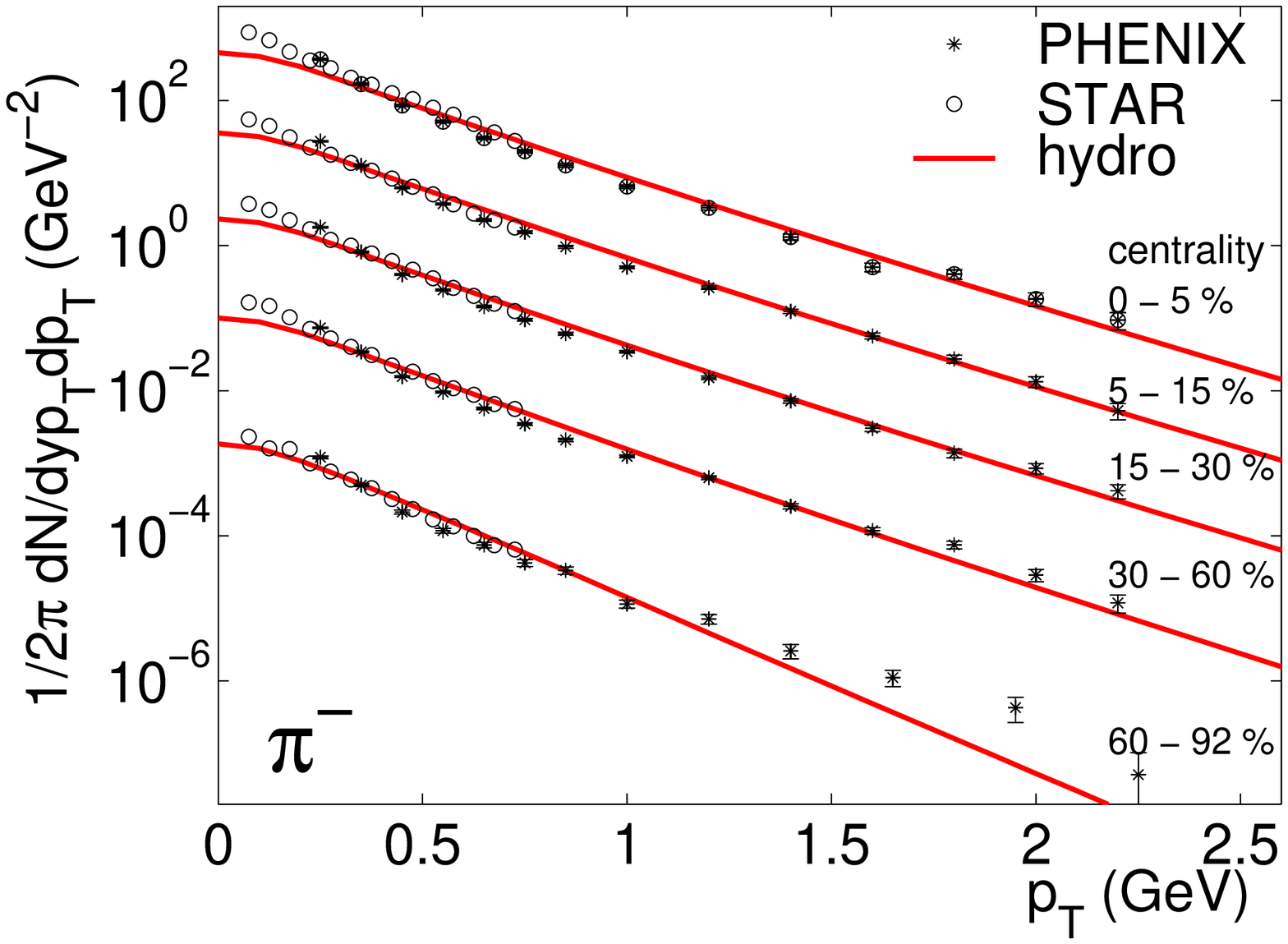,width=56mm,height=50mm}
\\
\epsfig{file=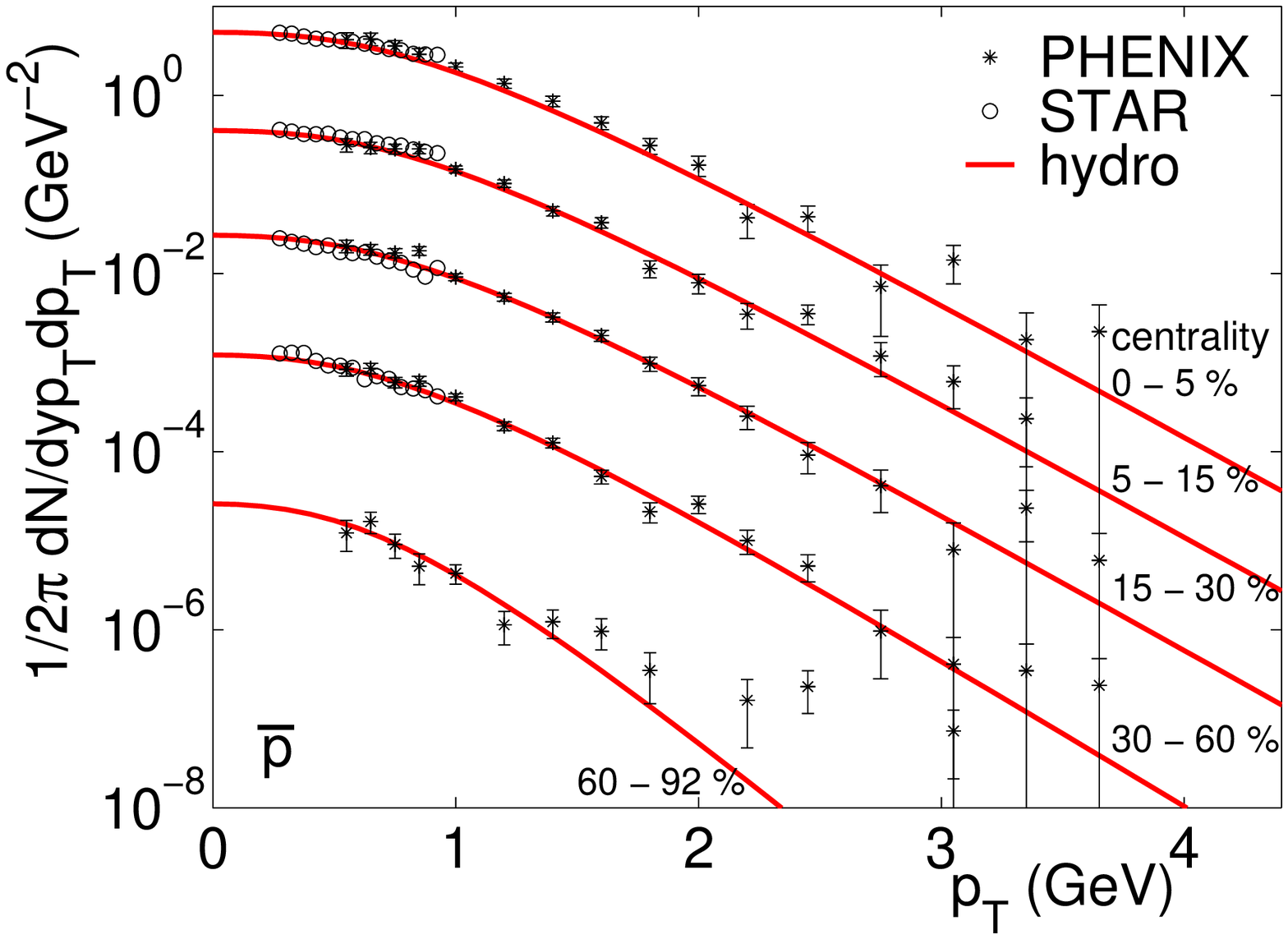,width=56mm,height=50mm}
\epsfig{file=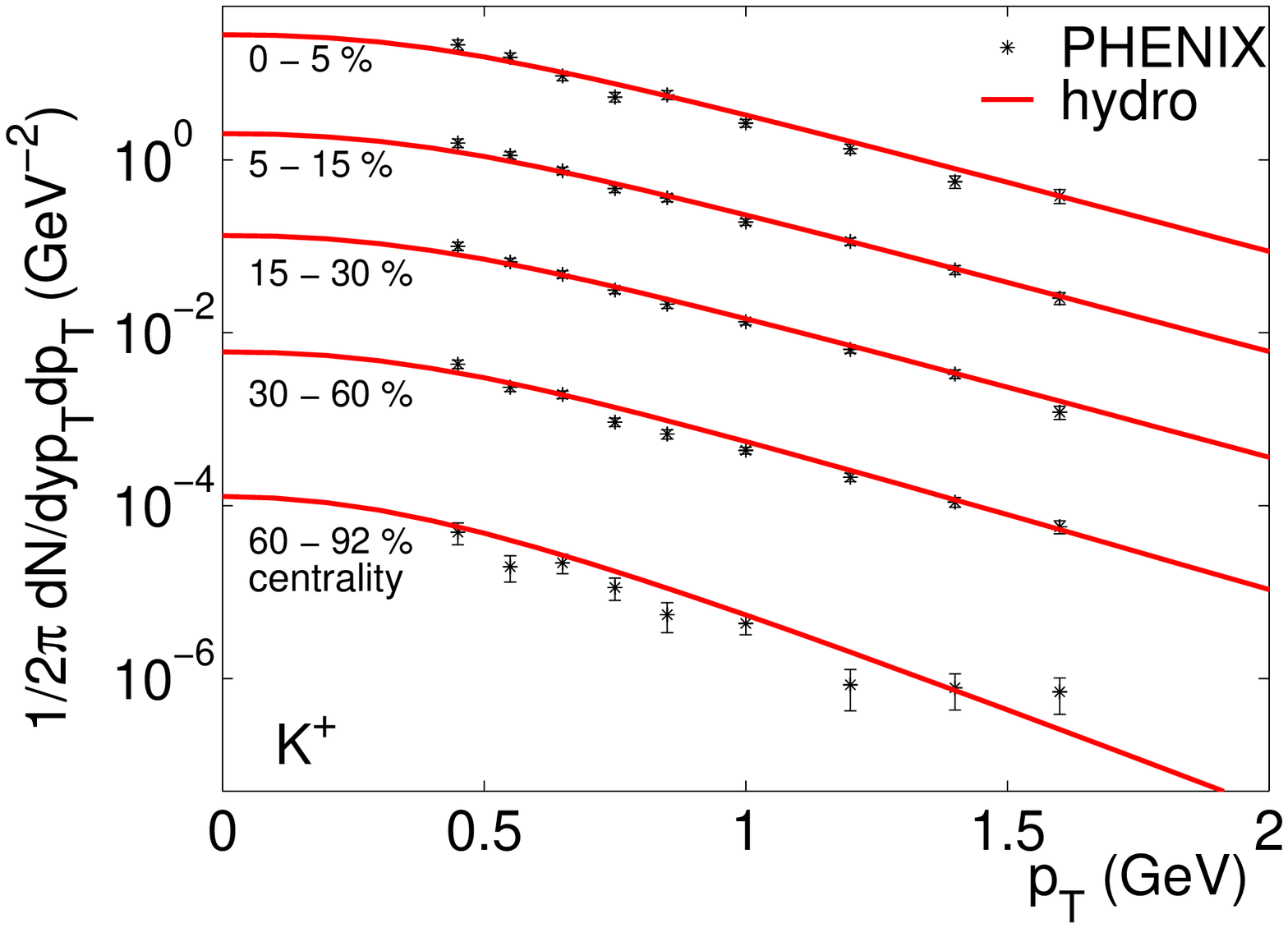,width=56mm,height=50mm}
\\[-5mm]
\end{center}
%\vspace*{-5mm}
\caption{\label{fig:spectra130}
         Identified pion, antiproton and kaon spectra for 
         $\sqrt{s_{\rm NN}}=130$~GeV from the 
         PHENIX \protect\cite{PHENIX01spec,PHENIX02spec} and 
         STAR \protect\cite{STAR01spec,Sanchez02} collaborations
         in comparison with results from an ideal fluid dynamical 
         calculation \protect\cite{HK02WWND}.
         The top left panel shows pion and (anti-)proton spectra from
         central collisions. Shown in the other panels are spectra
         of five different centralities: from most central (top) to 
         the most peripheral (bottom). The spectra are successively
         scaled by a factor 0.1 for clarity.
}
\vspace*{-4mm}
\end{figure}
%%%%%%%%%%%%%%%%%%%%%%%%%%%%%%%%%%%%%%%%%%%%%%%%%%%%%%%%%%%%%%%%%%%%%%%%%%%%%

%
The top left panel of Fig.~\ref{fig:spectra130} shows the ideal hydrodynamic 
fit \cite{HK02WWND} to the transverse momentum spectra of positive pions 
and antiprotons, as measured by the PHENIX and STAR collaborations in 
central ($b\eq0$) Au+Au collisions at $\sqrt{s}\eq130\,A$\,GeV
\cite{PHENIX01spec,PHENIX02spec,STAR01spec,Sanchez02}.
The fit yields an initial central entropy density $s_\equ\eq95$~fm$^{-3}$ 
at an equilibration time $\tau_\equ\eq0.6$~fm.
This corresponds to an initial temperature of $T_\equ\eq340$~MeV and
an initial energy density $e\eq25$\,GeV/fm$^3$ in the fireball center.
(Note that these parameters satisfy the ``uncertainty relation''
$\tau_\equ\cdot T_\equ \approx 1$.)
Freeze-out was implemented on a hypersurface of constant energy density 
with $e_\dec\eq0.075$~GeV/fm$^3$.
%

%%%%%%%%%%%%%%%%%%%%%%%  Table 1 %%%%%%%%%%%%%%%%%%%%%%%%%%%%%%%%%%%%%%%%%%%%
\begin{table}[htdp]
\vspace*{-2mm}
\begin{center}
\begin{tabular}{|c|c|c|c|}
\hline
                       &  SPS   & RHIC\,1     & RHIC\,2    \\
$\scm$ (GeV)           &  17    &   130     &  200      \\
\hline
$s_\equ$ (fm$^{-3}$)   &  43    &    95     &  110      \\
$T_\equ$ (MeV)         &  257   &   340     &  360      \\
$\tau_\equ$ (fm/$c$)   &  0.8   &   0.6     &  0.6      \\
\hline
\end{tabular}
\end{center}
\begin{tabnote}
         Table 1. Initial conditions for SPS and RHIC energies used to 
         fit the particle spectra from central Pb+Pb or Au+Au collisions.
         $s_\equ$ and $T_\equ$ refer to the maximum values at $\tau_\equ$ 
         in the fireball center. 
\end{tabnote}
\label{tab:initialconditions}
\vspace*{-2mm}
\end{table}
%%%%%%%%%%%%%%%%%%%%%%%%%%%%%%%%%%%%%%%%%%%%%%%%%%%%%%%%%%%%%%%%%%%%%%%%%

%
The fit in the top left panel of Fig.~\ref{fig:spectra130} was 
performed with EOS~Q which assumes chemical equilibrium in the HRG 
phase all the way down to $T_\mathrm{dec}$. 
The chemical equilibrium value for the $\bar p/\pi$ ratio at this 
temperature does not agree with experiment which indicates chemical
freeze-out at $T_c$ (see Fig.~\ref{F4}). 
The normalization of the other hadron spectra relative to that of the
pions must thus be adjusted by hand.
The information that is required to fix the initial and freeze-out 
conditions resides, however, in the shape (and not in the normalization)
of the pion and proton spectra.
After these conditions have been fixed, the shapes of other hadron 
spectra in central collisions are model predictions.  
Once their normalizations have been adjusted to reflect the measured
abundances in central collisions, the spectra of all hadron species
(shapes and normalizations) in non-central collisions are predicted
by the model without additional parameters.
The remaining three panels of Fig.~\ref{fig:spectra130} show the 
predicted transverse momentum spectra of pions, kaons and antiprotons 
in five different centrality bins, compared with measurements by the 
PHENIX \cite{PHENIX01spec,PHENIX02spec} and 
STAR \cite{STAR01spec,Sanchez02} collaborations. 
For all centrality classes, except the most peripheral one, the ideal
fluid dynamical predictions (solid lines) agree quite well with the data.
The kaon spectra are reproduced almost perfectly, but for pions the 
model consistently underpredicts the data at low $\pt$.
This has now been understood to be largely an artifact of having employed 
in these calculations a chemical equilibrium equation of state all the
way down to kinetic freeze-out.
Later calculations \cite{KR03} with a chemical non-equilibrium 
HRG equation of state, which will be compared to 
$\sqrt{s}\eq200\,A$\,GeV data below, show that, as the system cools 
below the chemical freeze-out point $T_{\rm chem}{\,\approx\,}\Tcrit$, 
a significant positive pion chemical potential builds up, emphasizing 
the concave curvature of the spectrum from Bose effects and increasing 
the feeddown corrections from heavier resonances at low $\pt$.
The inclusion of non-equilibrium baryon chemical potentials to avoid 
baryon-antibaryon annihilation further amplifies the resonance feeddown
for pions.
Significant discrepancies are also seen at large impact parameters
and large transverse momenta $\pt\,\gapp\,2.5$\,GeV/$c$.
This is not surprising since high-$\pt$ particles require more 
rescatterings to thermalize and escape from the fireball before doing so. 
This is in particular true in more peripheral collisions where the 
reaction zone is smaller.
For the calculations shown in Fig.~\ref{fig:spectra130} the same value
$\edec$ was used for all impact parameters. 
Recent work \cite{Heinz:2006ur} using the kinetic freeze-out criterium 
(\ref{focrit}) has shown that in peripheral collisions the fireball 
decouples at somewhat higher temperatures than in central collisions,
in agreement with the data shown in Fig.~\ref{F4}.
The consequences of this for the centrality dependence of the $\pt$-spectra
have not yet been explored within the hydrodynamic model.
Without transverse flow, thermal spectra exhibit 
{\em $\mt$-scaling} \cite{HR68}, i.e. after appropriate rescaling of 
the yields all spectra collapse onto a single curve. 
Transverse collective flow breaks this scaling at low $\pt\lapp m_0$ 
(i.e. for non-relativistic transverse particle velocities) by an amount 
which increases with the particle rest mass $m_0$ \cite{SR79,LHS90,SSH93}.
When plotting the spectra against $\pt$ instead of $\mt$, any breaking 
of $\mt$-scaling is at least partially masked by a kinematic effect at 
low $\pt$ that, unfortunately, again increases with the rest mass $m_0$.
To visualize the effects of transverse flow on the spectral shape thus
requires plotting the spectrum logarithmically as a function of $\mt$ 
or $\mt{-}m_0$.
Such plots \cite{STARlambdas,STAR_mtspec,STAR_Xi,STAR03omega} show a clear 
tendency of the heavier hadron spectra to curve and to begin to develop a 
shoulder at low transverse kinetic energy $\mt{-}m_0$, as expected from 
transverse flow.
%

%
%%%%%%%%%%%%%% Fig. 15 %%%%%%%%%%%%%%%%%%%%%%%%%%%%%%%%%%%%%%%%%%%%%%%%%%
%\vspace*{-3mm}
\begin{figure}[htb]
\centerline{\epsfig{file=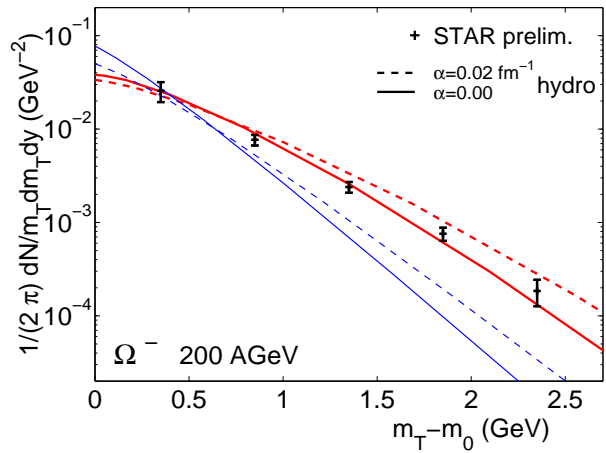,width=8cm,height=6cm}}
%\hspace*{4.5mm}
\caption{\label{fig:omegaspec}
   Transverse mass spectrum of $\Omega$ hyperons from central 
   200\,$A$\,GeV Au+Au collisions at RHIC \protect\cite{STAR03omega}.   
   The curves are ideal hydrodynamic calculations with different initial
   and freeze-out conditions: Solid lines correspond to the default
   of no initial transverse flow at $\tau_\equ$, dashed lines assume
   a small but non-zero radial flow, $v_r=\tanh(\alpha  r)$ with 
   $\alpha=0.02$~fm$^{-1}$, already at $\tau_\equ$. The lower (thin) 
   set of curves assumes $\Omega$-decoupling at $\Tcrit\eq164$\,MeV,
   the upper (thick) set of curves decouples the $\Omega$ together
   with the pions and protons at $T_\dec\eq100$\,MeV \protect\cite{KR03}.
}
\end{figure}
%\vspace*{-3mm}
%%%%%%%%%%%%%%%%%%%%%%%%%%%%%%%%%%%%%%%%%%%%%%%%%%%%%%%%%%%%%%%%%%%%%%%%%
%

\begin{figure} 
\vspace*{-2mm}
\centerline{
%\fbox{\rule[-15mm]{0cm}{3cm}{\em To create a place-holder}}
            \epsfig{file=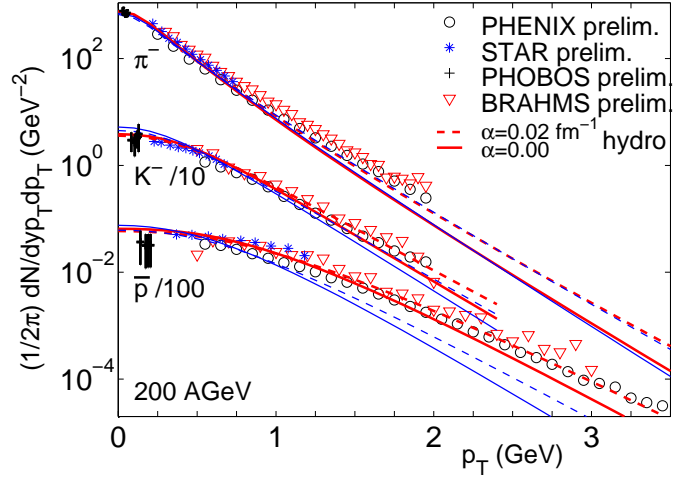,width=9cm} 
            }  
\vspace*{-2mm}
\caption{Particle spectra of $\pi^-$, $K^-$ and antiprotons at 
         $\scm=200$~GeV as measured by the four large experiments 
         at RHIC \protect\cite{PHENIX03spec200,STAR03spec200,%
         PHOBOS03spec200,BRAHMS03spec200}. The lines show hydrodynamic 
         results under various considerations (see text) \protect\cite{KR03}.
\label{fig:spectra200GeV} 
} 
\vspace*{-2mm}
\end{figure}

%
One such example is shown in Fig.~\ref{fig:omegaspec} where 
$\Omega$ hyperons spectra \cite{STAR03omega} are compared with 
hydrodynamic predictions. 
For this comparison the original calculations for 130\,$A$\,GeV Au+Au 
collisions \cite{HKHRV01} were repeated with RHIC2 initial conditions 
and a chemical non-equilibrium equation of state in the hadronic 
phase \cite{KR03}. 
The solid lines are based on default parameters (see 
Table~1) without any initial transverse flow at $\tau_\equ$.
(The dashed lines will be discussed further below.) 
Following a suggestion that $\Omega$ hyperons, being heavy and not 
having any known strong coupling resonances with pions, should not be 
able to participate in any increase of the radial flow during the 
hadronic phase and thus decouple early \cite{HSN98}, we show two solid
lines, the steeper one corresponding to decoupling at 
$\edec\eq0.45$\,GeV/fm$^3$, i.e. directly after hadronization at $\Tcrit$,
whereas the flatter one assumes decoupling together with pions and
other hadrons at $\edec\eq0.075$\,GeV/fm$^3$.
The data clearly favor the flatter curve, suggesting intense rescattering
of the $\Omega$'s in the hadronic phase.
The microscopic mechanism for this rescattering is still unclear.
However, without hadronic rescattering the hydrodynamic model, in spite
of its perfect local thermalization during the early expansion stages, 
is unable to generate enough transverse flow to flatten the $\Omega$
spectra as much as required by the data. 
Partonic hydrodynamic flow alone can not explain the $\Omega$ spectrum.
We now illustrate the effects on the ideal fluid dynamic particle 
spectra caused by correctly accounting for the non-equilibrium chemistry 
in the hadronic phase \cite{HT02,Rapp02,Teaney02,KR03,Hirano:2005wx,%
Huovinen:2007xh}. 
Figure~\ref{fig:spectra200GeV} shows a compilation of experimental pion, 
kaon and (anti-)proton spectra for 200\,$A$\,GeV Au+Au collisions from the 
four RHIC collaborations \cite{PHENIX03spec200,STAR03spec200,%
PHOBOS03spec200,BRAHMS03spec200}, compared with results from ideal 
hydrodynamics.
%
%%%%%%%%%%%%%%%%%%%%%%%%%% Fig. 16 %%%%%%%%%%%%%%%%%%%%%%%%%%%%%%%%%%%%%%%%%
%%%%%%%%%%%%%%%%%%%%%%%%%%%%%%%%%%%%%%%%%%%%%%%%%%%%%%%%%%%%%%%%%%%%%%%
%
%
The calculations (shown as thick solid red lines in 
Fig.~\ref{fig:spectra200GeV}) use the same decoupling energy density 
$e_\dec\eq0.075$\,GeV/fm$^3$ as before.
As discussed in Sec.~\ref{sec:nucleareos}, this corresponds to the
same flow strength as with the chemically equilibrated EOS, but a 
significantly lower freeze-out temperature 
$T_\dec{\,\approx\,}100$\,MeV \cite{Teaney02,HT02,KR03}.
The thin solid (blue) lines in the Figure, shown for comparison, were 
calculated by assuming kinetic freeze-out already at hadronization,
$\Tcrit\eq165$\,MeV.
The data clearly favor the additional radial boost resulting from the
continued buildup of radial flow in the hadronic phase.
Still, even at $e_\dec\eq0.075$\,GeV/fm$^3$, the spectra are still
steeper than the data and the previous calculations with a chemical
equilibrium equation of state shown in Fig.~\ref{fig:spectra130},
reflecting the combination of the same flow pattern with a lower 
freeze-out temperature.
Somewhat unexpectedly, the authors of the study \cite{KR03} were unable 
to significantly improve the situation by reducing $e_\dec$ even further:
The effects of a larger radial flow at lower $e_\dec$ were almost 
completely compensated by the accompanying lower freeze-out temperature,
leading to only modest improvements for kaons and protons and almost
none for the pions.   
The reason for this is \cite{Hirano:2005wx} the steep drop of $T$ with 
decreasing $e_\dec$ for the chemically non-equilibrated EOS (PCE) shown 
in the lower right panel of Fig.~\ref{F5}. 
This motivated the authors \cite{KR03} to introduce a small but
non-vanishing transverse ``seed'' velocity already at the beginning 
of the hydrodynamic stage (see also more recent work 
\cite{Broniowski:2008qk,Chojnacki:2004ec} and the Appendix of P. Kolb's 
thesis\cite{PFKthesis02}).
The dashed lines in Fig.~\ref{fig:spectra200GeV} (and also earlier in 
Fig.~\ref{fig:omegaspec}) show hydroynamic calculations
with an initial transverse flow velocity profile given by 
$v_r(r,\tau_\equ)\eq\tanh(\alpha \,r)$ with $\alpha=0.02$~fm$^{-1}$.
This initial transverse kick is seen to significantly improve the
agreement with the pion, kaon and antiproton data up to 
$\pt\gapp1.5-2$\,GeV/$c$ for pions and kaons and up to 
$\pt\gapp3.5$\,GeV/$c$ for (anti)protons \cite{KR03}.
It can be motivated by invoking some collective (although not ideal 
hydrodynamic) transverse motion of the fireball already during the 
initial thermalization stage.
However, this is not the only possible solution to the problem.
As discussed below, viscous effects in the late hadronic stage are
strong and contribute significantly \cite{Hirano:2007ei} to the 
required flattening of the spectra.
Such effects are not captured by an ideal fluid dynamical approach.
A great initial surprise at RHIC was the observation that
the antiproton/pion ratio increases with $\pt$ and actually exceeds 1
above $\pt\sim 2-2.5$\,GeV/$c$ \cite{PHENIX02spec,PHENIX03spec200}. 
Thermal momentum distributions boosted by hydrodynamical radial flow, 
combined with the small baryon chemical potential at RHIC, provide a 
natural explanation of this so-called ``$\bar p/\pi^-{\,>\,}1$ 
anomaly'' \cite{Kolb:2003dz}.
%

%
%%%%%%%%%%%%%%%%%%%% Fig. 17 %%%%%%%%%%%%%%%%%%%%%%%%%%%%%%%%%%%%%%%%%%%%%%%
\begin{figure}[htb]
%\vspace*{-4mm} 
\centerline{
            \epsfig{file=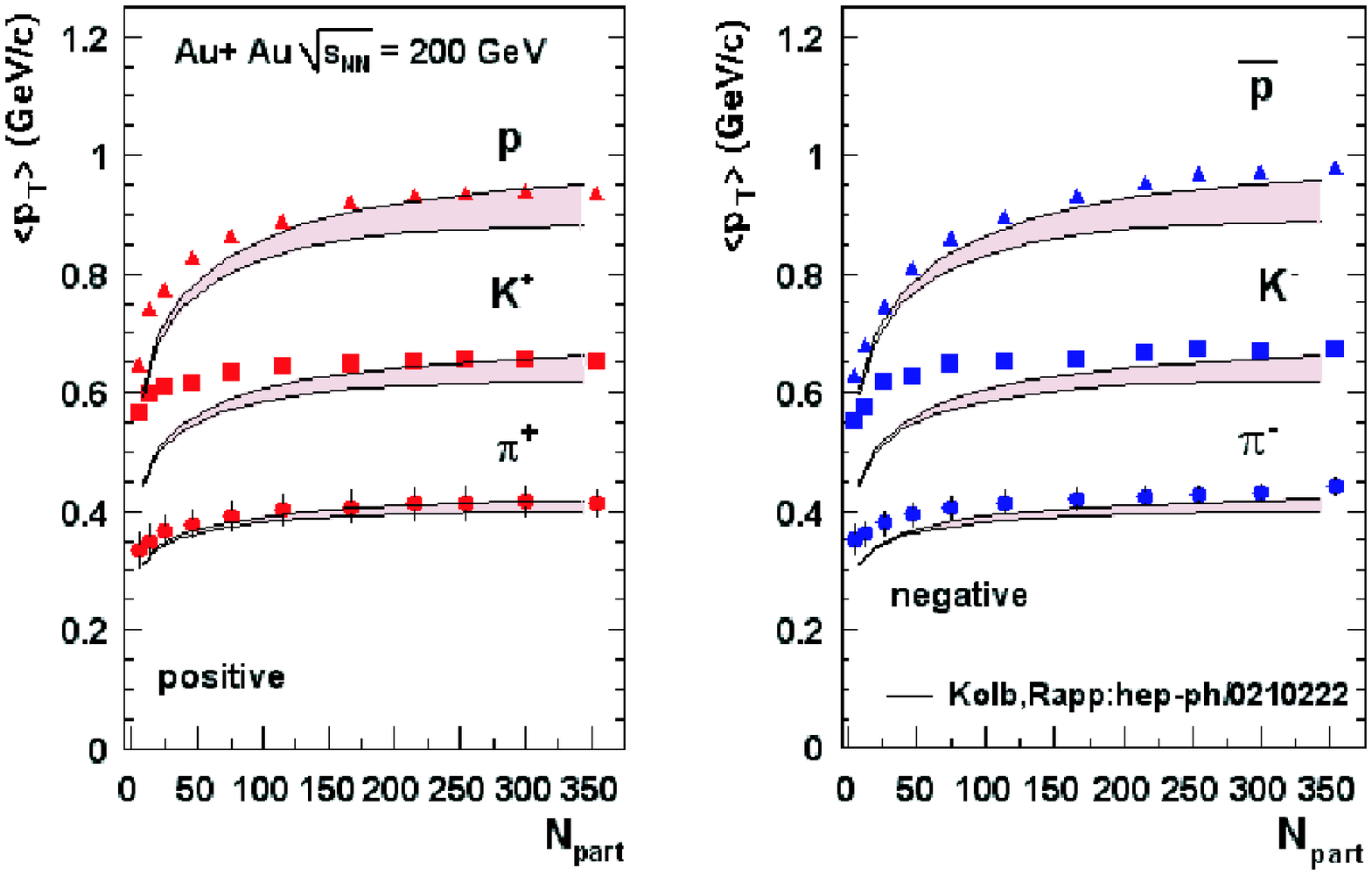,width=8.5cm,height=5.8cm,clip=}
            \epsfig{file=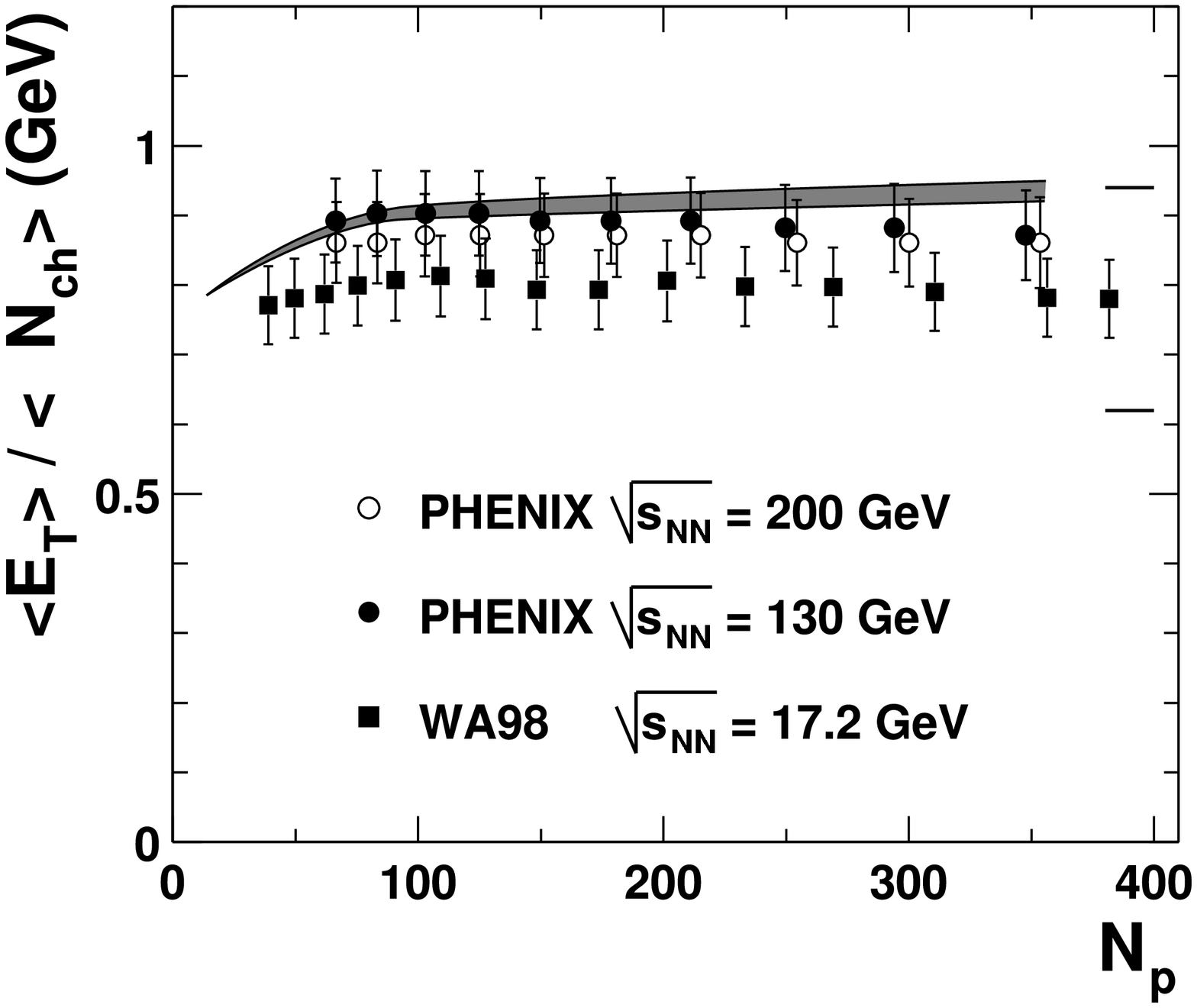,width=6.7cm,height=5.7cm}
            }
\vspace*{-1mm} 
\caption{{\em Left panels:} 
         Mean transverse momenta of pions, kaons and (anti)protons
         from 200\,$A$\,GeV Au+Au collisions \protect\cite{PHENIX03spec200,%
         Velkovsky:2004ce,Adler:2003cb}.
         Ideal fluid dynamic results are included as bands 
         whose lower ends reflect an initialization without initial 
         transverse flow while the upper ends correspond to an initial 
         transverse flow field $v_r\eq\tanh (\alpha r)$ with 
         $\alpha=0.02$~fm$^{-1}$ \protect\cite{KR03}.
         {\em Right panel:}
         Transverse energy per charged hadron as a function of 
         collision centrality, for Au+Au and Pb+Pb collisions at 
         three different beam energies \protect\cite{WA98-01ET,%
         PHENIX01ET,PHENIX03ET}. Superimposed on the original experimental 
         Figure \protect\cite{PHENIX03ET} are hydrodynamic results for
         Au+Au collisions at $\sqrt{s_{_{NN}}}\eq200$\,GeV \protect\cite{KR03}.
         The lower end of the band results from an initialization without 
         initial transverse flow, the upper end reflects an initial 
         transverse flow field $v_r\eq\tanh (\alpha r)$
         with $\alpha=0.02$~fm$^{-1}$.
\label{F17} 
} 
\vspace*{-3mm}
\end{figure} 
%%%%%%%%%%%%%%%%%%%%%%%%%%%%%%%%%%%%%%%%%%%%%%%%%%%%%%%%%%%%%%%%%%%%%%%
%

%
We close this subsection with a brief discussion of the 
centrality dependence of mean transverse momenta per particle, 
$\langle\pt\rangle$, and the average transverse energy per 
charged hadron, $\langle E_T\rangle/N_\mathrm{ch}$. 
Figure~\ref{F17} shows a comparison of $\la\pt\ra$ for identified pions, 
kaons, protons and antiprotons measured by PHENIX in 200\,$A$\,GeV 
Au+Au collisions \cite{PHENIX03spec200,Velkovsky:2004ce,Adler:2003cb} 
with the hydrodynamic results \cite{KR03}.
The bands reflect the theoretical variation resulting from possible
initial transverse flow already at the beginning of the hydrodynamic 
expansion stage, as discussed at the end of the previous subsection.
The figure shows some discrepancies between hydrodynamics and the data
for peripheral collisions (small $N_{\rm part}$) which are strongest for
the kaons whose spectra are flatter at large impact parameters than 
predicted by the model.

The right panel in Figure~\ref{F17} shows the total transverse energy 
per emitted charged hadron as a function of collision centrality.
Although both the charged particle multiplicity and total transverse 
energy vary strongly with the number of participating nucleons and 
collision energy, the transverse energy {\em per particle} is essentially 
independent of these parameters.
The superimposed band in Figure~\ref{F17} reflects ideal 
hydrodynamic calculations for Au+Au collisions at 
$\sqrt{s}\eq200\,A$\,GeV with and without initial transverse 
flow.
The slight rise of the theoretical curves with increasing $N_{\rm part}$ 
can be attributed to the larger average transverse flow developing
in more central collisions, resulting from the higher initial energy 
density and the somewhat longer duration of the expansion until 
freeze-out \cite{KHHET01}.
Successful reproduction of the data requires a correct treatment of the
chemical composition at freeze-out (by using a chemical non-equilibrium 
hadron equation of state below $\Tcrit$).
If one instead assumes chemical equilibrium of the hadron resonance gas 
down to kinetic freeze-out, ideal fluid dynamics overpredicts the transverse 
energy per particle by about 15-20\% \cite{KHHET01}.
%

%%%%%%%%%%%%%%%%%%%%%%%%%%%%%%%%%%%%%%%%%%%%%%%%%%%%%%%%%%%%%%%%%%%%%%%%%
% Section : MOMENTUM ANISOTROPIES AS EARLY FIREBALL SIGNATURES
% last updated: 1/27/09
%%%%%%%%%%%%%%%%%%%%%%%%%%%%%%%%%%%%%%%%%%%%%%%%%%%%%%%%%%%%%%%%%%%%%%%%%
\suse{Anisotropic transverse momentum spectra from deformed fireballs}
\label{anisotropies}
%%%%%%%%%%%%%%%%%%%%%%%%%%%%%%%%%%%%%%%%%%%%%%%%%%%%%%%%%%%%%%%%%%%%%%%%%
\sususe{Elliptic flow as an early fireball signature}
\label{sec:ellipticflow}
%%%%%%%%%%%%%%%%%%%%%%%%%%%%%%%%%%%%%%%%%%%%%%%%%%%%%%%%%%%%%%%%%%%%%%%%%
%
In non-central nuclear collisions, or if the colliding nuclei are deformed,
the nuclear overlap region is initially spatially deformed (see 
Fig.~\ref{fig:anisotropies}). 
Interactions among the constituents of the matter formed in that zone
transfer this spatial deformation onto momentum space.
Even if the fireball matter does not interact strongly enough to reach
and maintain almost instantaneous local equilibrium, and a hydrodynamic 
description therefore fails, any kind of re-interaction among the fireball 
constituents will still be sensitive to the anisotropic density gradients 
in the reaction zone and thus redirect the momentum flow preferably into 
the direction of the strongest density gradients (i.e. in the ``short'' 
direction) \cite{Sorge97,Sorge99,ZGK99,MG02,HL99}.
The result is a momentum-space anisotropy, with more momentum flowing 
into the reaction plane than out of it.
Such a ``momentum-space reflection'' of the initial spatial deformation
is a unique signature for re-interactions in the fireball and, when 
observed, proves that the fireball matter has undergone significant
nontrivial dynamics between creation and freeze-out.
Without rescattering, the only other mechanism with the ability to map
a spatial deformation onto momentum space is the quantum mechanical 
uncertainty relation.
For matter confined to smaller spatial dimensions in $x$ than in $y$ 
direction it predicts $\Delta p_x > \Delta p_y$ for the corresponding 
widths of the momentum distribution.
However, any momentum anisotropy resulting from this mechanism
is restricted to momenta $p \sim 1$/(size of the overlap zone) 
which for a typical fireball radius of a few fm translates into
a fraction of 200\,MeV/$c$.
This is the likely mechanism for the momentum anisotropy 
observed \cite{KNV03} in calculations of the classical dynamical 
evolution of a postulated deformed ``color glass condensate''
created initially in the collision.
Unlike the experimental data, this momentum anisotropy is
concentrated around relatively low $\pt$ \cite{KNV03}.
Whatever the detailed mechanism responsible for the observed momentum 
anisotropy, the induced faster motion into the reaction plane than
perpendicular to it (``elliptic flow'') rapidly degrades the initial 
spatial deformation of the matter distribution and thus eliminates the 
driving force for any further increase of the anisotropic flow.
Elliptic flow is therefore ``self-quenching'' \cite{Sorge97,Sorge99},
and any flow anisotropy measured in the final state must have been
generated early when the collision fireball was still spatially 
deformed (see Fig.~\ref{fig:anisoovertau}).
If elliptic flow does not develop early, it never develops at all.
It thus reflects the pressure and stiffness of the equation of state
during the earliest collision stages \cite{Sorge97,Sorge99,ZGK99,KSH00,%
KSH99}, but (in contrast to many other early fireball signatures) it can 
be easily measured with high statistical accuracy since it affects 
{\em all} final state particles.
Microscopic kinetic models show that, for a given initial spatial 
deformation, the induced momentum space anisotropy is a monotonically 
rising function of the strength of the interaction among the matter 
constituents \cite{ZGK99,MG02,HL99}.
The maximum effect should thus be expected if their mean free path
approaches zero, i.e. in the ideal fluid limit \cite{KSH00,HK02}. 
Viscous effects associated with finite mean free paths reduce the
elliptic flow \cite{HW02}, especially at larger $\pt$ \cite{MG02,%
Teaney:2003kp}.
Within the ideal fluid limit, the magnitude of the elliptic flow shows 
some sensitivity to the nuclear equation of state in the early collision 
stage, but the variation is not very large (see Fig.~34 in 
Ref.~\cite{Voloshin:2008dg}).
To the extent that the initial spatial fireball deformation is
known (see Fig.~\ref{fig:anisotropies} -- the average impact parameter 
can be determined geometrically from the ratio of the observed 
multiplicity in the event to the maximum multiplicity from all events), 
the observed magnitudes of the momentum anisotropies, and in particular 
their dependence on collision centrality \cite{HL99,VP99}, provide valuable 
measures for the degree of thermalization reached early in the collision.
Experimentally this program was first pursued at the SPS in 158\,$A$\,GeV
Pb+Pb collisions \cite{NA49-98v2}. 
These data still showed significant sensitivity to details of the
analysis procedure \cite{BDO00} and thus remained somewhat 
inconclusive \cite{KHHH01}.
Qualitatively, the SPS data (where the directed and elliptic flow 
coefficients, $v_1$ and $v_2$, can both be measured) confirmed 
Ollitrault's 1992 prediction \cite{Ollitrault92} that near midrapidity 
the preferred flow direction is {\em into} the reaction plane, 
supporting the conclusions from earlier measurements in Au+Au  
collisions at the AGS \cite{E895-99v2} where a transition from 
out-of-plane to in-plane elliptic flow had been found between 4 
and 6\,$A$\,GeV beam energy.
A comprehensive quantitative discussion of elliptic flow became first
possible with RHIC data, because of their better statistics and 
improved event plane resolution (due to the larger event multiplicities)
and also as a result of improved analysis techniques 
\cite{VP99,Voloshin:2008dg}.
In the meantime the latter have also been re-applied to SPS data
and produced very detailed results from Pb+Pb collisions at this 
lower beam energy \cite{NA49-03v2,CERES02v2,Agakichiev:2003gg}. 
Ideal fluid dynamical predictions for the spectra and differential elliptic 
flow $v_2(\pt)$ of pions and protons are now available for collision
energies ranging all the way from the AGS to LHC \cite{Eskola:2005ue,%
Kestin:2008bh,Niemi:2008ta}.
%

\begin{figure}[htb] 
\centerline{
%\fbox{\rule[-15mm]{0cm}{3cm}{\em To create a place-holder}}
            \epsfig{file=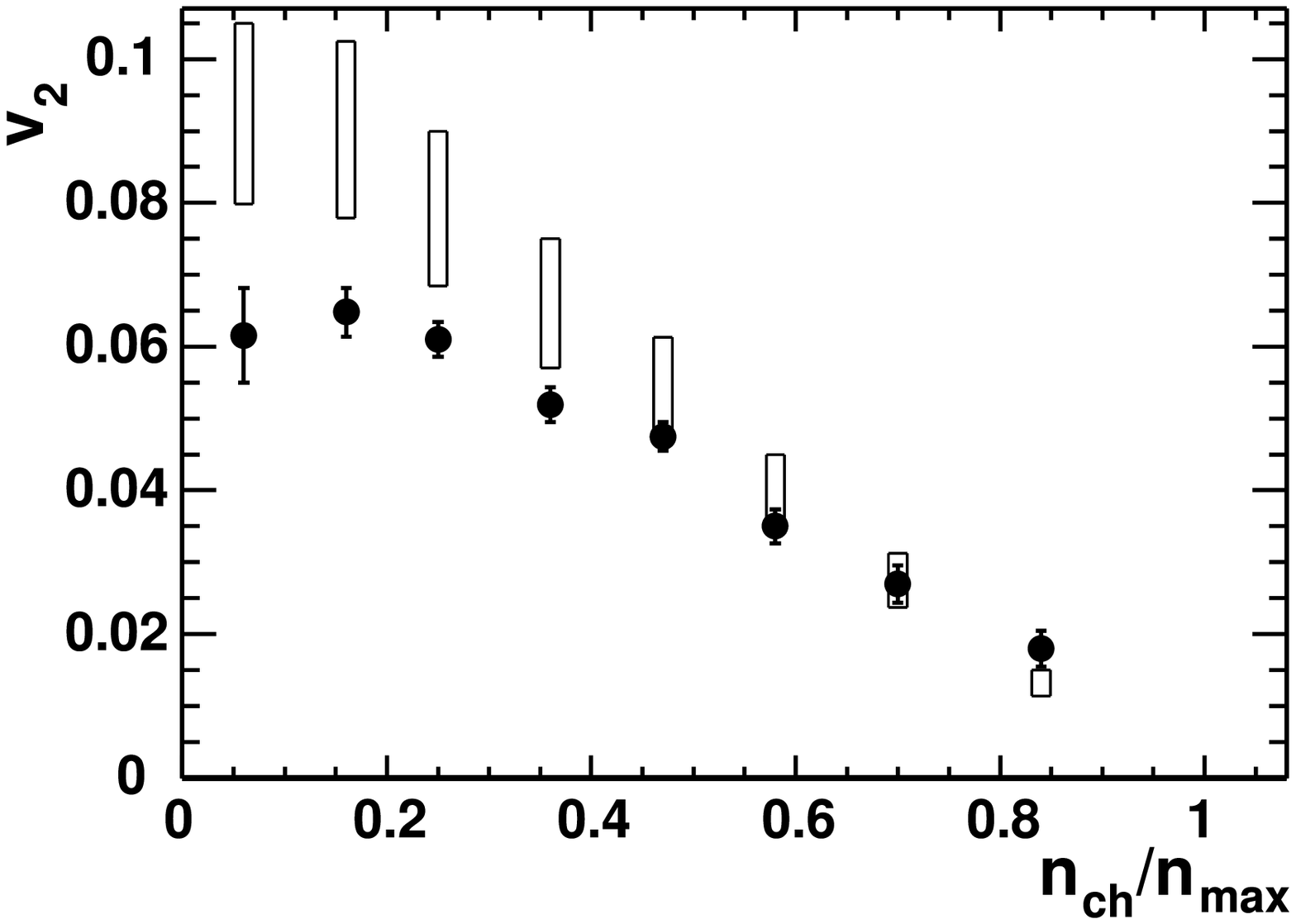,width=5.5cm,height=5cm}
\hspace*{1cm}
            \epsfig{file=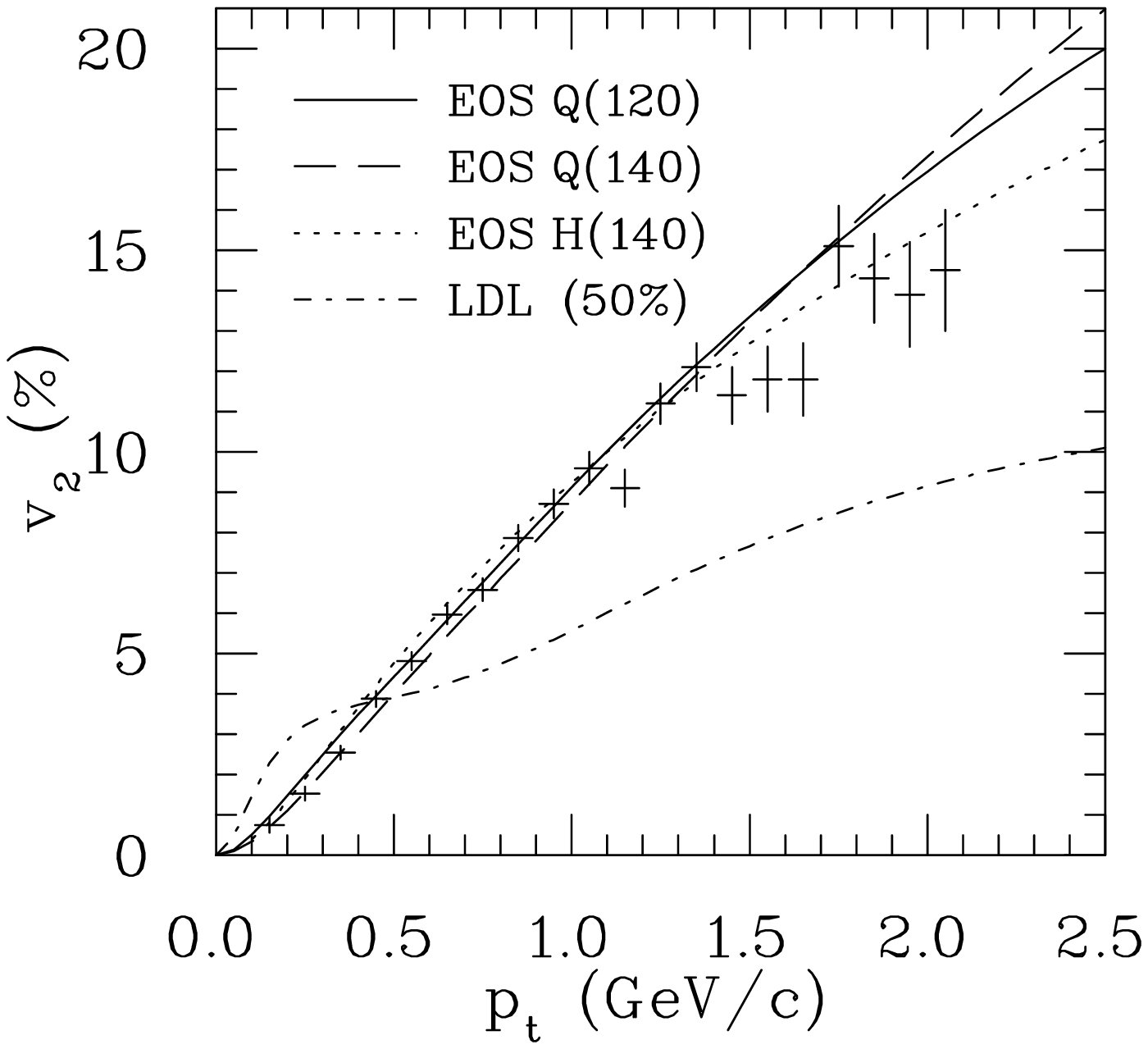,width=5.59cm}
            }  
\caption{Elliptic flow of unidentified charged particles in 
         130\,$A$\,GeV Au+Au collisions, integrated over $\pt$ as 
         function of centrality (left) and for minimum bias collisions
         as a function of $\pt$ (right). Both data sets (symbols with 
         error bars) are from the original STAR 
         publication \protect\cite{STAR01v2}. The vertical bars in the 
         left panel \protect\cite{STAR01v2} indicate the range of earlier 
         hydrodynamic predictions for a variety of equations of state 
         and initial conditions \protect\cite{KSH00}. The top three 
         curves in the right panel \protect\cite{KHHH01} represent 
         hydrodynamic predictions for semiperipheral collisions with 
         initial conditions tuned to the observed \protect\cite{PHOBOS02dN130} 
         total charged multiplicity in central collisions where $v_2$ 
         vanishes. Different curves correspond to different equations 
         of state and freeze-out temperatures \protect\cite{KHHH01}.
\label{fig:starv2nch} 
} 
\end{figure}

%%%%%%%%%%%%%%%%%%%%%%%%%%%%%%%%%%%%%%%%%%%%%%%%%%%%%%%%%%%%%%%%%%%%%%%%%
% Section : ELLIPTIC FLOW AT RHIC
% last updated: 1/27/09 UH
%%%%%%%%%%%%%%%%%%%%%%%%%%%%%%%%%%%%%%%%%%%%%%%%%%%%%%%%%%%%%%%%%%%%%%%%%
\sususe{Elliptic flow at RHIC}
\label{sec:ellipticflowdetails}
%%%%%%%%%%%%%%%%%%%%%%%%%%%%%%%%%%%%%%%%%%%%%%%%%%%%%%%%%%%%%%%%%%%%%%%%%
%
The second published and still among the most important results from 
Au+Au collisions at RHIC was the centrality and $\pt$ dependence of 
the elliptic flow coefficient at midrapidity \cite{STAR01v2}.
For central to midperipheral collisions and for transverse momenta
$\pt\lapp1.5$~GeV/$c$ the data were found to be in stunning agreement
with hydrodynamic predictions \cite{KSH00,KHHH01}, as seen in  
Fig.~\ref{fig:starv2nch}.
In the left panel, the ratio $n_{\rm ch}/n_{\rm max}$ of the charged 
particle multiplicity to the maximum observed value is used to 
characterize the collision centrality, with the most central collisions
towards the right near 1.
$n_{\rm ch}/n_{\rm max}\eq0.45$ corresponds to an impact parameter 
$b{\,\approx\,}7$\,fm \cite{STARPRC02v2}.
Up to this value the observed elliptic flow $v_2$ is found to track 
very well the increasing initial spatial deformation $\epsilon_x$
of the nuclear overlap zone \cite{STARPRC02v2}, as predicted by 
ideal fluid dynamics \cite{KSH00}.   
%

%%%%%%%%%%%%%%%%%%%%%%%%% Fig. 18 %%%%%%%%%%%%%%%%%%%%%%%%%%%%%%%%%%%%%%%%
%%%%%%%%%%%%%%%%%%%%%%%%%%%%%%%%%%%%%%%%%%%%%%%%%%%%%%%%%%%%%%%%%%%%%%%
%

\begin{figure} 
\centerline{\epsfig{file=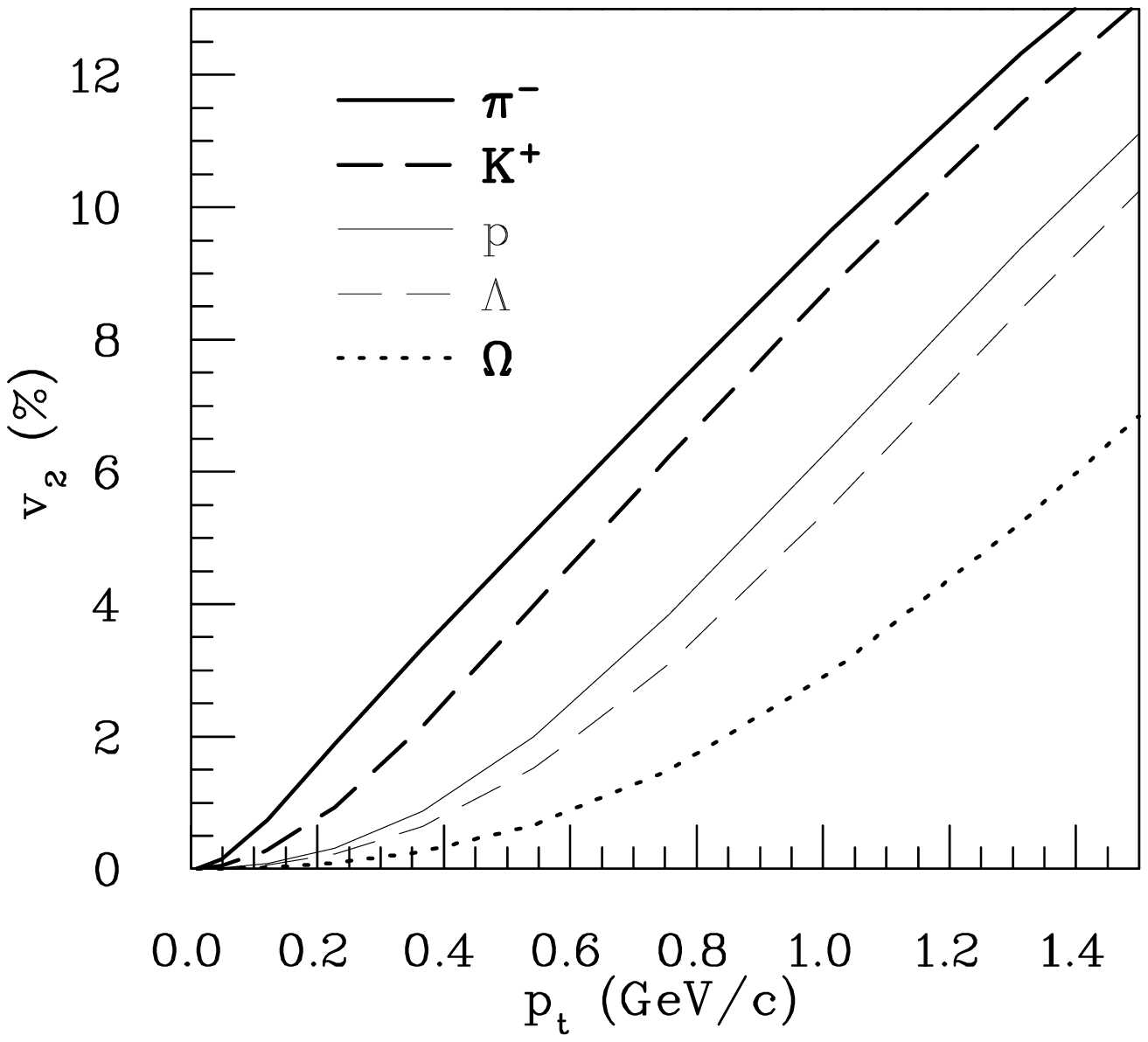,width=5.5cm} \hspace*{1cm}
            \epsfig{file=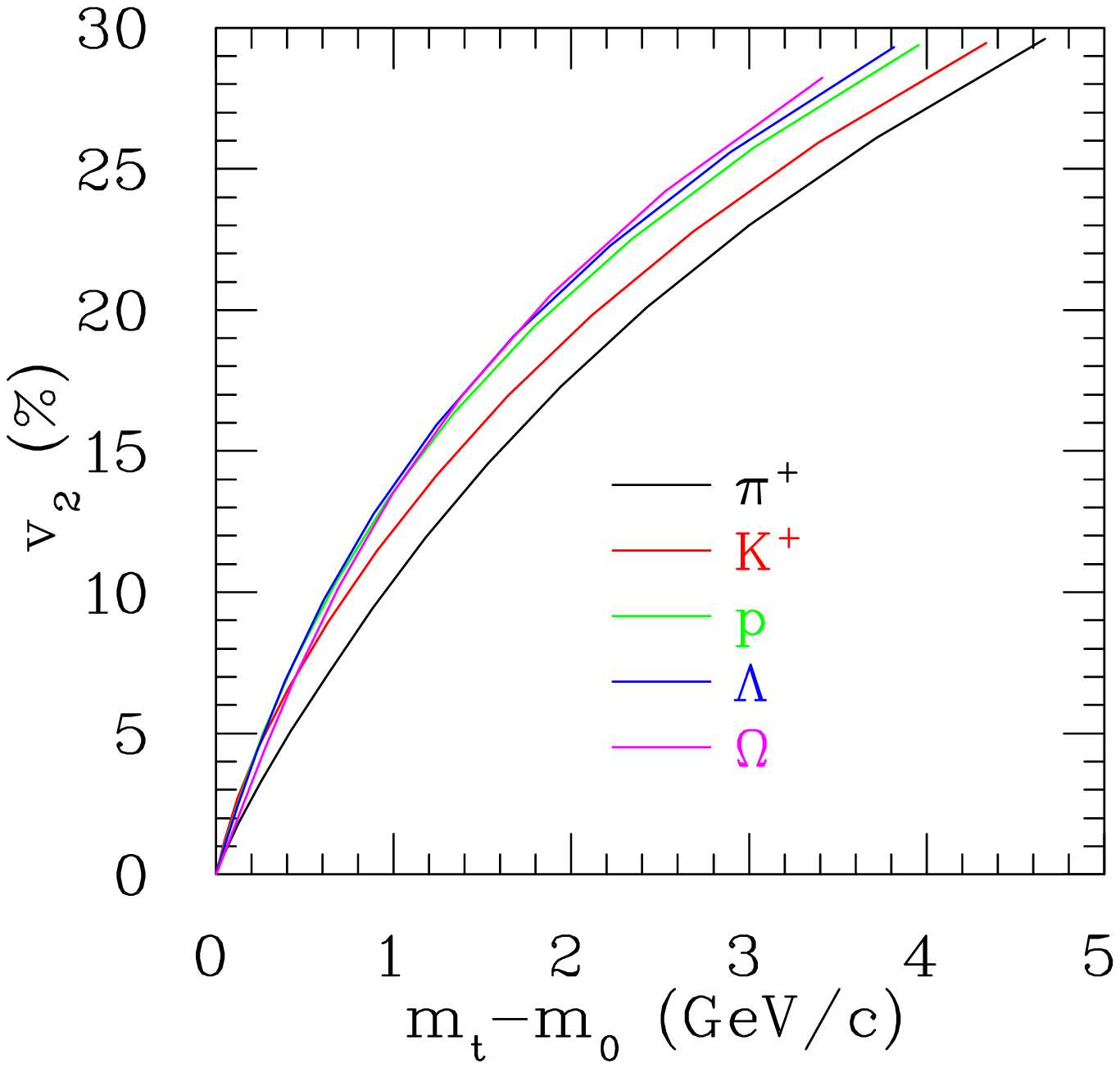,width=5.5cm}}
\caption{{\sl Left:} Ideal fluid dynamical predictions for the 
         differential elliptic flow $v_2(\pt)$ in minimum bias 
         Au+Au collisions at $\sqrt{s}\eq130\,A$\,GeV, for different 
         identified hadron species \protect\cite{HKHRV01}.
         The hydrodynamical simulations use EOS~Q.
         {\sl Right:} The same curves redrawn as functions of the 
         transverse kinetic energy $\KET\equiv\mt-m_0$ 
         \protect\cite{pasipriv}.
\label{F19} 
} 
%\vspace*{-4mm}
\end{figure}

%
An important prediction of the hydrodynamic model is the characteristic 
dependence of the differential elliptic flow $v_2(\pt)$ on the particle
rest mass, shown in the left panel of Fig.~\ref{F19} \cite{HKHRV01}.
It arises primarily from the assumption of local thermal equilibrium
on which hydrodynamics is based.
Thermal hadron spectra exhibit $m_T$-scaling which is exact in the 
absence of flow and slightly broken at low $\pt$ by radial flow (see 
discussion in Sec.~\ref{ptdist}) \cite{LHS90,SSHPRC93}.
When plotted as a function of $\pt$, an exponential function in $m_T$
exhibits a shoulder at low $\pt$ that becomes broader and flatter with
increasing particle rest mass.
This flattening of the single-particle spectra at low $\pt$ is the 
primary reason \cite{HKHRV01} for the flattening of $v_2(\pt)$ at low 
$\pt$ with increasing rest mass seen in Fig.~\ref{F19}, left panel.
Additional scale-breaking effects from radial flow exist but are 
of less importance.
For this reason, the  rest mass dependence of the differential elliptic 
flow can be eliminated almost completely by replotting $v_2$ as a function 
of the transverse kinetic energy $\KET=m_T{-}m_0$ instead of 
$\pt$.
This is shown in the right panel of Fig.~\ref{F19}.
%

%
%%%%%%%%%%%%%%%%%%%%%%% Fig. 19 %%%%%%%%%%%%%%%%%%%%%%%%%%%%%%%%%%%%%%%%%%%%%
%%%%%%%%%%%%%%%%%%%%%%%%%%%%%%%%%%%%%%%%%%%%%%%%%%%%%%%%%%%%%%%%%%%%%%%
%

%
%%%%%%%%%%%%%%%%%%%%%%%%%%% Fig. 20 %%%%%%%%%%%%%%%%%%%%%%%%%%%%%%%%%%%
\begin{figure}[bt]
\begin{center}
  \epsfig{file=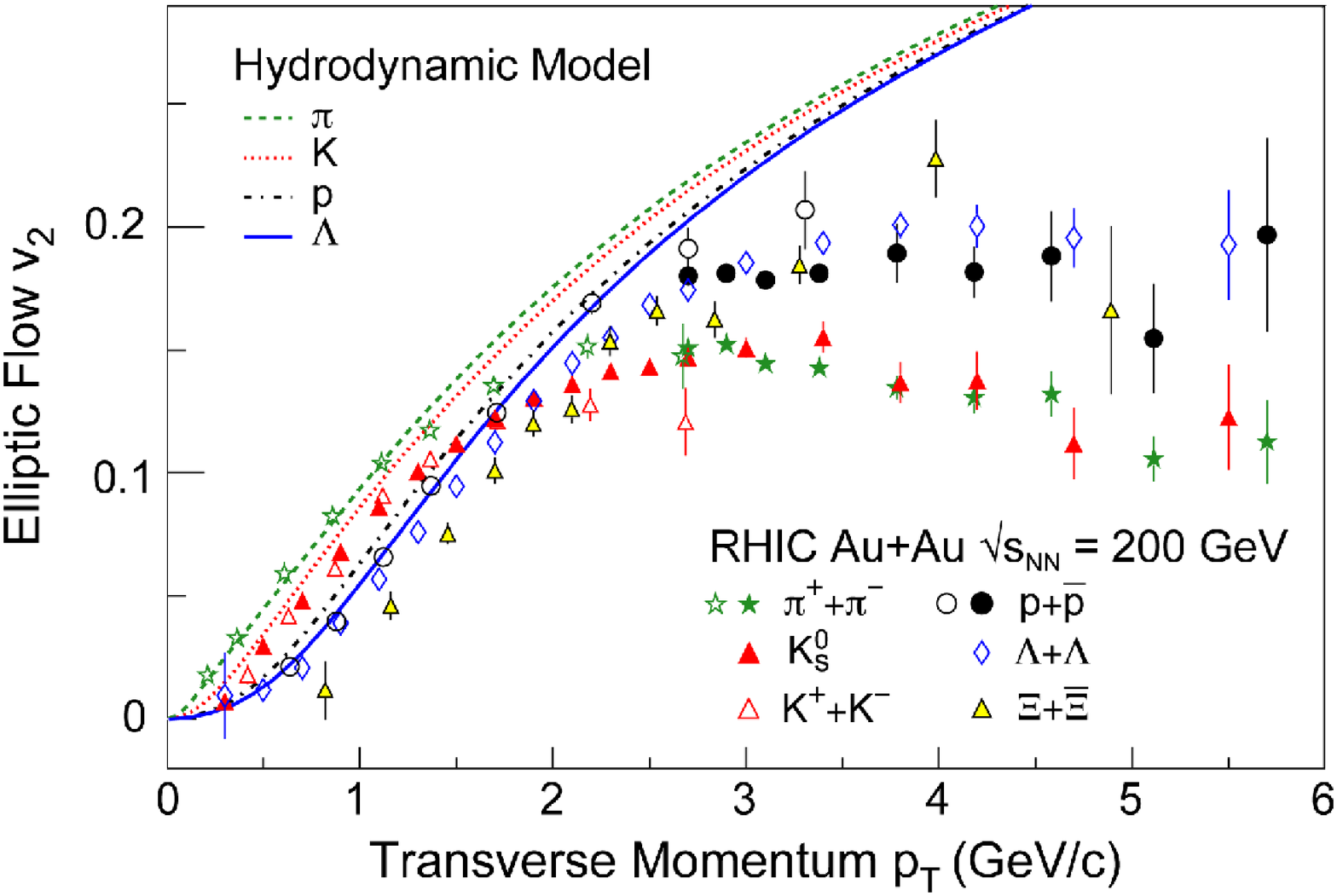,width=0.463\linewidth,clip=}
  \epsfig{file=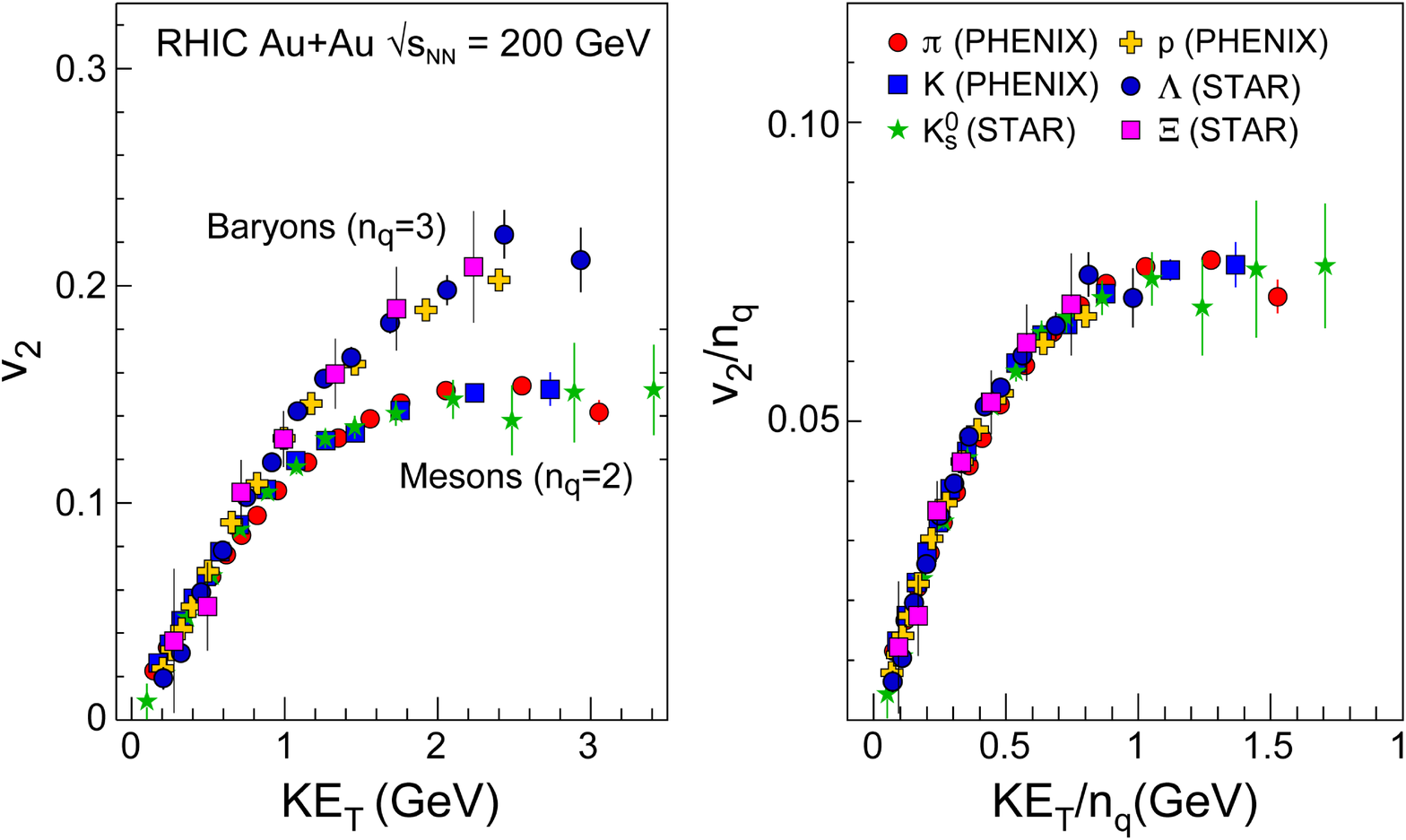,width=0.527\linewidth,clip=}
\end{center}
\caption{
{\sl Left:} Up to $\pt\sim1.5$\,GeV/$c$, the differential elliptic flow
$v_2(\pt)$ follows the hydrodynamical predictions for an ideal fluid
almost perfectly \protect\cite{STARv2}. 
Note that $>99\%$ of all final hadrons have $\pt<1.5$\,GeV/$c$.
{\sl Middle:} When plotted against transverse kinetic energy, the 
differential elliptic flow follows different universal curves for
mesons and baryons. 
{\sl Right:} When scaled by the number of valence quarks, the differential 
elliptic flow per quark follows the same universal curve {\em for all 
hadrons and for all values of (scaled) transverse kinetic energy}
\protect\cite{Adare:2006ti}.
\label{F20}}
\end{figure} 
%%%%%%%%%%%%%%%%%%%%%%%%%%%%%%%%%%%%%%%%%%%%%%%%%%%%%%%%%%%%%%%%%%%%%%%
%

%
Figure~\ref{F20} shows that these predictions of the hydrodynamic
model are nicely borne out in the experimental data \cite{STAR01v2piKp,%
PHENIX02v2overeps,STAR02v2KLambda,Adler:2003kt,Sorensen:2003wi,STARv2,%
Adare:2006ti}.
The left panel shows the differential elliptic flow as a function of $\pt$
for five different hadron species.
Up to transverse momenta of $\pt\sim1.5$\,GeV/$c$ the data show a clear
tendency of $v_2(\pt)$ to decrease with increasing rest mass, and
they agree even quantitatively with the hydrodynamic predictions.
(Remember that radial flow plays only a subdominant role in this
mass hierarchy, so this should be taken as support for approximate
local thermal equilibrium, but not necessary for {\em ideal}
fluid dynamics.)
Since the majority of hadrons ($>99\%$) have transverse momenta 
below 1.5\,GeV/$c$, the bulk of the fireball is seen to be well 
described by the hydrodynamic model.
At larger transverse momenta ($\pt\gapp1.5$\,GeV/$c$ for mesons,
$\pt\gapp2.3$\,GeV/$c$ for baryons), the measured elliptic flow 
lags behind the ideal fluid dynamical predictions.
This is expected if one accepts that the ideal fluid assumption of
instantaneous thermalization is unrealistic and allows for a finite
mean free path of the particles.
The latter leads to viscous corrections which manifest themselves more
strongly as $\pt$ increases\cite{MG02,Teaney:2003kp} (see Eq.~(\ref{viscf})).   
What is not expected is that, above these breakaway points from 
ideal hydrodynamics, the elliptic flow curves appear to cluster
into two groups which, instead of being arranged by mass, are ordered
according to whether the hadron is a meson or a baryon. 
This is more clearly seen in the middle panel of Fig.~\ref{F20}
where the differential elliptic flow is replotted as a function
of transverse kinetic energy.
Hydrodynamics predicts that then all curves should approximately collapse
onto a single line (right panel of Fig.~\ref{F19}).
This indeed happens at low $\KET$, where the left panel in Fig.~\ref{F20} 
has shown that the data agree with hydrodynamics, but at larger 
$\KET$, where the data break away from the fluid dynamical prediction,
$v_2(\KET)$ splits into two curves for baryons and mesons.
Clearly, hydrodynamics provides no explanation for this behaviour, 
since the splitting only happens where the hydrodynamic model ceases 
to be valid.
The observations can be explained in a quark coalescence 
model\cite{Fries:2003vb,Greco:2003xt,Molnar:2003ff,Hwa:2003ic}
which postulates that at intermediate transverse momenta
(i.e. above the point where hydrodynamics breaks down and below
the range where hard jet fragmentation dominates the hadron yield)
hadron production proceeds through the coalescense of valence
quarks.
This model predicts a scaling of $v_2$ with the number of valence
quarks $n_q$ inside the hadron \cite{Molnar:2003ff}:
$v_2^h(\pt)= n_q\,v_2^q\left(\frac{\pt}{n_q}\right)$. 
Where this scaling holds, it should yield a universal curve if 
one plots $\frac{v_2^h}{n_q}$ against $\frac{\pt}{n_q}$.
At high $\pt$ where rest masses can be neglected, a plot against
$\KET/n_q$ should be equally good. 
At low $\pt$ where $v_2$ agrees with hydrodynamics which predicts a 
linear dependence of $v_2$ on $\KET$, a rescaling of both axes by
$n$ has no effect on the shape of the curve.
Hence, a plot of $\frac{v_2}{n_q}\left(\frac{\KET}{n_q}\right)$ should 
yield a universal curve both at low $\KET$ where hydro works and at 
intermediate $\KET$ where the coalescence model applies.
As seen in the right panel of Fig.~\ref{F20}, the experimental data 
confirm this expectation beautifully.
This collapse of all hadronic elliptic flow data onto a single
universal curve works at both $\scm=63$ and 200\,GeV and at all 
centralities \cite{:2008ed}, even though both the slope of the curve
at low $\KET$ and the saturation values at intermediate $\KET$ change
with collision centrality (due to the changing initial fireball 
eccentricity).
We should not leave this discussion without pointing out an important
{\em caveat}:
The comparison between experimental data and ideal fluid dynamics
shown in the left panel of Fig.~\ref{F20} is based on simulations 
with EOS~Q which do not implement the correct non-equilibrium 
chemical composition in the HRG stage.
When this deficiency is corrected, the $\pt$-slope of the pion elliptic
flow $v_2^\pi(\pt)$ increases by about 30\% \cite{HT02,KR03,Huovinen:2007xh},
hence the good agreement between theory and experiment shown in 
Fig.~\ref{F20} is a bit deceptive.
We will pursue this theme further in Section~\ref{viscosity}.
%

%%%%%%%%%%%%%%%%%%%%%%%%%%%%%%%%%%%%%%%%%%%%%%%%%%%%%%%%%%%%%%%%%%%%%%%%%%%
\sususe{Implications: Rapid thermalization and ``strongly coupled 
quark-gluon plasma'' (sQGP)}
\label{implications}
%%%%%%%%%%%%%%%%%%%%%%%%%%%%%%%%%%%%%%%%%%%%%%%%%%%%%%%%%%%%%%%%%%%%%%%%%%%

%
The apparent success of the ideal fluid dynamical picture in describing
bulk hadron emission from relativistic heavy-ion fireballs not just 
on a superficial qualitative level, but in many aspects even 
quantitatively had tremendous implications for the heavy-ion community's 
view of hot QCD matter.
It let to a genuine paradigm shift, away from the idea of the QGP as a 
weakly interacting gas of quarks and gluons and towards that of a
strongly coupled plasma with liquid behaviour \cite{HK02,Gyulassy:2004vg,%
Gyulassy:2004zy}. 
This shift has both motivated and survived the quantitative refinements 
of the picture that will be discussed in Section~\ref{viscosity}.
It generated strong interest outside the field of nuclear physics, in 
particular in the area of cold atoms (see discussion around 
Fig.~\ref{fig:atomtrap}) and in superstring theory.
In experiments with cold atoms one has the unique ability to continuously
change the interaction strength among the particles by dialling
an external magnetic field, moving the atoms onto or away from a 
Feshbach resonance \cite{OHGGT02}.
In superstring theory, Maldacena's AdS/CFT correspondence 
\cite{Maldacena:1997re} between strongly coupled conformal 
field theories (CFT) and weakly coupled (classical) gravitational theories
in curved 5-dimensional Anti-de-Sitter (AdS) spaces opened a window
for performing analytical calculations for quantum field theories in 
the strong-coupling limit, by solving classical differential equations 
for strings moving in appropriately curved space-times.
Even though QCD is not itself a conformal field theory, such studies have
established several ``universal'' results, such as a lower limit for
the shear viscosity to entropy density ratio $\eta/s \geq \hbar/(4\pi k_B)$
(the so-called ``KSS bound''\cite{Policastro:2001yc,Kovtun:2004de})
that applies to a large class of conformal field theories including
a supersymmetric version of QCD.
While it is presently unclear whether the almost perfect liquid behaviour 
of the QGP, as indicated by the results presented above, really signals
a breakdown of perturbative QCD, and one may remain doubtful whether in 
the end superstring theoretical methods based on the AdS/CFT correspondence
will really lead to a more efficient and quantitative understanding of QGP 
properties than appropriately refined perturbative methods in QCD (resummed
pQCD), it is an undeniable fact that this cross-fertilization has generated
unprecedented productivity in practically relevant situations within field
that has long been plagued by doubts that it can ever be confirmed or 
falsified experimentally. 
Furthermore, AdS/CFT--based calculations have permanently reshaped the 
way theorists look at strongly coupled field theoretical systems where
conventional particle-based pictures break down absolutely.
%

%
%%%%%%%%%%%%%%%%%%%%%%%%%% Fig. 21 %%%%%%%%%%%%%%%%%%%%%%%%%%%%%%%%%%%
\begin{figure}[ht] 
\vspace*{-3mm}
\centerline{
%\fbox{\rule[-15mm]{0cm}{3cm}{\em To create a place-holder}}
            \epsfig{file=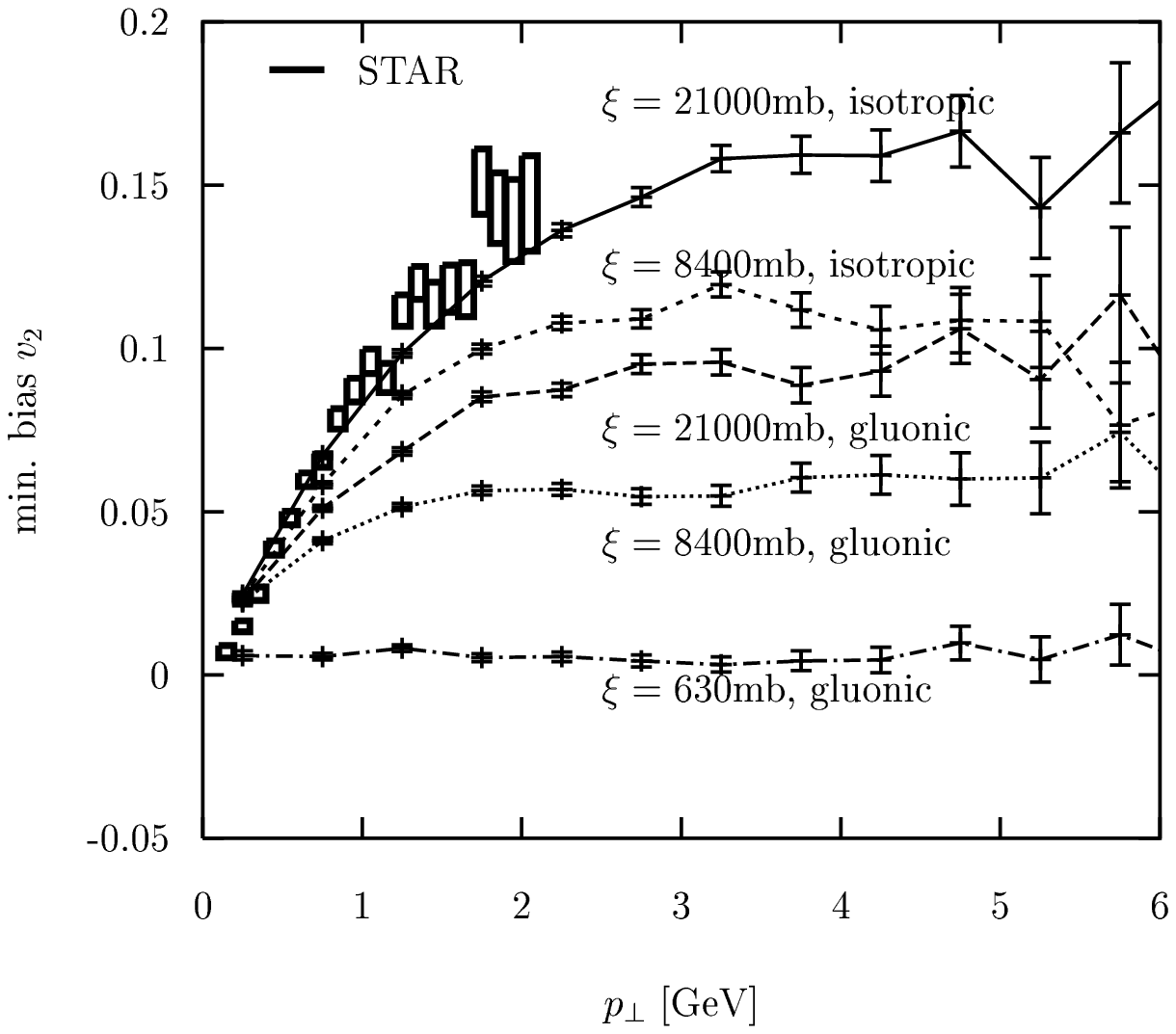,width=7.5cm,height=5.9cm}
            \epsfig{file=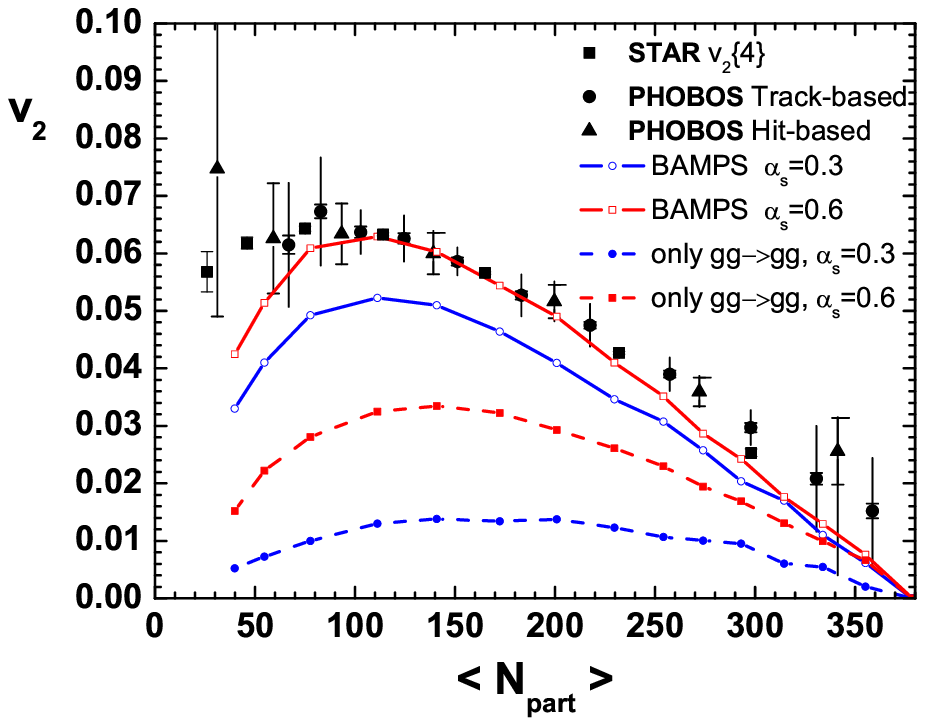,width=7.5cm,clip=} 
            }  
\vspace*{-2mm}
\caption{{\sl Left:} Impact parameter averaged elliptic flow as a 
         function of transverse momentum, for Au+Au collisions at 
         $\scm=130$\,GeV. Experimental results from 
         STAR\protect\cite{STAR01v2} are compared with parton cascade  
         calculations\protect\cite{MG02} based on 2-body collisions 
         with varying transport opacities $\xi$. 
         {\sl Right:} $\pt$-integrated elliptic flow as a function
         of collision centrality, for Au+Au collisions at $\scm=200$\,GeV.
         Experimental results from the STAR \protect\cite{Adams:2004bi}
         and PHOBOS \protect\cite{Back:2004mh} collaborations are compared
         with parton cascade simulations \protect\cite{Xu:2007jv} including 
         both two- and three-body interactions \protect\cite{Xu:2004mz} 
         with varying values for the strong coupling constant $\alpha_s$.
\label{F21}
} 
\vspace*{-4mm}
\end{figure} 
%%%%%%%%%%%%%%%%%%%%%%%%%%%%%%%%%%%%%%%%%%%%%%%%%%%%%%%%%%%%%%%%%%%%%%%
%

%
After the discovery of strong elliptic flow at RHIC it was 
quickly realized \cite{MG02} that the measured \cite{STAR01v2} 
almost linear rise of the charged particle (i.e. predominantly 
pionic) elliptic flow with $\pt$ requires strong rescattering among
the fireball constituents.
The left panel of Figure~\ref{F21} shows the results from microscopic 
simulations which describe the dynamics of the early expansion stage
by solving a Boltzmann equation with a 2-body collision term for 
colliding on-shell partons \cite{MG02}.
The different curves are parametrized by the transport opacity 
$\xi =\sigma_0 dN_g/d\eta$ involving the product of the parton 
rapidity density and cross section in the early collision stage.
As the opacity is increased, the elliptic flow is seen to approach
the data (and the hydrodynamic limit) monotonically {\em from below}.
Whereas the hydrodynamic limit predicts a continuous rise of 
$v_2(\pt)$, the elliptic flow from the parton cascade saturates
at high $\pt$, as also seen in the data \cite{STAR03v2highpt}.
This is due to incomplete equilibration at high $\pt$. 
The critical $\pt$ at which the cascade results cease to follow the 
hydrodynamic rise shifts to higher (lower) values as the transport 
opacity is increased (decreased), corresponding to a decrease (increase)
of viscous corrections to the distribution function (see 
Eq.~(\ref{viscf})).
From that figure it was concluded \cite{MG02} that the measured elliptic 
flow requires very large transport opacities, at least as long as only 
two-body collisions are included, exceeding perturbative expectations 
by a factor 15-30 \cite{MG02}. 
Xu and Greiner \cite{Xu:2004mz} pointed out that the inclusion of 
radiative collision processes $gg\leftrightarrow ggg$ changes this 
conclusion.
The right panel in Fig.~\ref{F21} shows that these radiative processes 
strongly accelerate the thermalization process, and that a perturbative 
description of the experimental data does not appear to be entirely excluded
(although $\alpha_s=0.6$ may be an uncomfortably large coupling constant
for a perturbative QCD approach).
What remains true, aside from all ongoing discussions about details,
is that the experimental data require high interaction rates and
short scattering time scales, i.e. rapid thermalization.
Using the BAMPS cascade \cite{Xu:2004mz} Xu and Greiner showed that 
even a shear viscosity to entropy ratio close to the KSS bound
$\eta/s\geq1/4\pi$ can be reached for $\alpha_s=0.6$ \cite{Xu:2007jv}.
These microscopic simulations complement the empirical observations
in cold atom systems (see Fig.~\ref{fig:atomtrap}) in demonstrating
the importance of rapid thermalization for a successful description 
of the elliptic flow data. 
But there is an additional, quite general argument that further 
reinforces this point \cite{KSH00,Kolb:2003dz}. 
As mentioned earlier, the hydrodynamically predicted elliptic
flow is proportional to the initial spatial eccentricity 
$\epsilon_x(\tau_\equ)$ at the beginning of the hydrodynamic evolution.
If thermalization is slow, the matter will start to evolve in the 
transverse directions before $\tau_\equ$ is reached, following its 
initial locally isotropic transverse momentum distribution.
Even if no reinteractions among the produced particles occur, this
radial free-streaming motion dilutes the spatial deformation, but 
without generating any momentum anisotropy.
Thus, if thermalization and hydrodynamic behavior set in late,
they will be able to build only on a significantly reduced spatial 
eccentricity $\epsilon_x$, and the resulting elliptic flow response
will be correspondingly smaller.
To reach a certain measured value of $v_2$ at a given impact parameter
thus requires thermalization to set in before free radial motion has
reduced the spatial deformation so much that even perfect hydrodynamic
motion can no longer produce the measured momentum anisotropy.
This consideration yields a {\em rigorous upper limit for the 
thermalization time $\tau_\equ$}.
The dilution of the spatial eccentricity by collisionless radial
free-streaming is easily estimated \cite{KSH00,PFKthesis02}, using the 
analytic solution of the collisionless Boltzmann equation for the 
distribution function $f(\br,\bm{p}_T,\tau)$ of initially produced 
approximately massless partons (we only consider their transverse motion):
\beq{equ:freestr}
  f(\br,\bp_T,\tau{+}\Delta\tau) = 
  f\left(\br - c\Delta\tau\,\be_p, \bp_T, \tau\right)\,.
\end{equation}
Here $\be_p$ is a unit vector in direction of $\bp_T$. With 
Eq.~(\ref{equ:freestr}) it is straightforward to compute the
time-dependence of the spatial eccentricity:
\bea{equ:extau}
 &&\epsilon_x(\tau_0{+}\Delta\tau) = 
   \frac{\int dx\,dy(y^2{-}x^2) 
         \int d^2p_T\,f(\br - c\Delta\tau\,\be_p,\bp_T,\tau_0)}
        {\int dx\,dy(y^2{+}x^2) 
         \int d^2p_T\,f(\br - c\Delta\tau\,\be_p,\bp_T,\tau_0)}
\\\nonumber
 &&=
   \frac{\int dxdy\,\pt d\pt d\phi_p\,
         [(y{+}c\Delta\tau\sin\phi_p)^2 -(x{+}c\Delta\tau\cos\phi_p)^2]\,
         f(\br,\bp_T,\tau_0)}
        {\int dxdy\,\pt d\pt d\phi_p\,
         [(y{+}c\Delta\tau\sin\phi_p)^2 + (x{+}c\Delta\tau\cos\phi_p)^2]\,
         f(\br,\bp_T,\tau_0)}.
\eea
The initial distribution at $\tau_0$ is even in $x$ and $y$, and the 
initial transverse momentum distribution can be assumed to be locally 
isotropic. 
From this it follows directly that
\beq{equ:epsilondepletion}
  \frac{\epsilon_x(\tau_0{+}\Delta\tau)}{\epsilon_x(\tau_0)}
  = \left[ 
  1+\frac{(c\,\Delta\tau)^2}{\la\br^2\ra_{\tau_0}} \right]^{\!-1}\,,
\end{equation}
where $\la\br^2\ra_{\tau_0}$ is the azimuthally averaged initial
transverse radius squared of the reaction zone.
Inserting typical values for, say, Au+Au collisions at $b\eq7$~fm
one finds that a delay of thermalization by $\Delta t\eq2.5$\,fm/$c$ 
(3.5\,fm/$c$) leads to a decrease of the spatial eccentricity by 
30\% (50\%), without generating any momentum anisotropy.
The elliptic flow signal resulting from subsequent hydrodynamic expansion
would then be degraded by a similar percentage.
If we assume, for the sake of the argument, that the RHIC data exhaust 
at least 2/3 of the ideal fluid limit calculated with the {\em full 
initial eccentricity} $\epsilon_x(\tau_0)$, the thermalization 
time $\tau_\equ$ can therefore not be larger than about 2.5\,fm/$c$.
%

%%%%%%%%%%%%%%%%%%%%%%%%%%%%%%%%%%%%%%%%%%%%%%%%%%%%%%%%%%%%%%%%%%%%%
% Subsection: Viscous corrections to the ideal fluid
%%%%%%%%%%%%%%%%%%%%%%%%%%%%%%%%%%%%%%%%%%%%%%%%%%%%%%%%%%%%%%%%%%%%%
\suse{Signs of viscosity}
\label{viscosity}
%%%%%%%%%%%%%%%%%%%%%%%%%%%%%%%%%%%%%%%%%%%%%%%%%%%%%%%%%%%%%%%%%%%%%
\sususe{Spectra and elliptic flow at midrapidity}
\label{midrapidity}
%%%%%%%%%%%%%%%%%%%%%%%%%%%%%%%%%%%%%%%%%%%%%%%%%%%%%%%%%%%%%%%%%%%%%

\begin{figure}[ht] 
\vspace*{-3mm}
\begin{center}
%\fbox{\rule[-15mm]{0cm}{3cm}{\em To create a place-holder}}
            \epsfig{file=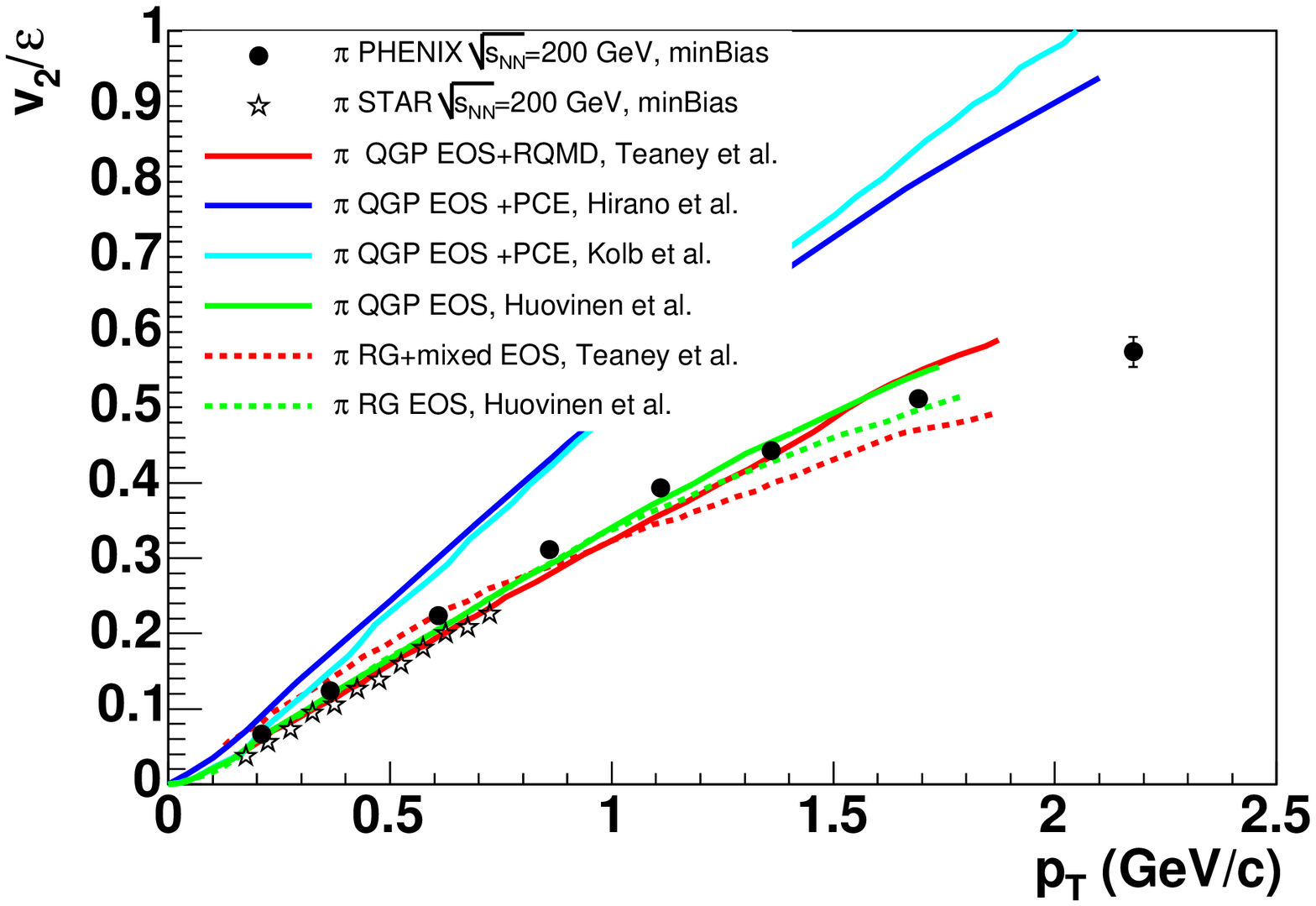,width=7.5cm,clip=}
            \epsfig{file=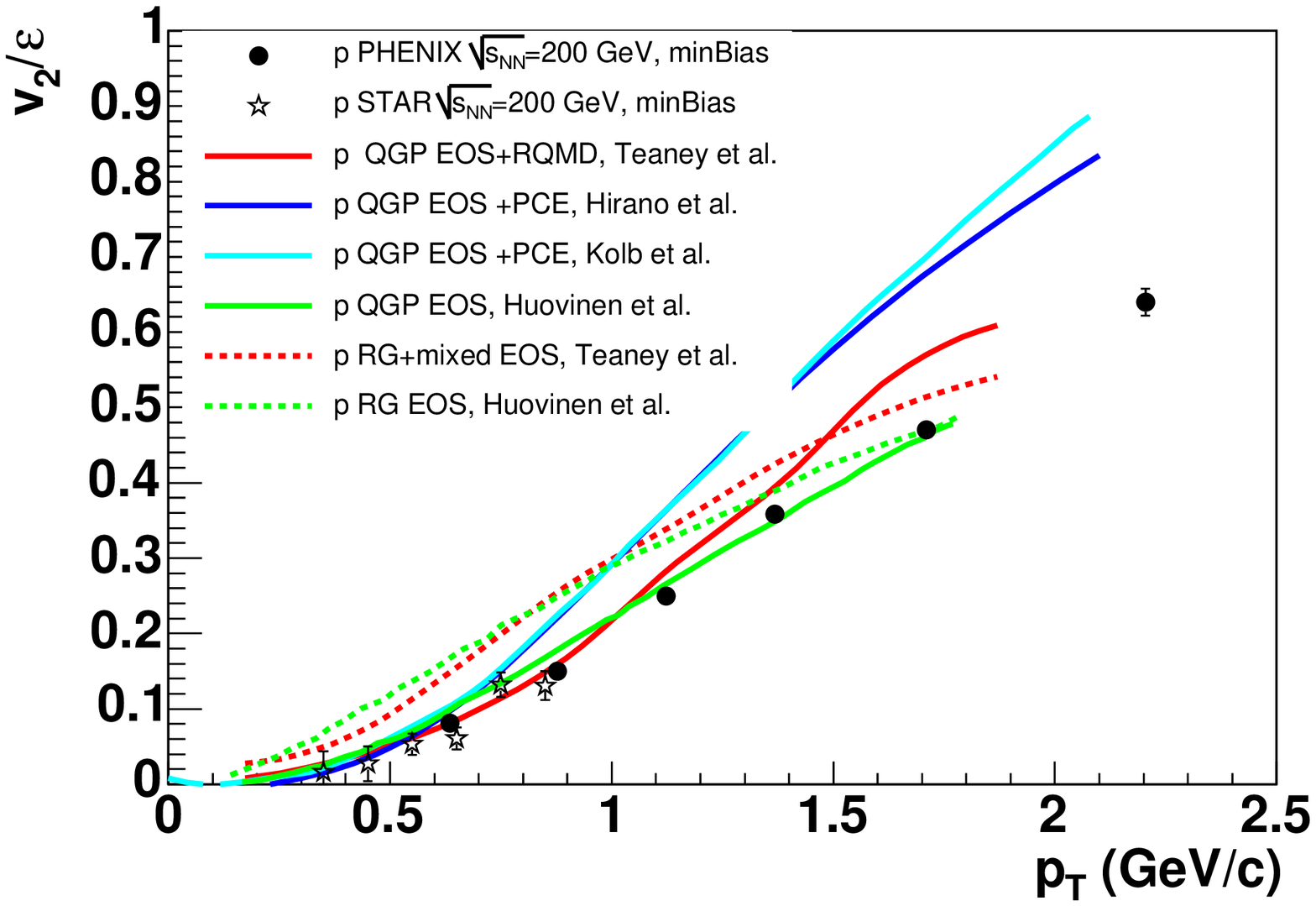,width=7.5cm,clip=} \\
            \epsfig{file=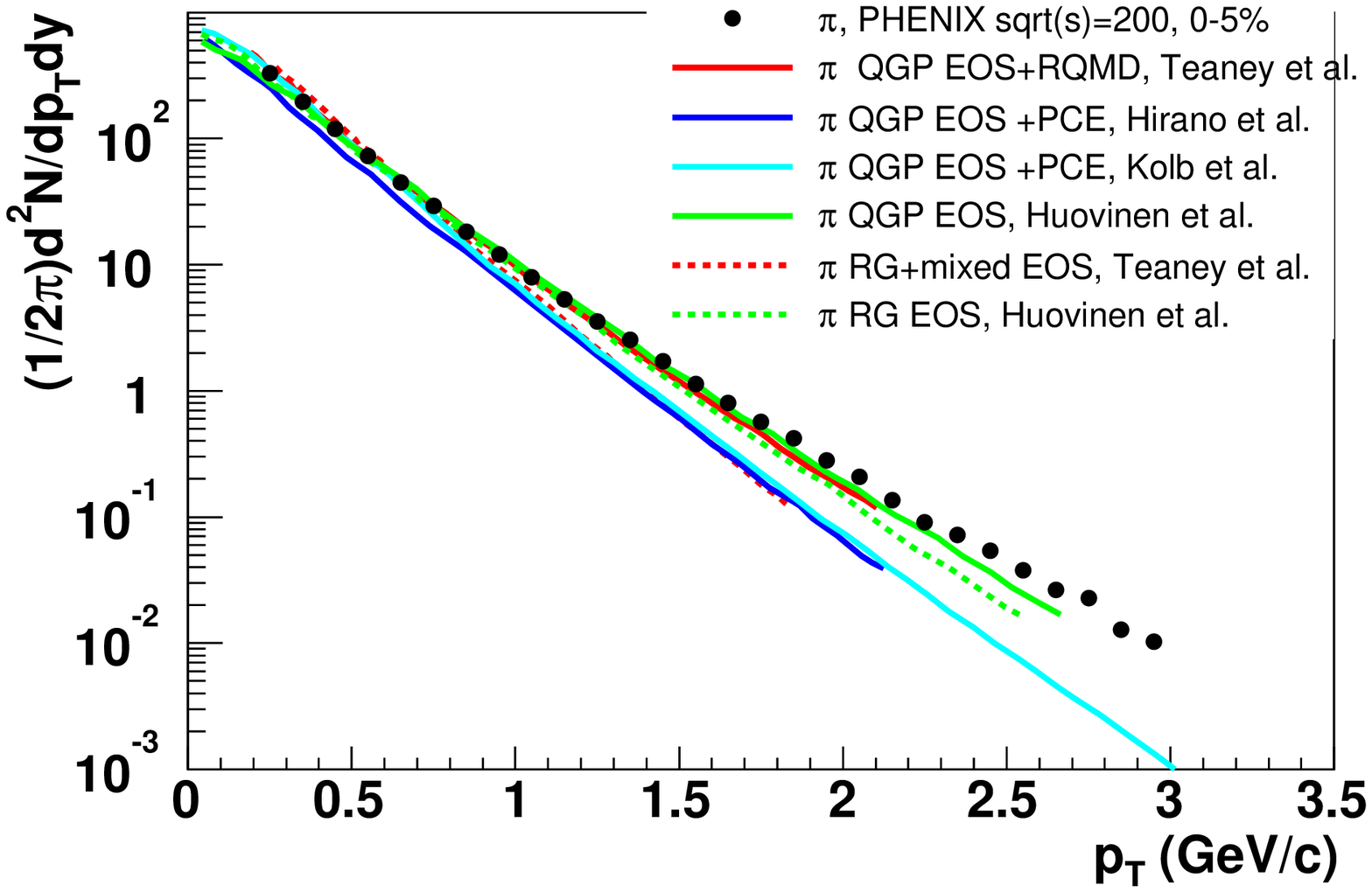,width=7.5cm,clip=}
            \epsfig{file=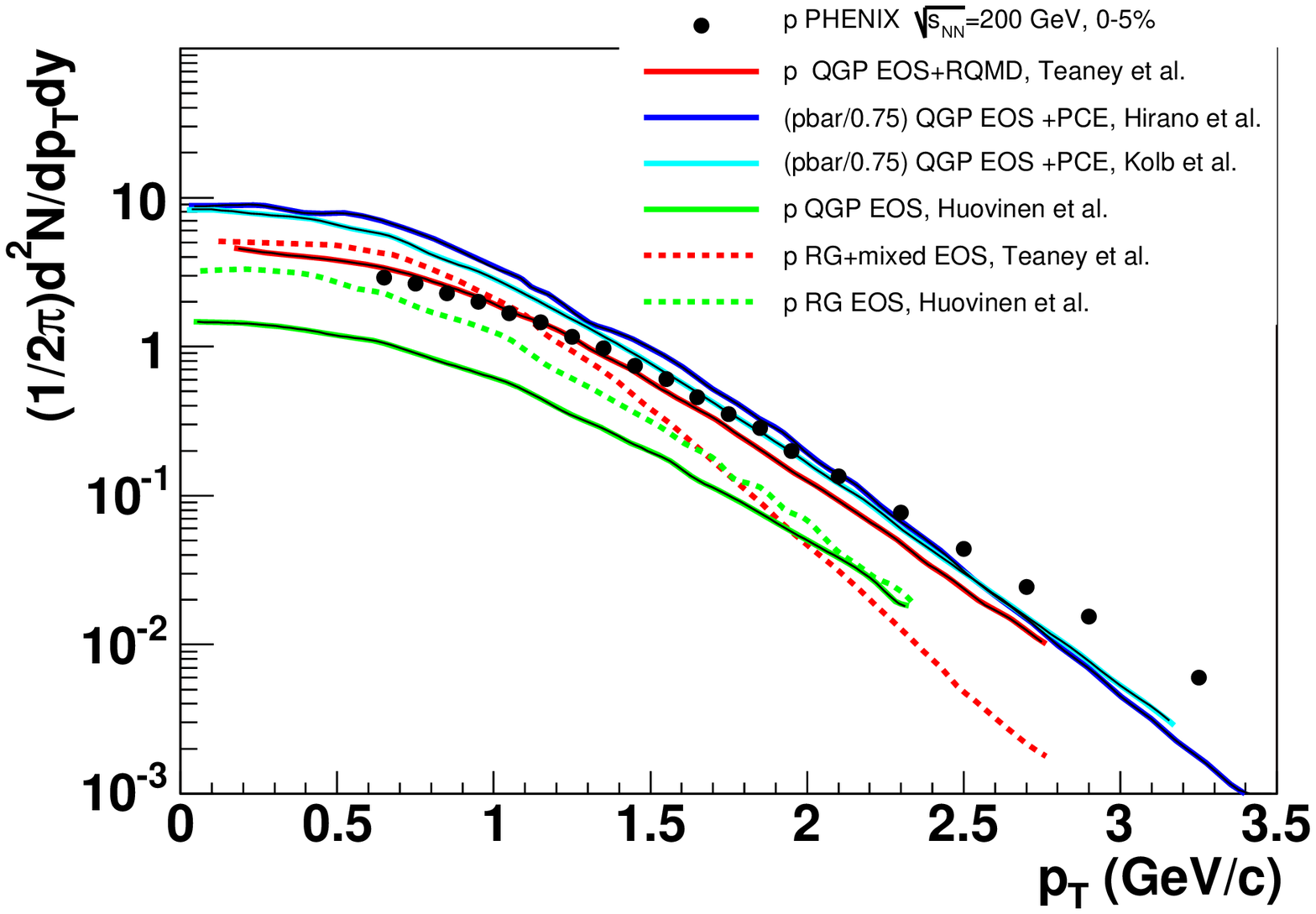,width=7.5cm,clip=} 
\end{center}
\caption{Compilation minimum bias elliptic flow (top row)
         and central collision $\pt$-spectra (bottom row)
         for pions (left column) and protons (right column) from 
         Au+Au collisions at $\scm=200$\,GeV \protect\cite{Adcox:2004mh}. 
         The experimental data are from the PHENIX Collaboration, the 
         theoretical curves are from a variety of ideal fluid dynamical 
         calculations (with Cooper-Frye freeze-out or coupled to a 
         hadronic RQMD cascade) with different equations of state. See 
         text for discussion and original paper \protect\cite{Adcox:2004mh} 
         for references.  
\label{F22}
} 
%\vspace*{-4mm}
\end{figure}

Figure~\ref{F22} shows a comparison of experimental transverse momentum
spectra and elliptic flow measurements for pions and protons from
$200\,A$\,GeV Au+Au collisions at RHIC with a compilation of theoretical
predictions based on ideal fluid dynamics \cite{Adcox:2004mh}.
All theoretical curves treat the QGP stage of the expansion as an ideal 
fluid in thermal and chemical equilibrium (albeit with varying equations 
of state).
The differences between the predictions arise mostly from how they
deal with the hadronic phase.
Except for the solid red lines, all calculations use sudden Cooper-Frye 
freeze-out at some decoupling energy density of order 
$e_\dec\approx 0.075$\,GeV/fm$^3$.
For them the fireball matter is assumed to expand as an ideal fluid until 
it reaches the decoupling point.
The solid red lines represent a calculation \cite{TLS01} that couples ideal 
fluid dynamics for the QGP above $T_c$ to an RQMD hadron cascade below 
$T_c$. 
It allows freeze-out to happen gradually, by following the hadronic 
scattering processes microscopically.
%

%
%%%%%%%%%%%%%%%%%%%%%%%%%% Fig. 22 %%%%%%%%%%%%%%%%%%%%%%%%%%%%%%%%%%%
%%%%%%%%%%%%%%%%%%%%%%%%%%%%%%%%%%%%%%%%%%%%%%%%%%%%%%%%%%%%%%%%%%%%%%%
%

%
A careful study of the figure shows that none of the purely hydrodynamic
simulations yields a good description of all the experimental data.
Calculations that do not allow for a phase transition to quark-gluon plasma
and treat the matter as a chemically equilibrated hadron gas reproduce
the pion elliptic flow but miss that of the protons at well as the shape 
of either the pion or proton spectrum.
Using an EOS featuring a quark-hadron phase transition allows to reproduce
the elliptic flow of both pions {\em and} protons, but {\em only if} the HRG 
phase is assumed to be in chemical equilibrium all the way down to $e_\dec$.
As discussed before, this assumption yields an incorrect $p/\pi$ ratio; 
correspondingly the relative normalization between pion and proton
spectra comes out wrong.
The curves labelled ``PCE'' correct this deficiency by implementing
non-equilibrium chemical potentials in the HRG phase that ensure the
correct chemical composition in the final state, as measured in experiment.
Now the pion and proton spectra are (roughly) correctly normalized, but 
they are too steep, because with the PCE EOS freeze-out 
happens at a lower temperature (see Fig.~\ref{F5}) while the radial 
flow remains the same.
Even worse, the $\pt$-dependent pion elliptic flow now has a slope that 
is about 30\% larger than in experiment.
The reason is two-fold: 
(i) To absorb the given total momentum anisotropy that was generated 
during the early hydrodynamic evolution, steeper single particle spectra 
require a faster rise of $v_2(\pt)$ with $\pt$.
This is seen to affect both pions and protons (blue curves in the
upper row of Fig.~\ref{F22}). 
(ii) In the PCE case pions constitute a smaller fraction of the total 
because baryon-antibaryon pairs are prohibited from annihilating. 
Each pion must then carry a larger share of the total hydrodynamic 
momentum anisotropy.
This explains the stronger effect on pions (upper left panel)
than protons (upper right).
The only curve that gives a reasonable description of all data 
simultaneously is Teaney's hydro+RQMD hybrid calculation \cite{TLS01}.
By switching from fluid dynamics to a hadron cascade at $T_c$, it
suppresses chemical reactions (which are slow in RQMD) and thus 
correctly reproduces the observed chemical freeze-out at 
$T_\mathrm{chem}\approx T_c$.
By allowing the hadrons to rescatter quasi-elastically, it generates 
additional radial flow below $T_c$ that is not too different
from the one generated by hydrodynamics.
But viscous effects in the RQMD cascade further flatten the $\pt$ 
spectra \cite{Hirano:2007ei}, compensating for the lower temperature
in the chemically non-equilibrated hadronic environment and bringing
the spectra in line with experiment.
At the same time they suppress the buildup of elliptic flow in the
hadronic stage \cite{Hirano:2005xf}, thereby reducing the slope 
of $v_2(\pt)$ from the PCE hydrodynamic calculations and bringing
it also back in line with the data.
Even though it took several years to fully understand these mechanisms,
Teaney's work \cite{TLS01} was the first to exhibit the important viscous
effects inherent in the non-equilibrium hadronic cascade dynamics 
during the late hadronic stage and freeze-out.
Figure~\ref{F22} shows that hadronic viscosity plays a key role
for both transverse momentum spectra and elliptic flow, but its most
dramatic effect is the reduction of elliptic flow that it causes.
%

%%%%%%%%%%%%%%%%%%%%%%%%%%%%%%%%%%%%%%%%%%%%%%%%%%%%%%%%%%%%%%%%%%%%%
\sususe{Centrality and rapidity dependence of elliptic flow}
\label{cent_rap}
%%%%%%%%%%%%%%%%%%%%%%%%%%%%%%%%%%%%%%%%%%%%%%%%%%%%%%%%%%%%%%%%%%%%%

%
Effects of hadronic viscosity become much more prominent in peripheral 
collisions and at forward rapidities. 
As one selects larger impact parameters or moves away from midrapidity,
the charged multiplicity density per unit overlap area, 
$(1/S)(dN_\mathrm{ch}/dy)$, decreases, corresponding to a decrease
of the initial entropy density $s(\bm{x}_\perp,\tau_0)$ \cite{Heinz:2004et}.
Correspondingly the system reaches the phase transition sooner and spends
a larger fraction of its evolution in the viscous hadronic phase.
%

%
%%%%%%%%%%%%%%%%%%%%%%%%%% Fig. 23 %%%%%%%%%%%%%%%%%%%%%%%%%%%%%%%%%%%
\begin{figure}[ht] 
\vspace*{-3mm}
\begin{center}
%\fbox{\rule[-15mm]{0cm}{3cm}{\em To create a place-holder}}
            \epsfig{file=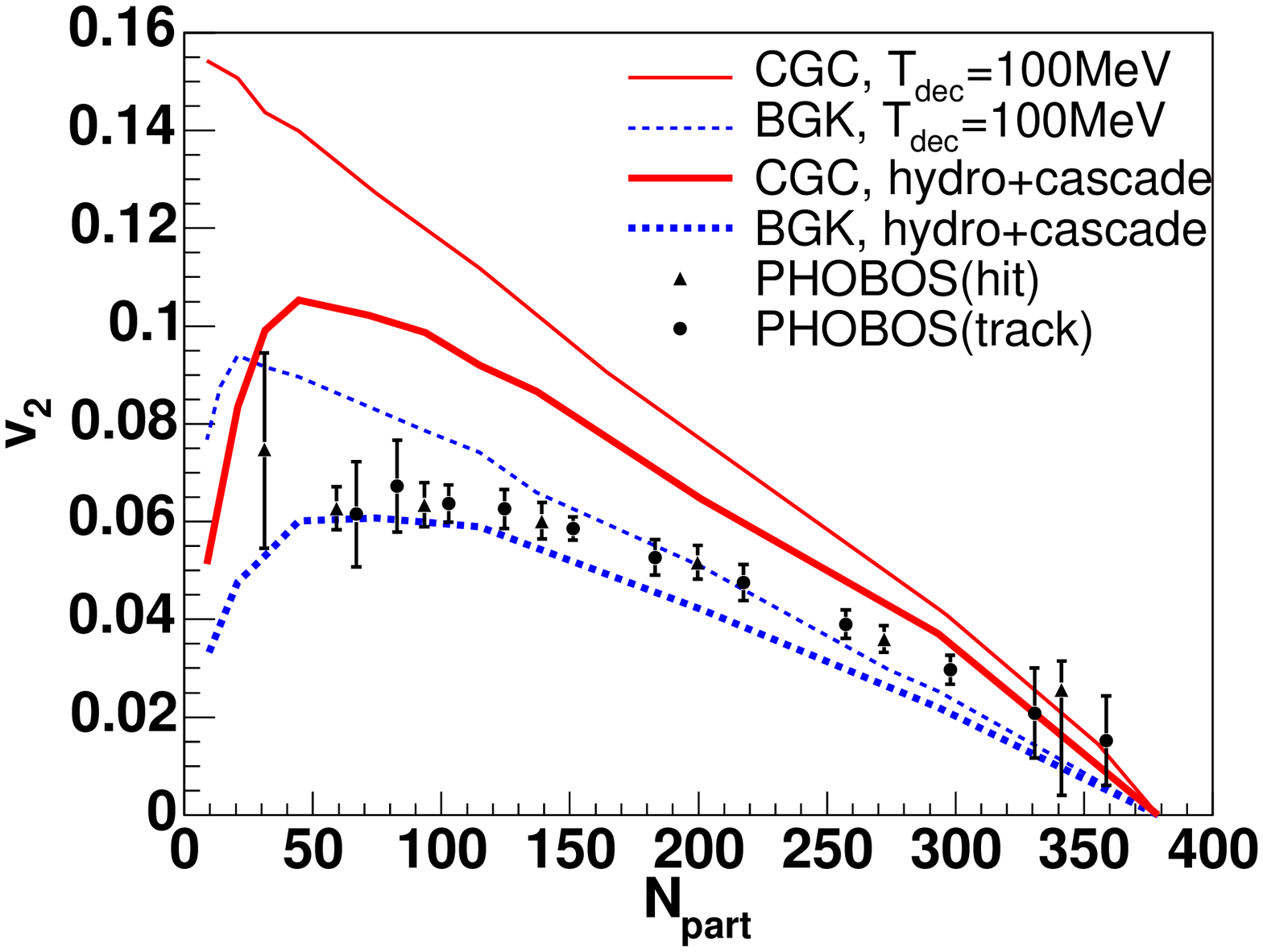,width=7.5cm,clip=}
            \epsfig{file=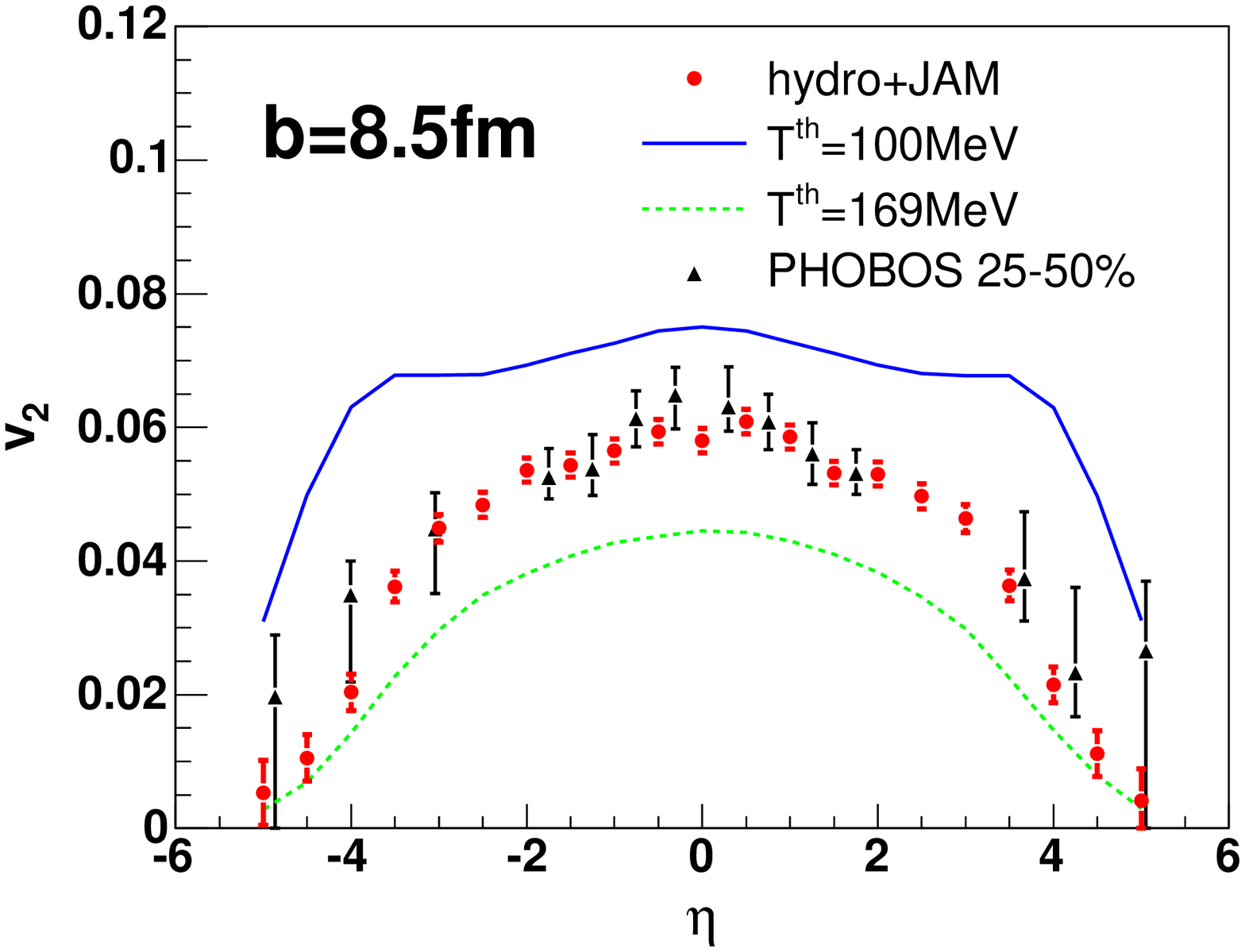,width=7.3cm,clip=} 
\end{center}
\caption{$\pt$-integrated elliptic flow as a function of centrality (left)
         and rapidity (right) for Au+Au collisions at $\scm=200$\,GeV.
         Experimental data (black symbols) are from the PHOBOS 
         experiment \protect\cite{Back:2004mh}. The curves and red symbols 
         show (3+1)-d ideal fluid dynamical calculations with Cooper-Frye 
         freeze-out or coupled to a hadronic cascade 
         (JAM) \protect\cite{Hirano:2005xf}. See text for discussion.
\label{F23}
} 
%\vspace*{-4mm}
\end{figure} 
%%%%%%%%%%%%%%%%%%%%%%%%%%%%%%%%%%%%%%%%%%%%%%%%%%%%%%%%%%%%%%%%%%%%%%%
%

%
Figure~\ref{F23} shows the effect this has on the centrality 
and rapidity dependence of elliptic flow \cite{Hirano:2005xf}.
The solid lines show hydrodynamic calculations that treat the hadronic
phase as a (chemically non-equilibrated) ideal fluid.
They overpredict the elliptic flow in peripheral collisions and
at forward rapidities. 
The hydrodynamic elliptic flow generated during the QGP stage (dotted
green line in the right panel of Fig.~\ref{F23}) is not enough to
explain the data, so {\em some} hadronic contribution to $v_2$ is
required.
But an ideal fluid overdoes it -- it produces too much elliptic flow.
Replacing hydrodynamics by a viscous hadron cascade below $T_c$ gets
the elliptic flow just right -- at least for Glauber model 
initial conditions (see Sec.~\ref{Glauber}) which were used in the right
panel of Fig.~\ref{F23} and for the blue (``BGK'') lines in the left
panel.
For such initial conditions, the assumption of an ideal (zero viscosity) 
QGP fluid followed by a viscous microscopiuc hadron cascade works 
beautifully, at all collision centralities and all rapidities --
in straightforward generalization of the observations made in Fig.~\ref{F22}.
Unfortunately, this is not the happy end of the story.
The thin and thick red solid lines in the left panel of Fig.~\ref{F23}
show that this conclusion becomes untenable once one allows for
alternate models of the initial state.
As shown in the right panel of Fig.~\ref{fig:anisotropies}, the CGC model
produces up to 50\% larger initial source eccentricities than the Glauber
model, which hydrodynamics transforms into correspondingly larger
elliptic flow coefficients -- see the thin solid line in the left panel 
of Fig.~\ref{F23}.
Hadronic viscosity reduces this, but not enough to agree with the data 
(thick solid line in the same panel). 
The measured elliptic flow in peripheral collisions is still overpredicted
by about 50\%.
Does this imply a large shear viscosity during the early QGP stage?
The answer is ``No!''. 
While 50\% looks like a large effect, we will see in Sec.~\ref{QGPvisc}
that even a small amount of QGP viscosity can cause $v_2$ to decrease
by 30\% and thus eliminate the discrepancy.
The inconvenient truth is, however, that a 50\% uncertainty in the
initial source eccentricity translates in some sense into an infinite 
uncertainty about the specific shear viscosity $\eta/s$:
The difference between zero QGP viscosity (seemingly compatible with 
the data for Glauber initial conditions) and even a small nonzero QGP
viscosity (required in the case of CGC initial conditions) cannot
be reasonably expressed in percent.
%  
    
%%%%%%%%%%%%%%%%%%%%%%%%%%%%%%%%%%%%%%%%%%%%%%%%%%%%%%%%%%%%%%%%%%%%%
\sususe{Multiplicity scaling of elliptic flow}
\label{scaling}
%%%%%%%%%%%%%%%%%%%%%%%%%%%%%%%%%%%%%%%%%%%%%%%%%%%%%%%%%%%%%%%%%%%%%

%
Before pursuing the issue of quantifying the QGP viscosity further, let 
us discuss one more piece of qualitative evidence for the increasing 
importance of viscous effects as the energy density and temperature 
of QCD matter decrease.
%

%
%%%%%%%%%%%%%%%%%%%%%%%%%% Fig. 24 %%%%%%%%%%%%%%%%%%%%%%%%%%%%%%%%%%%
\begin{figure}[ht] 
\vspace*{-3mm}
\begin{center}
%\fbox{\rule[-15mm]{0cm}{3cm}{\em To create a place-holder}}
            \epsfig{file=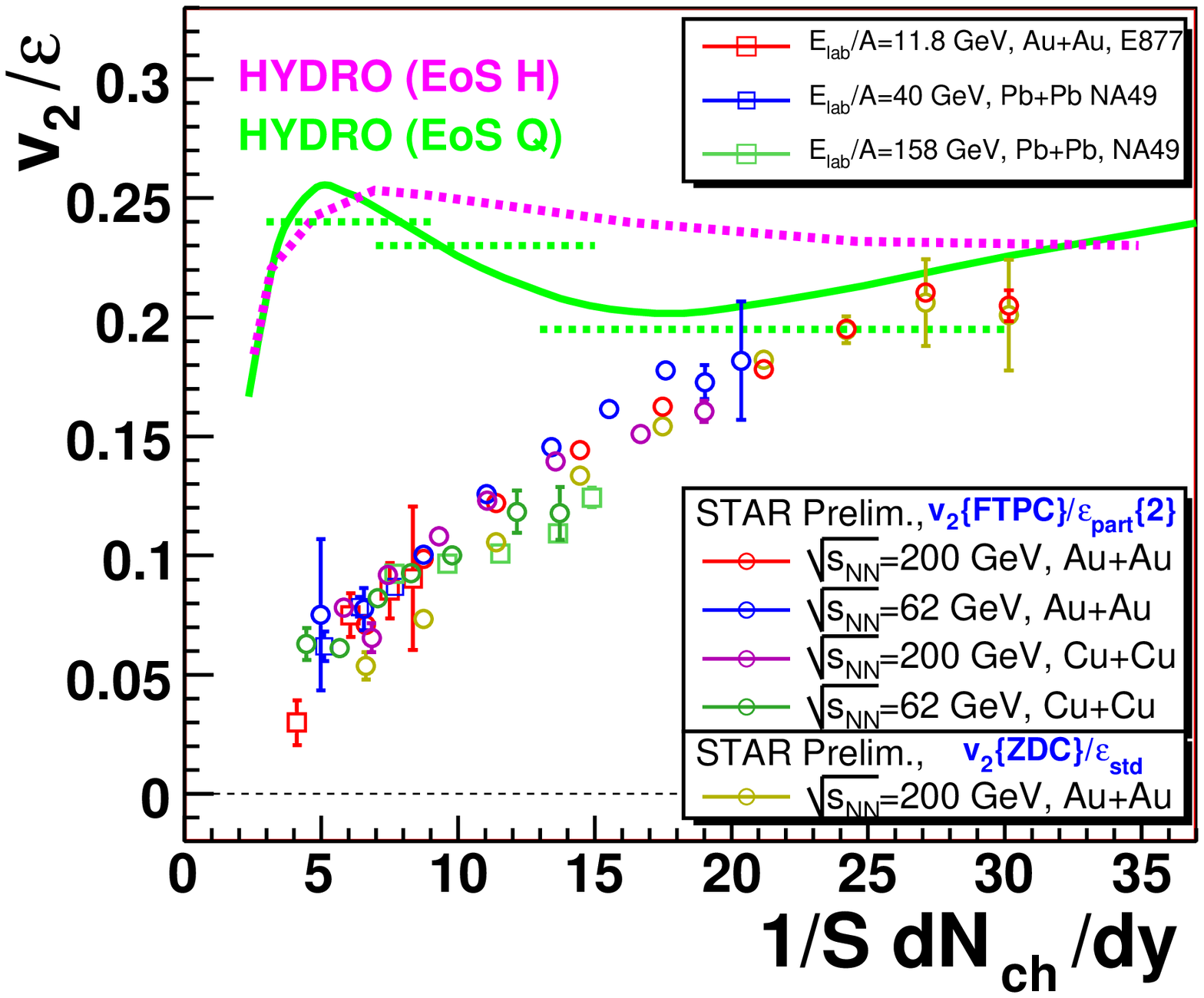,width=7.5cm,height=7cm,clip=}
            \epsfig{file=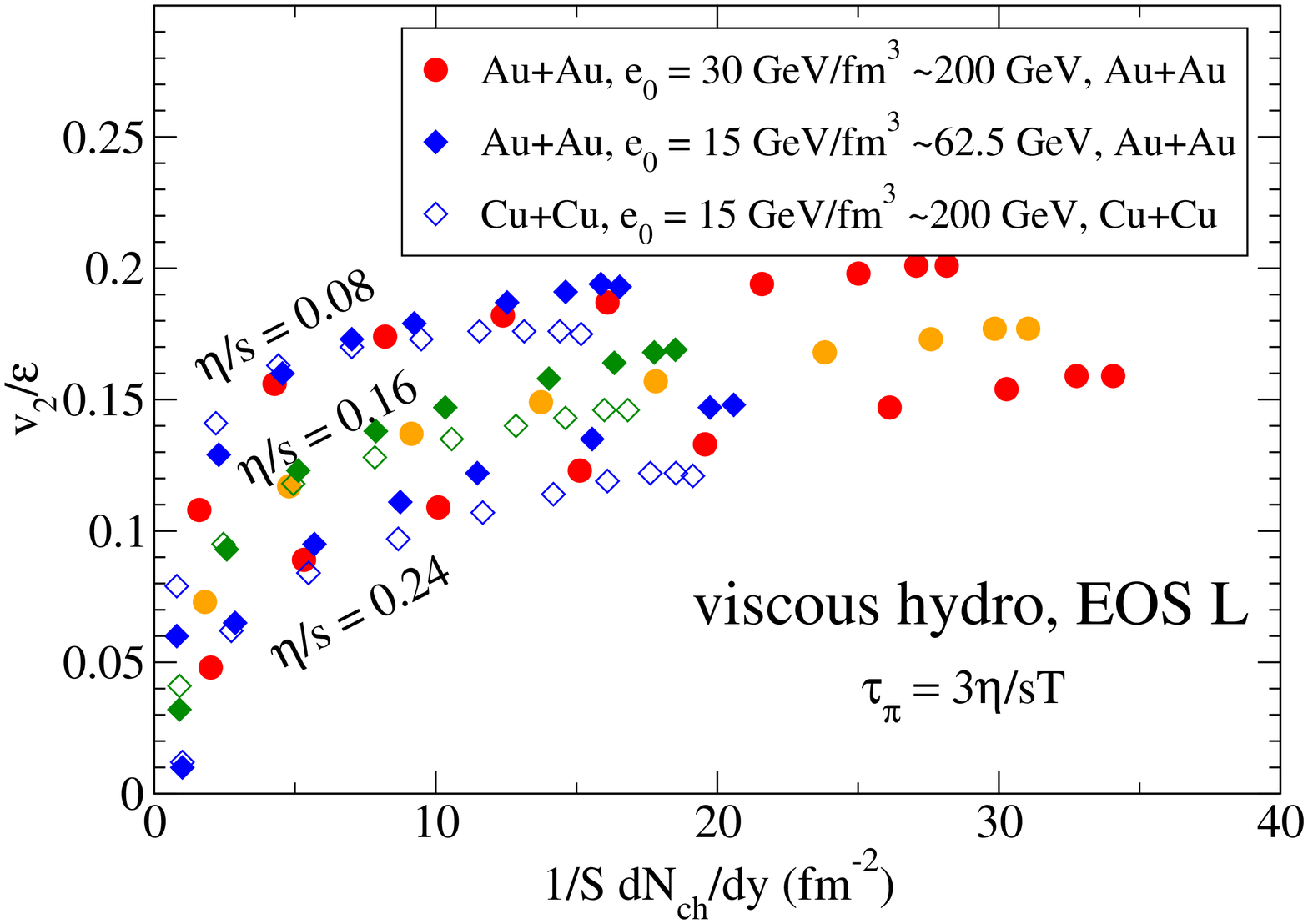,width=7.5cm,clip=} \\
\end{center}
\caption{Multiplicity scaling of elliptic flow. Shown is the charged 
         hadron elliptic flow normalized by the initial source 
         eccentricity as a function of charged hadron multiplicity
         per unit rapidity and nuclear overlap area at midrapidity.
         The left panel shows experimental data from AGS, SPS and 
         RHIC \protect\cite{NA49-03v2,VoloshinQM06} together with lines 
         indicating the results from ideal fluid dynamical calculations 
         with EOS~Q and a pure hadron gas equation of state without
         phase transition (EOS~H). The right panel shows results from
         (2+1)-d viscous hydrodynamics with EOS~L (see Fig.~\ref{F5}),
         for three fixed values for the specific shear viscosity 
         $\eta/s$ as indicated \protect\cite{Song:2008si}. See text for
         discussion.
\label{F24}
} 
\end{figure} 
%%%%%%%%%%%%%%%%%%%%%%%%%%%%%%%%%%%%%%%%%%%%%%%%%%%%%%%%%%%%%%%%%%%%%%%
%

%
The left panel of Fig.~\ref{F24} shows an empirical systematics that 
has become known as ``multiplicity scaling of elliptic 
flow'' \cite{NA49-03v2,VoloshinQM06}. 
The horizontal axis is the charged hadron multiplicity density
per unit rapidity and overlap area, which is proportional to the
initial entropy density \cite{Heinz:2004et}.
The vertical axis shows the $\pt$-integrated elliptic flow normalized
by the initial eccentricity.
The scale invariance of the ideal fluid dynamic equations implies that 
this ratio depends only on the squared speed of sound $c_s^2$
(see Eq.~(\ref{eq14})) if the elliptic flow is allowed to fully
develop and the fireball doesn't freeze out before.
This is reflected in the solid and dashed ``HYDRO'' lines indicated in 
the left panel.
Except for the steep drop on the left side which is due to premature
freeze-out when the initial entropy density is not large enough to
let the elliptic flow develop to saturation, these curves vary only
because the effective stiffness of the EOS probed during expansion 
depends on the initial entropy density and temperature \cite{KSH00}.
The dip in the HYDRO curve corresponding to EOS~Q arises from the dip
in the speed of sound near the quark hadron phase transition
(see Fig.~\ref{F5}).
The left diagram shows that the experimental data approach the ideal 
fluid dynamic limit at high multiplicity, but stay well below that limit 
at low multiplicity. 
They show an almost linear scaling with the charged multiplicity 
density which differs dramatically from the ideal fluid prediction.
In particular, the data show no sign of any structure related to 
the dip in the speed of sound near the phase transition.
The agreement of the experimental data with ideal fluid dynamics 
at high multiplicities relies on the fact that the measured elliptic
flow $v_2$ has been scaled with initial eccentricities calculated
from the Glauber model.
If the larger eccentricities predicted by the CGC model had been used, 
the experimental data would stay significantly below the ideal fluid 
prediction even at the highest multiplicities.
The right panel shows predictions for the eccentricity-scaled elliptic 
flow from {\em viscous} hydrodynamics \cite{Song:2008si}.
The calculations where done with constant specific shear viscosities
$\eta/s$, ranging from the minimal value $\eta/s=1/4\pi=0.08$ suggested 
by the KSS bound \cite{Kovtun:2004de} to three times that value,
as indicated.
For each of these values, one observes approximate ``multiplicity scaling'',
just as in the data: to first approximation, all dependence of 
$v_2/\epsilon$ on system size, collision energy, and impact parameter 
is through the multiplicity density $(1/S)(dN_\mathrm{ch}/dy)$ associated 
with these parameters. 
One sees that inclusion of viscous effects brings the theoretical
predictions closer to the data on the left than ideal fluid dynamics.
But it is also obvious that with a constant ratio $\eta/s$ agreement
with experiment cannot be achieved.
The data require larger $\eta/s$ values at low multiplicity densities
(low initial entropy densities) and smaller ratios at higher multiplicity
densities.
This indicates less specific shear viscosity in the hot QGP phase 
than at lower temperatures, especially at temperatures below $T_c$.
With Glauber eccentricities, as used in both panels of Fig.~\ref{F24},
``minimal'' shear viscosity $\eta/s=0.08$ seems to work fine at the
highest multiplicities.
However, as explained before, this would be quite different if the CGC 
model eccentricities were true:
They would lower $v_2/\epsilon$ by about 30\%, allowing for $\eta/s$ values
of up to 3 times the minimal value near the right end of the graph.
The viscous hydrodynamic calculations in the right panel of Fig.~\ref{F24} 
were done with EOS~Q which assumes chemical equilibrium in the HRG phase.
All the earlier caveats about hydrodynamic simulations of elliptic flow 
that do not correctly account for the non-equilibrium chemical composition 
in the hadron phase therefore apply.
Before this is corrected, the numbers quoted above should not be taken 
too seriously.
They do, however, give a feeling for the rough approximate size of 
the QGP shear viscosity to be expected from future quantitative comparisons
with experiment.
Clearly, we are not talking about shear viscosities that exceed the KSS 
bound by a factor 10 or more.
Indeed, it would be surprising if $(\eta/s)_\mathrm{QGP}$ turned out to be 
larger than about 3-5 times the KSS value once all physical effects are
properly included.
%

%%%%%%%%%%%%%%%%%%%%%%%%%%%%%%%%%%%%%%%%%%%%%%%%%%%%%%%%%%%%%%%%%%%%%
\sususe{Towards extracting the QGP viscosity}
\label{QGPvisc}
%%%%%%%%%%%%%%%%%%%%%%%%%%%%%%%%%%%%%%%%%%%%%%%%%%%%%%%%%%%%%%%%%%%%%

\begin{figure}[ht] 
\vspace*{-3mm}
\begin{center}
%\fbox{\rule[-15mm]{0cm}{3cm}{\em To create a place-holder}}
\hspace*{-5mm} 
            \epsfig{file=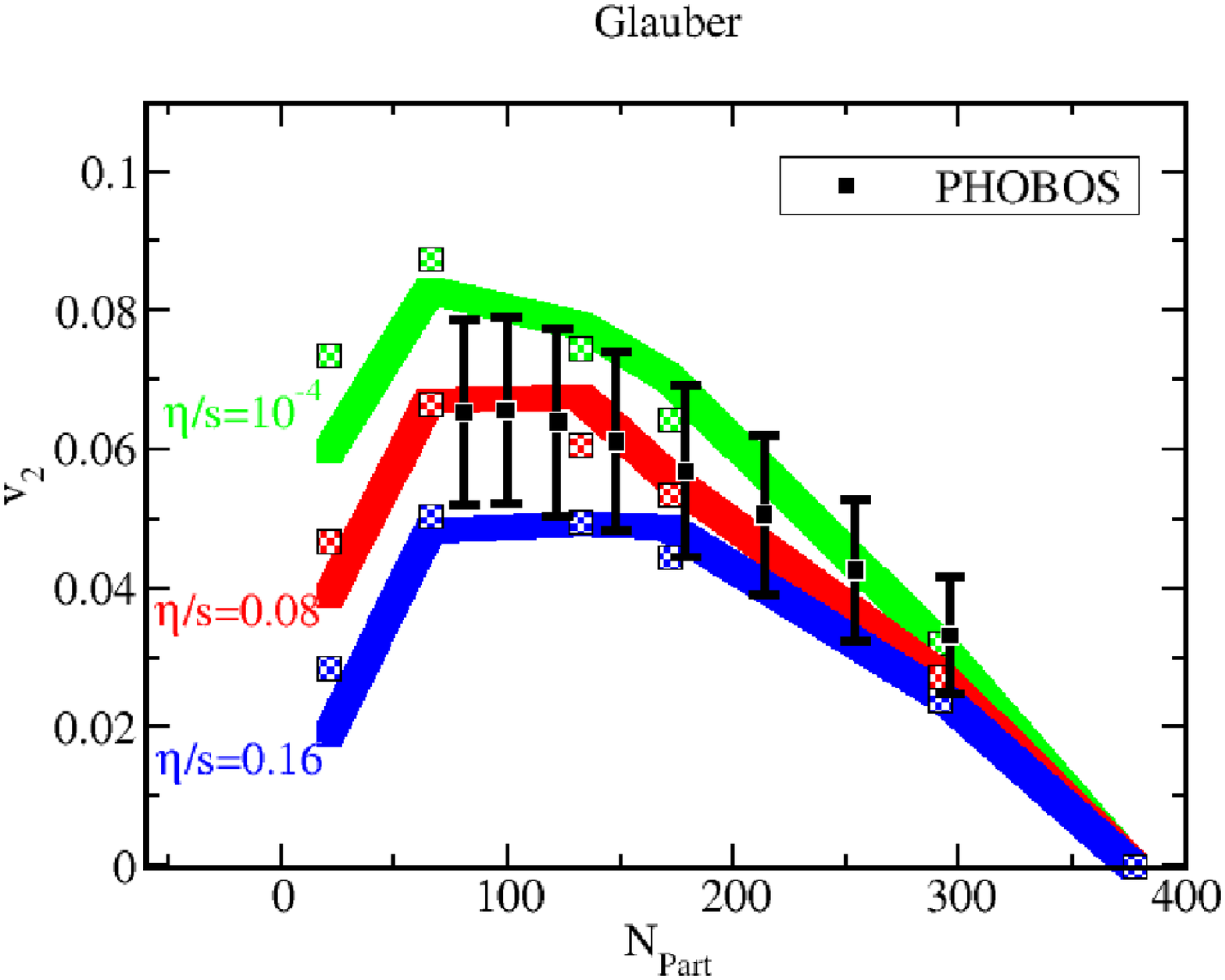,width=7.5cm,clip=}
            \epsfig{file=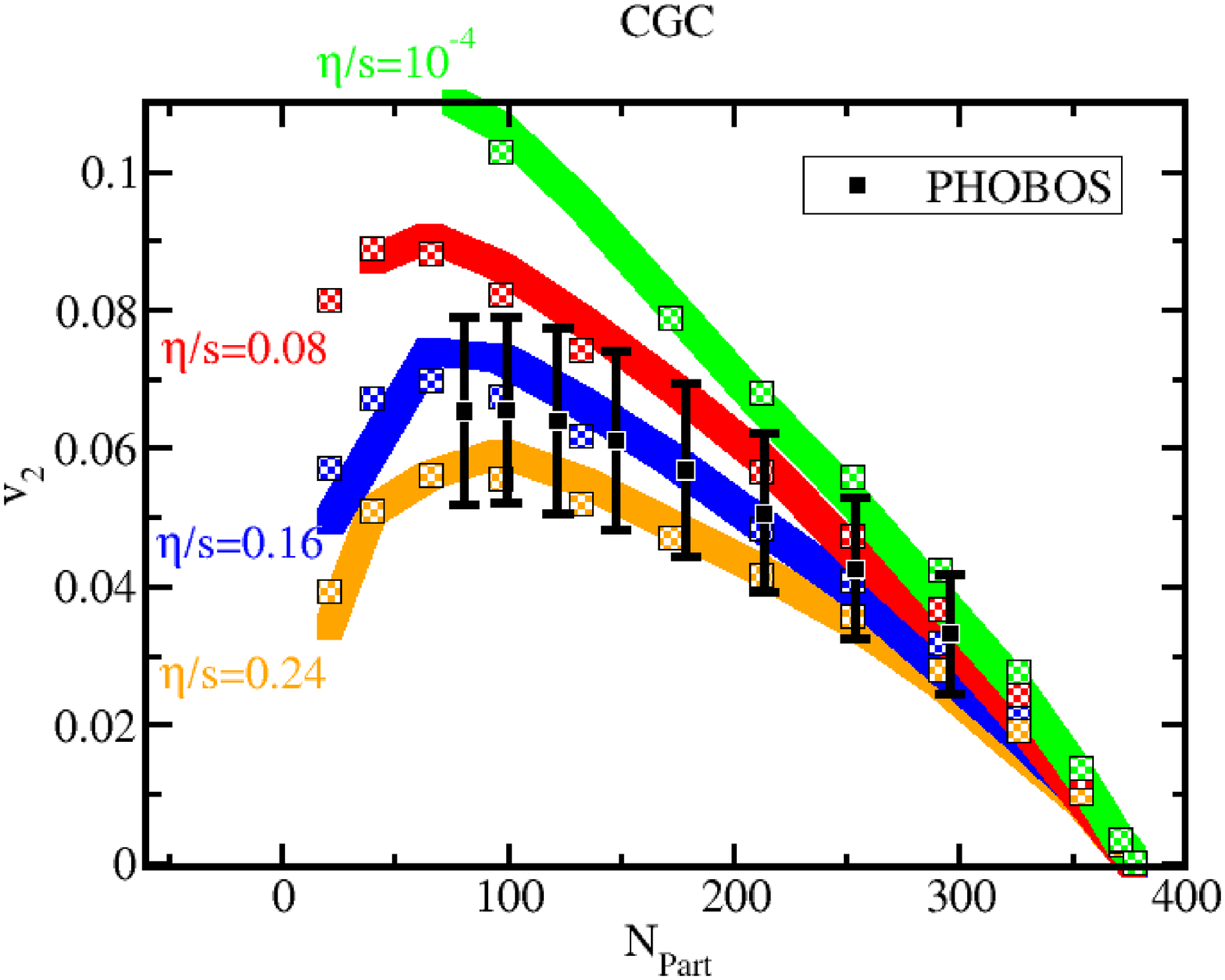,width=7.5cm,clip=} \\
            \epsfig{file=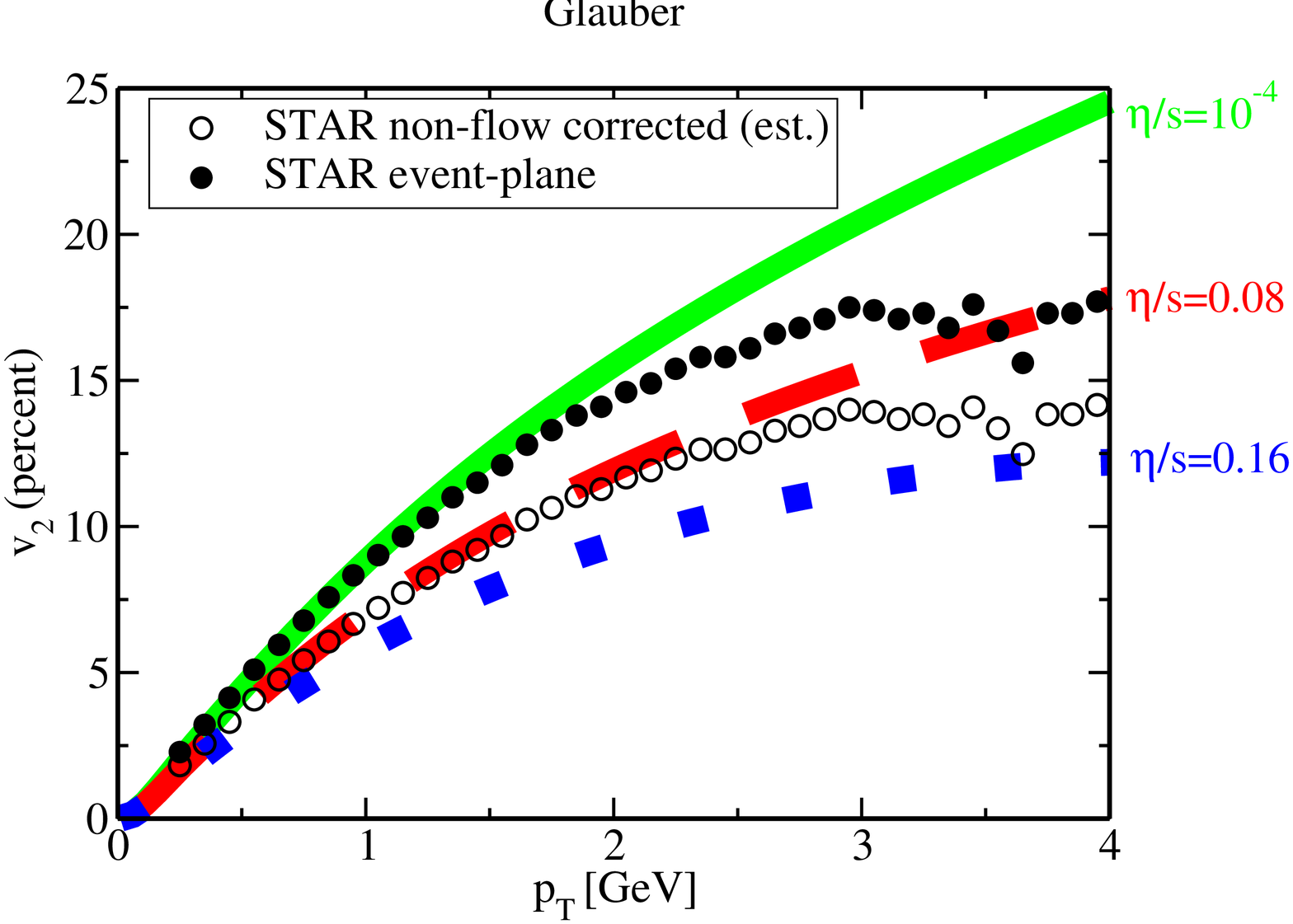,width=7.5cm,clip=}
            \epsfig{file=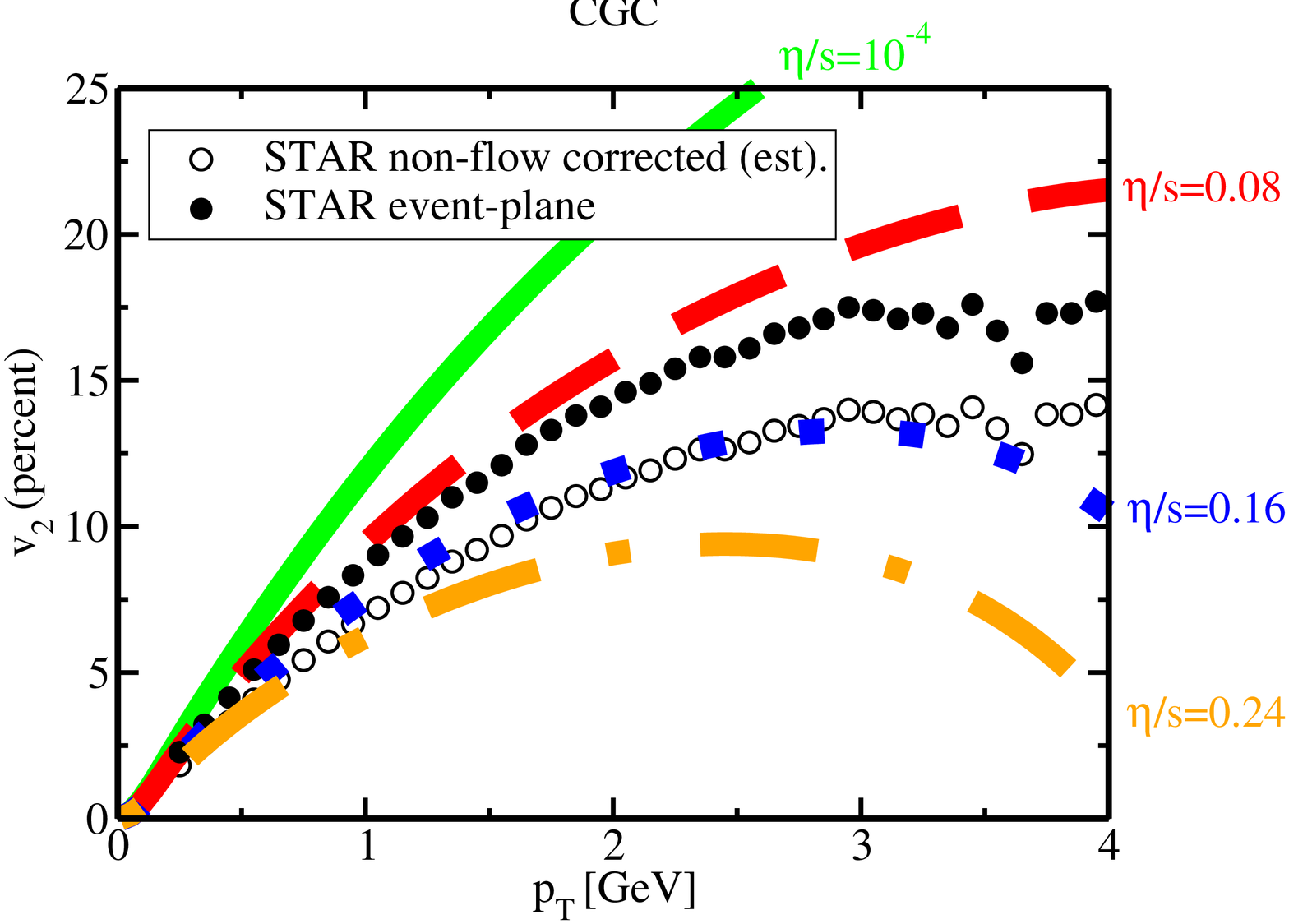,width=7.5cm,clip=} \\
\end{center}
\caption{Elliptic flow as a function of centrality (top row)
         and of transverse momentum (bottom row) for Glauber model
         (left column) and CGC (right column) initial conditions.
         Experimental points are from the PHOBOS \protect\cite{Alver:2007qw}
         and STAR \protect\cite{:2008ed} experiments for Au+Au collisions 
         at $\scm=200$\,GeV. The curves show (2+1)-d viscous hydrodynamic 
         simulations \protect\cite{Luzum:2008cw} with specific shear 
         viscosities $\eta/s$ as indicated.
\label{F25}
} 
%\vspace*{-4mm}
\end{figure}

%
We close this review with a discussion of a recent attempt to estimate
the QGP shear viscosity from midrapidity elliptic flow data in 
$200\,A$\,GeV Au+Au collisions, as a function of centrality 
and $\pt$ \cite{Luzum:2008cw}.
Figure~\ref{F25} shows (2+1)-d viscous hydrodynamic calculations
of charged hadron elliptic flow with Glauber model (left) and CGC (right)
initial conditions, for several constant values of $\eta/s$ as indicated.
The simulations used a lattice QCD equation of state above $T_c$ matched 
to a chemically equilibrated hadron gas below $T_c$.
For $v_2(\pt)$ two sets of data are shown: the originally published
data from the STAR Collaboration \cite{:2008ed} and a set of points
were all values were reduced by 20\% to approximately account for
``non-flow'' contributions in the data \cite{:2008ed}. 
%

%
%%%%%%%%%%%%%%%%%%%%%%%%%% Fig. 25 %%%%%%%%%%%%%%%%%%%%%%%%%%%%%%%%%%%
%%%%%%%%%%%%%%%%%%%%%%%%%%%%%%%%%%%%%%%%%%%%%%%%%%%%%%%%%%%%%%%%%%%%%%%
%

%
Depending on the assumed initial eccentricities and which set of data
one prefers, the comparison indicates a preferred range of
$0 < \eta/s < 0.2$.
In particular, for Glauber eccentricities and vanishing non-flow 
contributions in the data, there seems to be no room left for any 
non-zero shear viscosity at all (even though the simulation treats 
even the hadronic phase as an almost ideal fluid, which is 
known \cite{TLS01,Hirano:2005xf} to be incorrect)!
This is presumably an artifact of the incorrect chemistry of the hadronic 
phase assumed here.
But even after accounting for this, the window for QGP shear viscosity
is not large.
Even for the larger CGC eccentricities, $\eta/s$ values larger than about 
three times the KSS bound quickly become incompatible with the experimental 
data.
This conclusion gets stronger when one allows for additional effects from 
bulk viscosity which appear to further reduce the hydrodynamically predicted
elliptic flow \cite{Song:2008hj}.
%

%%%%%%%%%%%%%%%%%%%%%%%%%%%%%%%%%%%%%%%%%%%%%%%%%%%%%%%%%%%%%%%%%%%%%%%%%%%%
\section{Epilogue}
%%%%%%%%%%%%%%%%%%%%%%%%%%%%%%%%%%%%%%%%%%%%%%%%%%%%%%%%%%%%%%%%%%%%%%%%%%%%
%
Clearly, this is only the beginning of the story of the QGP shear 
viscosity, and its ending must be told in a future review.
But it is the story of trying to answer a question that 10 years
ago we didn't even know how to ask!
As the last century came to a close, the heavy-ion community was focussed 
on discovering the quark-gluon plasma; now, as we are about to complete
the first decade of the 21$^\mathrm{st}$ century, we are in the middle of 
a process of quantitatively extracting its transport properties.
In this endeavour, relativistic hydrodynamics of viscous fluids plays a 
key role.
With the advent of RHIC, hydrodynamics has found a permanent place
in the dynamical modelling of heavy-ion collisions.
For the first time in the history of high-energy physics, it has 
proven to be able to deliver quantitative explanations for experimental
observations.
Relativistic fluid dynamics is and will be the work horse of all future 
efforts to describe the dynamics of heavy-ion collisions.
For precise predictions, it must be carefully stitched together with
a reliable dynamical theory of the very early pre-equilibrium stage,
covering the first 1\,fm/$c$ or so, and a realistic hadronic rescattering
cascade for the late hadronic scattering and freeze-out stage, covering
the last few fm/$c$. 
During the 10-15\,fm/$c$ that lie in between these two points, hydrodynamics
rules.
%  

%%%%%%%%%%%%%%% ACKNOWLEDGEMENTS %%%%%%%%%%%%%%%

\section*{Acknowledgments}

This work was supported by the U.S. Department of Energy under 
grant DE-FG02-01ER41190. Fruitful discussions with and valuable 
comments from K. Dusling, K. Eskola, P. Huovinen, P. Kolb, H. Niemi,
S. Pratt, D. Rischke, P. Romatschke and H. Song, as well as the 
persistent encouragement by Reinhard Stock and the enduring patience 
of Christiane, Matthias and Michael (my wife and sons) are gratefully 
acknowledged.

%
%\include{expansion}
%\include{observables}
%\include{conclusions}

%%%%%%%%%%%%%%%%%%%%%%%%%%%%%%%%%%%%%%%%%%%%%%%%%%%%%%%%%%%%%%%%%%%%%%%%%%%%
% bibliography.tex
% last edited on 1/2/2009
%%%%%%%%%%%%%%%%%%%%%%%%%%%%%%%%%%%%%%%%%%%%%%%%%%%%%%%%%%%%%%%%%%%%%%%%%%%%

%
\end{document}